\documentclass[preprint,10pt]{elsarticle}

\usepackage{amsmath, amssymb, amsfonts}
\usepackage{cancel}
\usepackage{bm}
\usepackage{booktabs}
\usepackage{geometry}
\usepackage{graphicx}
\usepackage{jcomp}
\usepackage{framed,multirow}
\usepackage[utf8]{inputenc}
\usepackage{amsmath}
\usepackage{mathrsfs}
\usepackage{epsfig, setspace}
\usepackage{url}
\usepackage{graphicx, transparent, color}
\usepackage{algorithm}
\usepackage{algorithmicx}
\usepackage{etex}
\usepackage[bookmarks=true,hidelinks]{hyperref}
\usepackage{cleveref}
\usepackage{pdflscape} % Rotates single page into Landscape mode
\usepackage{silence}
\usepackage{ulem,cancel}

\newtheorem{example}{Example}[section]
\usepackage{float}
\usepackage{subfigure}% subcaptions for subfigures
\usepackage{subfigmat}% matrices of similar subfigures,

\usepackage{tikz}
\usepackage{capt-of}
\usepackage{setspace,lipsum}
\usepackage{lineno}
\modulolinenumbers[1]

\usepackage{listings}
\usepackage{xcolor}
\definecolor{aquamarine}{rgb}{0.5, 1.0, 0.83}
\definecolor{OliveGreen}{rgb}{0,0.6,0}
\newcommand{\Rone}[1]{#1}
\newcommand{\Rthree}[1]{#1}
\newcommand{\Rfour}[1]{#1}
\usepackage{colortbl,tcolorbox}
\definecolor{darkblue}{HTML}{1B3A5C}
\definecolor{midblue}{HTML}{2B6299}
\definecolor{lightblue}{HTML}{C4DAF0}
\definecolor{verylight}{HTML}{EDF4FB}
\definecolor{phase1bg}{HTML}{E8EEF5}
\definecolor{phase1dark}{HTML}{2B4A6F}
\definecolor{phase2bg}{HTML}{E5F0E0}
\definecolor{phase2dark}{HTML}{2F5420}
\definecolor{rowgray}{HTML}{F5F5F5}
\definecolor{headertext}{HTML}{FFFFFF}
\definecolor{ratingblue}{HTML}{1B3A5C}
\definecolor{mutedgray}{HTML}{999999}
\definecolor{tier1}{HTML}{1B5E20}
\definecolor{tier2}{HTML}{E65100}
\definecolor{tier3}{HTML}{757575}
\definecolor{critbg}{HTML}{FFF8E1}
\definecolor{critborder}{HTML}{F9A825}
\definecolor{keybg}{HTML}{E8F5E9}
\definecolor{keyborder}{HTML}{2E7D32}
\tcbuselibrary{skins, breakable}

\definecolor{codegreen}{rgb}{0,0.6,0}
\definecolor{codegray}{rgb}{0.5,0.5,0.5}
\definecolor{codepurple}{rgb}{0.58,0,0.82}
\definecolor{backcolour}{rgb}{0.95,0.95,0.92}

\lstdefinestyle{mystyle}{
    backgroundcolor=\color{backcolour},   
    commentstyle=\color{codegreen},
    keywordstyle=\color{magenta},
    numberstyle=\tiny\color{codegray},
    stringstyle=\color{codepurple},
    basicstyle=\ttfamily\footnotesize,
    breakatwhitespace=false,         
    breaklines=true,                 
    captionpos=b,                    
    keepspaces=true,                 
    numbers=left,                    
    numbersep=5pt,                  
    showspaces=false,                
    showstringspaces=false,
    showtabs=false,                  
    tabsize=2}
\begin{document}

\begin{frontmatter}

\title{Wave-Appropriate Reconstruction of Compressible Multiphase and Multicomponent Flows: Fully Conservative and Semi-Conservative Eigenstructures}

\author{Amareshwara Sainadh Chamarthi\corref{cor1}}
\ead{sainath@caltech.edu}
\cortext[cor1]{Corresponding author}
\address{Division of Engineering and Applied Science, California Institute of Technology, Pasadena, CA 91125, USA}

\begin{abstract}
Compressible multiphase and multicomponent solvers require an accurate representation of material interfaces without introducing spurious pressure oscillations. At these interfaces, pressure and velocity remain physically continuous, whereas density and the equation of state exhibit abrupt discontinuities. The standard methodology involves reconstructing primitive variables or their associated characteristic variables, which directly capture the continuity of these properties. While effective, this approach does not elucidate the failure mechanisms of conservative or semi-conservative reconstruction and forgoes the wave-decoupling advantages of characteristic decomposition. The present study addresses this issue by deriving the complete eigenstructure of the Allaire five-equation model using two distinct sets of variables. In the fully conservative (FC) formulation,  $\mathbf{U} = [\alpha_1\rho_1,\,\alpha_2\rho_2,\,\rho u,\,\rho v,\,\rho E,\, \alpha_1]^T$, the eigenvectors include a thermodynamic jump term $\Psi$ that compensates for compressibility mismatches between the two fluids at material contacts, thereby enforcing $dp = 0$ and $du = 0$ algebraically. In the semi-conservative (SC) formulation, $\mathbf{V} = [\alpha_1\rho_1,\, \alpha_2\rho_2,\,\rho u,\,\rho v,\,p,\,\alpha_1]^T$, pressure replaces total energy while the conserved momenta $\rho u$ and $\rho v$ are maintained; the volume-fraction eigenvector carries a structural zero in the pressure slot,  enforcing equilibrium without any thermodynamic correction. For one- and two-dimensional scenarios governed by stiffened-gas thermodynamics, explicit expressions for the left and right eigenvectors are derived for both variable sets. Both formulations directly satisfy Abgrall's equilibrium condition within the conservative framework, provided reconstruction is carried out in characteristic space. Reconstruction of any non-primitive variable in physical space leads to $\mathcal{O}(1)$ errors in pressure and velocity at material interfaces, irrespective of the variable set. A further consequence of the eigenvector structure is that the shear wave is structurally decoupled from all thermodynamic and interface fields in both FC and SC formulations, extending the central reconstruction argument for the shear characteristic, previously established for single-species flows, to compressible multiphase flows,  including gas-liquid configurations. One- and two-dimensional test cases confirm that FC and SC characteristic reconstruction provide oscillation-free, accurate results across a range of gas-gas and gas-liquid benchmarks.
\end{abstract}


\end{frontmatter}

%% ============================================================
%% SECTION 1 -- INTRODUCTION
%% ============================================================
\section{Introduction}
\label{sec:introduction}

Compressible multiphase and multicomponent flows are encountered in a variety of challenging applications, including underwater explosions and cavitation, shock-driven aerobreakup in propulsion systems, blast mitigation involving shock-particle dynamics, and interface instabilities in inertial confinement fusion~\cite{saurel2018diffuse}. In these contexts, strong shocks coexist with significant density contrasts, acoustic-impedance mismatches, and stiff, phase-dependent equations of state. Accurate simulation demands that material interfaces be resolved without introducing spurious pressure or velocity oscillations, as even minor violations of this criterion can lead to catastrophic instabilities in extended multi-dimensional computations.

Karni~\cite{Karni1994} demonstrated that solving the multicomponent Euler equations in fully conservative form generates spurious pressure oscillations at contact discontinuities between gases with different thermodynamic properties. This issue results from a discrete thermodynamic inconsistency in the energy equation when the ratio of specific heats changes across the interface. To resolve this, Abgrall~\cite{abgrall1996prevent} proposed the quasi-conservative approach, which augments the conservative energy equation with a non-conservative advection equation for $1/(\gamma-1)$, discretized consistently with the other conservation laws. This formulation ensures that the numerical scheme maintains uniform pressure and velocity across a material interface, thereby satisfying equilibrium conditions. Shyue~\cite{Shyue1998} extended the quasi-conservative framework to the stiffened gas equation of state, enabling its application to liquid-gas interfaces. Abgrall and Karni~\cite{abgrall2001computations} later provided a comprehensive review and introduced the double-flux method. The seven-equation model of Baer and Nunziato~\cite{BaerNunziato1986}, originally developed for deflagration-to-detonation transitions in granular materials, provides a compressible multiphase framework with separate conservation laws and independent pressures and velocities for each phase. Saurel and Abgrall~\cite{SaurelAbgrall1999} reformulated this model within a Godunov-type framework that incorporates stiff velocity and pressure relaxation. Kapila et al.~\cite{Kapila2001} performed an asymptotic reduction of the Baer-Nunziato model under mechanical equilibrium assumptions, resulting in a five-equation model that includes a volume fraction equation with a compaction source term and the non-monotonic Wood sound speed. Allaire, Clerc, and Kokh~\cite{allaire2002five} introduced a five-equation diffuse-interface model with a simplified volume-fraction advection equation and an isobaric closure, which is now widely adopted. The present study employs the five-equation model of Allaire et al.~\cite{allaire2002five} for simulations.

Once the equations are obtained, the choice of variables to be reconstructed also plays a significant role in maintaining the equilibrium conditions. At material interfaces, reconstruction must preserve the continuous velocity and pressure while avoiding nonlinear mixing of distinct wave families. To mitigate numerical diffusion of material interfaces, various  interface sharpening techniques have been developed, including anti-diffusive  schemes~\cite{Kokh2010,So2012}, gradient-based  compression~\cite{shukla2010interface}, and the THINC scheme of Xiao et  al.~\cite{Xiao2005}, which was first applied to compressible two-phase flows  by Shyue and Xiao~\cite{shyue2014eulerian}. \Rone{The implementation in the present work uses the symmetry-preserving algebraic THINC interface-value formulation of Wakimura, Takagi, and Xiao~\cite{wakimura2022symmetry}, built on the original hyperbolic-tangent THINC reconstruction.} \Rthree{Characteristic reconstruction suppresses nonlinear reconstruction errors at contacts. The present work applies wave-appropriate reconstruction to the FC and SC five-equation eigensystems: acoustic characteristics use an upwind-biased scheme, material/interface characteristics use THINC when sharpening is required, and the shear characteristic uses a central scheme away from shocks~\cite{hoffmann2024centralized,chamarthi2025wave,chamarthi2025physics}. The reconstruction variables determine this wave-wise treatment in inviscid multiphase flow. This point concerns the reconstruction, or projection, step in the standard projection--evolution decomposition of upwind schemes~\cite{huynh1995accurate}: the reconstruction step chooses which data are approximated inside each cell, whereas the evolution step evaluates an upwind flux or approximate Riemann solver using the resulting face states. Harten, Engquist, Osher, and Chakravarthy~\cite{Harten1987} established local characteristic projection for high-order ENO schemes to prevent nonlinear scalar reconstruction from mixing wave families. In multicomponent flow, however, this principle does not by itself specify the state variables in which the projection should be performed. Allaire et al.~\cite{allaire2002five} derived a conservative wave structure for the evolution step, but their second-order MUSCL scheme reconstructs the variables \((z_1\rho_1,z_2\rho_2,u,P,z)\), explicitly described there as primitive variables, and then obtains conservative interface states from them. The earlier interface-capturing formulations of Karni~\cite{Karni1994} and Abgrall~\cite{abgrall1996prevent} likewise use primitive-variable reconstruction. Johnsen and Colonius~\cite{johnsen2006implementation} then explicitly advocated high-order reconstruction of characteristic variables projected from primitive variables; Coralic and Colonius~\cite{coralic2014finite} followed the same primitive-characteristic approach. Wong et al.~\cite{wong2021positivity} likewise project primitive variables to characteristic variables before WENO interpolation. In the finite-difference setting, Nonomura et al.~\cite{Nonomura2012} found that primitive-variable interpolation was required for oscillation-free multicomponent results. Chamarthi~\cite{chamarthi2023gradient} likewise used characteristic variables projected from primitive variables in a gradient-based reconstruction for single- and multicomponent flows. Zhang et al.~\cite{zhang2024hybrid} reconstruct primitive variables. Li and Fu~\cite{li2024family} use a hybrid choice: primitive variables in interfacial cells and characteristic variables projected from primitive variables away from interfaces. Collis et al.~\cite{collis2026robust} construct ENO-type interpolation from the primitive basis \(\mathbf{W}=[T,Y_{pc},u,v,w,P]^T\), project that primitive basis to characteristic variables, and recover conservative face states only after reconstruction before the HLLC solve. Thus, these methods reconstruct primitive variables or characteristic variables formed from primitive-variable increments, rather than characteristic variables of the FC or SC state vectors themselves. The FC and SC shear eigenvectors derived here have zero thermodynamic and interface components: a shear-wave update leaves phasic densities, pressure, and volume fraction unchanged. This algebraic property permits central reconstruction of the shear characteristic in inviscid gas--gas and gas--liquid calculations without a reconstruction-level pressure or interface perturbation.} \Rthree{The SC eigensystem and the reconstruction-level comparison of FC and SC characteristic bases are developed within the conservative finite-volume framework. The analysis also identifies the wave-decoupling change introduced by direct physical-space reconstruction of non-primitive variables.}

\Rthree{Hoffmann, Chamarthi, and Frankel~\cite{hoffmann2024centralized} showed for single-species compressible flow that selectively centralizing non-acoustic characteristics can reduce dissipation and that conservative-variable characteristic reconstruction can outperform its primitive-variable counterpart. Chamarthi~\cite{chamarthi2025wave} extended the wave-appropriate idea to the five-equation model by assigning different reconstructions to acoustic, vorticity, and interface/contact content. Chamarthi~\cite{chamarthi2025physics} applied physics-based central treatment of tangential velocity to viscous multicomponent flow using primitive variables. The FC and SC characteristic bases derived here provide the corresponding shear-wave structure for inviscid gas--gas and gas--liquid flows.}

This work addresses the issue by deriving the five-equation eigenstructure in two sets of variables that explicitly reveal the wave structure. In the fully conservative (FC) formulation, the volume-fraction right eigenvector contains a thermodynamic jump term $\Psi$ that provides the precise internal-energy increment required to offset the compressibility mismatch at a material interface, thereby satisfying Abgrall's equilibrium condition within the conservative framework without transforming to primitive variables. \Rthree{This term is consistent with the conservative eigenstructure of Allaire et al.; with the same volume-fraction convention, the present notation corresponds to the same thermodynamic compensation written here as $\Psi$ rather than as the product of the thermodynamic vector and mixture coefficient used in that work. The FC derivation is therefore a stiffened-gas specialization and reconstruction-oriented recasting of that 1D result. It makes the material-wave $\Psi$ increment and its matched cancellation in the acoustic left eigenvectors explicit in the state basis being reconstructed, thereby providing a reconstruction-level proof of discrete pressure-equilibrium preservation for FC characteristic reconstruction. The comparison with the SC basis then makes clear why this correction disappears when pressure is stored as a state variable.} In the semi-conservative (SC) formulation, pressure replaces total energy while the conserved momenta $\rho u$ and $\rho v$ are retained. It will be shown that characteristic-space reconstruction is essential for either eigensystem to maintain its equilibrium guarantee at the discrete level. Direct reconstruction of any non-primitive variable in physical space bypasses wave decoupling. The principal contributions of this study are as follows:

\begin{itemize}
    \item \Rthree{Explicit eigenvalues and left- and right-eigenvector expressions are presented for the FC and SC variable sets in one- and two-dimensional stiffened-gas flows. The FC expressions recast Allaire's conservative eigensystem in the reconstruction basis used here and yield an explicit proof of the FC reconstruction-level equilibrium cancellation. The SC eigensystem and the reconstruction-level comparison of the two bases are derived explicitly for direct implementation.}

    \item It is proven that both eigensystems satisfy Abgrall's equilibrium condition, and it is demonstrated that characteristic-space reconstruction is necessary for this to hold at the discrete level.
	\item The shear wave is shown to be structurally decoupled from all thermodynamic and interface fields in both FC and SC formulations, enabling central reconstruction of the shear characteristic in multiphase flows, including gas-liquid configurations. This extends the argument of~\cite{hoffmann2024centralized} beyond the single-species case. The SC shear characteristic is a density-weighted tangential-momentum perturbation, not the primitive tangential velocity; consequently, the two central reconstructions are not algebraically equivalent in variable-density flow. The extension is non-trivial because the additional volume-fraction wave could in principle couple to the shear characteristic; the eigenvector analysis shows algebraically that it does not. This is consistent, for the shear characteristic specifically, with the non-dissipative implementations of~\cite{jain2020conservative} and~\cite{jain2022kinetic}, which apply a fully central discretization to all operators in the five-equation model for compressible gas--liquid flows and demonstrate stability in turbulent and acoustic regimes; the acoustic characteristic may still require a dissipative bias in certain flow conditions~\cite{chamarthi2026wave}.
    \item Both formulations are validated on a suite of canonical one- and two-dimensional multiphase benchmarks, confirming that characteristic reconstruction in FC and SC achieves the accuracy of primitive-variable schemes while preserving the conservation properties of the finite-volume framework.
\end{itemize}

The structure of the paper is as follows. Section~\ref{sec:governing} reviews  the governing equations and the two state-vector formulations.  Section~\ref{sec:fc} derives the FC eigenstructure and establishes Abgrall  equilibrium preservation. Section~\ref{sec:sc} presents the SC eigenstructure  and the corresponding structural guarantee. Section~\ref{sec:comparison}  compares the two formulations. Section~\ref{sec:reconstruction} details the  wave-appropriate characteristic reconstruction algorithm.  Section~\ref{sec:results} presents the numerical results, and Section~\ref{sec:conclusions} provides concluding remarks.
%% ============================================================
%%  SECTION 2 -- GOVERNING EQUATIONS
%% ============================================================
\section{Governing Equations and Variable Formulations}
\label{sec:governing}

\subsection{The Five-Equation Model}

We consider two compressible immiscible phases $k = 1,2$ with volume fractions~$\alpha_k$, phasic densities~$\rho_k$, a common velocity field~$(u,v)$, and a common pressure~$p$~\cite{allaire2002five}, which is as follows:
\begin{align}
\frac{\partial(\alpha_1\rho_1)}{\partial t} + \nabla\cdot(\alpha_1\rho_1\mathbf{u}) &= 0, \label{eq:m1}\\
\frac{\partial(\alpha_2\rho_2)}{\partial t} + \nabla\cdot(\alpha_2\rho_2\mathbf{u}) &= 0, \label{eq:m2}\\
\frac{\partial(\rho\mathbf{u})}{\partial t} + \nabla\cdot(\rho\mathbf{u}\otimes\mathbf{u} + p\mathbf{I}) &= \mathbf{0}, \label{eq:mom}\\
\frac{\partial(\rho E)}{\partial t} + \nabla\cdot\bigl((\rho E + p)\mathbf{u}\bigr) &= 0, \label{eq:energy}\\
\frac{\partial\alpha_1}{\partial t} + \mathbf{u}\cdot\nabla\alpha_1 &= 0, \label{eq:alpha}
\end{align}
with constraints $\alpha_1+\alpha_2=1$, mixture density $\rho = \alpha_1\rho_1 + \alpha_2\rho_2$, and total energy $\rho E = \rho e + \tfrac{1}{2}\rho|\mathbf{u}|^2$. The 1D specialization retains Eqs.~\eqref{eq:m1}--\eqref{eq:alpha} with $v=0$ and all $y$-derivatives dropped. The volume-fraction equation~\eqref{eq:alpha} is non-conservative: it is written in advective rather than divergence form. The non-conservative advection form~\eqref{eq:alpha} is the specific choice that makes the five-equation system compatible with Abgrall's equilibrium condition~\cite{abgrall1996prevent}. Each phase obeys the stiffened-gas equation of state,
\begin{equation}\label{eq:eos}
p = (\gamma_k - 1)\rho_k e_k - \gamma_k\pi_{\infty,k},
\end{equation}
where $\gamma_k$ is the adiabatic exponent and $\pi_{\infty,k}$ is the reference pressure (zero for ideal gases). The mixture internal energy satisfies $\rho e = \alpha_1\rho_1 e_1 + \alpha_2\rho_2 e_2$, leading to the mixture parameters
\begin{equation}\label{eq:mixture}
\frac{1}{\gamma_m - 1} = \frac{\alpha_1}{\gamma_1-1} + \frac{\alpha_2}{\gamma_2-1},\qquad
\frac{\gamma_m\pi_{\infty,m}}{\gamma_m-1} = \frac{\alpha_1\gamma_1\pi_{\infty,1}}{\gamma_1-1} + \frac{\alpha_2\gamma_2\pi_{\infty,2}}{\gamma_2-1},
\end{equation}
where subscript $m$ denotes mixture quantities. The mixture sound speed is
\begin{equation}\label{eq:soundspeed}
c^2 = \frac{\gamma_m(p+\pi_{\infty,m})}{\rho} = (\gamma_m - 1)\Bigl(H - \tfrac{1}{2}|\mathbf{u}|^2\Bigr),
\end{equation}
where $H = (\rho E + p)/\rho$ is the specific total enthalpy and we define the enthalpy scaling $\chi = (\gamma_m - 1)/c^2 = 1/h$, with $h = c^2/(\gamma_m - 1)$. The quantity $\chi$ appears throughout the left eigenvector matrices as it converts pressure perturbations into entropy or acoustic wave amplitudes.

\subsection{Two State-Vector Formulations}

In this subsection, the two state-vector choices, FC and SC, and various notations are introduced. The state vector choice determines the structure of the flux Jacobian, the form of the eigenvectors, and whether pressure equilibrium at material interfaces is enforced algebraically or structurally. Both choices produce valid hyperbolic systems, are compatible with conservative finite-volume schemes, and differ only in how the thermodynamic information is encoded and in the resulting eigenvector structure.
\paragraph{Fully conservative (FC) formulation.}
\begin{equation}\label{eq:FC}
\mathbf{U} = \bigl[\alpha_1\rho_1,\; \alpha_2\rho_2,\; \rho u,\; \rho v,\; \rho E,\; \alpha_1\bigr]^T.
\end{equation}
All components are conserved quantities of the system Eqs.~\eqref{eq:m1}--\eqref{eq:energy}, plus $\alpha_1$ which is advected. The total energy $\rho E$ stores the complete thermodynamic state, and pressure must be recovered through the mixture EOS at each evaluation point. The advantage of this formulation is that it is fully conservative across shocks: the Rankine-Hugoniot jump conditions are automatically satisfied for all quantities. The challenge is that the volume fraction $\alpha_1$ enters the EOS nonlinearly, so the Jacobian $\partial \mathbf{F}/\partial \mathbf{U}$ contains a thermodynamic coupling term in the volume-fraction column. This term, the quantity $\Psi$ derived in Section~\ref{sec:fc_proof}, must be included exactly for the FC eigensystem to satisfy Abgrall's condition. 
\paragraph{Semi-conservative (SC) formulation.}
\begin{equation}\label{eq:SC}
\mathbf{V} = \bigl[\alpha_1\rho_1,\; \alpha_2\rho_2,\; \rho u,\; \rho v,\; p,\; \alpha_1\bigr]^T,
\end{equation}
in which the total energy is replaced by the thermodynamic pressure. We define partial densities $m_k = \alpha_k\rho_k$, mixture density $\rho = m_1 + m_2$, and mass fractions $Y_k = m_k/\rho$. The SC formulation retains the conserved momentum components $\rho u$ and $\rho v$ while replacing total energy with pressure. This retention is deliberate and has structural consequences. Table~\ref{tab:thermo} collects the thermodynamic variables used throughout the eigenstructure derivations. The specific total enthalpy $H$ and the enthalpy scaling $\chi$ appear repeatedly in the left eigenvector matrices for both FC and SC formulations. The relation $\chi = (\gamma_m - 1)/c^2$ converts a pressure perturbation into a characteristic wave amplitude and is entirely determined by the mixture thermodynamics.

\begin{table}[H]
\centering
\caption{Thermodynamic variables used in the eigenstructure derivations. $\kappa = \tfrac{1}{2}|\mathbf{u}|^2$ denotes specific kinetic energy.}
\label{tab:thermo}
\begin{tabular}{llll}
\toprule
Symbol & Name & Definition & Relation \\
\midrule
$e$ & Specific internal energy & Thermal energy per unit mass & $e = p/[\rho(\gamma_m-1)] + \pi_{\infty,m}/\rho$ \\
$E$ & Specific total energy & $e + \tfrac{1}{2}|\mathbf{u}|^2$ & $E = e + \kappa$ \\
$h$ & Specific static enthalpy & $e + p/\rho$ & $h = \gamma_m(p+\pi_{\infty,m})/[\rho(\gamma_m-1)]$ \\
$H$ & Specific total enthalpy & $E + p/\rho$ & $H = h + \kappa$ \\
$\chi$ & Enthalpy scaling & $(\gamma_m-1)/c^2$ & $\chi = 1/h$ \\
\bottomrule
\end{tabular}
\end{table}

In two dimensions, the characteristic decomposition is performed with respect to a cell-face normal direction. For a face with unit outward normal $\mathbf{n} = (n_x, n_y)$ and tangent $\boldsymbol{\ell} = (-n_y, n_x)$, define the normal and tangential velocity components and the total speed squared as
\begin{equation}\label{eq:aux2d}
u_n = u\,n_x + v\,n_y,\qquad u_t = -u\,n_y + v\,n_x,\qquad q^2 = u^2 + v^2.
\end{equation}
The normal velocity $u_n$ is the velocity component in the direction across the face and determines the eigenvalues $u_n \pm c$ and $u_n$ of the 2D system. The tangential velocity $u_t$ appears in the shear eigenvector, which carries tangential velocity information along the face. In the FC formulation, $u_t$ also appears in the energy row of the shear eigenvector but does not induce any pressure perturbation. In the SC formulation, $u_t$ appears only in the momentum rows of the shear eigenvector, making the structural isolation of the shear wave from the thermodynamic variables even more transparent.

\section{Eigenstructures}

%% ============================================================
%% SECTION 3 -- CONSERVATIVE EIGENSTRUCTURE
%% Reorganized: Psi is derived inside the proof where it is needed.
%% The Johnsen failure mode appears as a closing remark.
%% ============================================================
\subsection{Conservative Eigenstructure}
\label{sec:fc}

The eigenstructure of the five-equation model in the fully conservative variable set is more complex than that of the single-species Euler equations because the system has two additional degrees of freedom: the two phasic partial densities, which introduce two additional entropy waves at the repeated eigenvalue $\lambda = u$. The central new feature is the thermodynamic coupling term $\Psi$ that appears in the volume-fraction right eigenvector. The derivations below present the complete 1D and 2D eigensystems with explicit expressions for all left and right eigenvectors, suitable for direct implementation. The proof that these eigensystems satisfy Abgrall's condition~\cite{abgrall1996prevent} follows in Section~\ref{sec:fc_proof}.

\subsubsection{1D Conservative Eigenstructure}

For the 1D state vector
\[
\mathbf{U} = [\alpha_1\rho_1,\; \alpha_2\rho_2,\; \rho u,\; \rho E,\; \alpha_1]^T,
\]
the Jacobian $\partial\mathbf{F}/\partial\mathbf{U}$ has eigenvalues
\begin{equation}\label{eq:fc1d_evals}
\lambda_1 = u - c,\quad
\lambda_2 = \lambda_3 = \lambda_4 = u,\quad
\lambda_5 = u + c.
\end{equation}
The two acoustic eigenvalues $u \pm c$ correspond to right- and left-running acoustic waves. The three repeated eigenvalues $\lambda = u$ correspond to waves that are convected with the flow: one for each phasic partial density (the composition waves) and one for the volume fraction. Together, these three degenerate waves carry all of the material-interface physics, and the structure of their eigenvectors determines whether Abgrall's condition can be satisfied. The wave families at $\lambda = u$ comprise two partial-density (entropy) waves and one volume-fraction wave.

\paragraph{Right eigenvector matrix $\mathbf{R}$.}

The columns of $\mathbf{R}$ are ordered as $[\lambda_1,\,\mathrm{entr}_1,\,\mathrm{entr}_2,\,\mathrm{VF},\,\lambda_5]$.
\begin{equation}\label{eq:fc1d_R}
\mathbf{R} =
\begin{bmatrix}
Y_1 & 1 & 0 & 0 & Y_1 \\
Y_2 & 0 & 1 & 0 & Y_2 \\
u-c & u & u & 0 & u+c \\
H-uc & \dfrac{u^2}{2} & \dfrac{u^2}{2} & \Psi & H+uc \\
0 & 0 & 0 & 1 & 0
\end{bmatrix},
\end{equation}
where $\Psi$ is the thermodynamic jump term derived and explained in Section~\ref{sec:fc_proof}. Its appearance in the energy row of the volume-fraction eigenvector (column~4) is the important aspect of the conservative formulation. The acoustic columns (1 and 5) have the standard single-species Euler structure in their momentum and kinetic-energy entries, with the mass fractions $Y_k$ appearing in the partial-density rows. The entropy wave columns (2 and 3) carry a unit change in one phasic partial density at constant velocity, with the kinetic energy $u^2/2$ appearing in the energy row to reflect that the density change occurs at constant velocity. The volume-fraction column (4) carries a unit change in $\alpha_1$ at constant partial densities and velocity; its only non-zero entry besides the volume-fraction row is the thermodynamic term $\Psi$ in the energy row, which correctly represents the change in mixture internal energy at constant pressure.

\paragraph{Left eigenvector matrix $\mathbf{L}$.}

The left eigenvectors are the rows of $\mathbf{L}$ and satisfy $\mathbf{L}\mathbf{R} = \mathbf{I}$. The biorthogonality condition uniquely determines $\Psi$ in the volume-fraction column of $\mathbf{L}$: the acoustic rows must be orthogonal to the VF column of $\mathbf{R}$, which requires $-\chi\Psi/2$ in the volume-fraction position of each acoustic left eigenvector. This equal-and-opposite sign is what produces the exact cancellation that enforces pressure equilibrium, as demonstrated in Section~\ref{sec:fc_proof}.

\begin{equation}\label{eq:fc1d_L}
\mathbf{L} =
\begin{bmatrix}
\dfrac{1}{2}\!\left(\chi\dfrac{u^2}{2}+\dfrac{u}{c}\right) &
\dfrac{1}{2}\!\left(\chi\dfrac{u^2}{2}+\dfrac{u}{c}\right) &
-\dfrac{1}{2}\!\left(\chi u+\dfrac{1}{c}\right) &
\dfrac{\chi}{2} &
-\dfrac{\chi}{2}\Psi \\[4pt]
1-\chi Y_1\dfrac{u^2}{2} &
-\chi Y_1\dfrac{u^2}{2} &
\chi Y_1 u &
-\chi Y_1 &
\chi Y_1\Psi \\[4pt]
-\chi Y_2\dfrac{u^2}{2} &
1-\chi Y_2\dfrac{u^2}{2} &
\chi Y_2 u &
-\chi Y_2 &
\chi Y_2\Psi \\[4pt]
0 & 0 & 0 & 0 & 1 \\[4pt]
\dfrac{1}{2}\!\left(\chi\dfrac{u^2}{2}-\dfrac{u}{c}\right) &
\dfrac{1}{2}\!\left(\chi\dfrac{u^2}{2}-\dfrac{u}{c}\right) &
-\dfrac{1}{2}\!\left(\chi u-\dfrac{1}{c}\right) &
\dfrac{\chi}{2} &
-\dfrac{\chi}{2}\Psi
\end{bmatrix}.
\end{equation}
One verifies that $\mathbf{L}\mathbf{R} = \mathbf{I}$ and
$\mathbf{L}\mathbf{A} = \boldsymbol{\Lambda}\mathbf{L}$ with
$\boldsymbol{\Lambda} = \mathrm{diag}(u-c,\,u,\,u,\,u,\,u+c)$.

The key structural feature of $\mathbf{L}$ is the entry $-\chi\Psi/2$ in the last column (the volume-fraction column) of the acoustic rows (rows 1 and 5). This entry has the opposite sign to the corresponding entry $+\Psi$ in the energy row of the volume-fraction column of $\mathbf{R}$. When a material-interface jump is projected onto the acoustic characteristics, these two entries cancel exactly, giving zero acoustic wave amplitude. The cancellation is not numerical coincidence: it is imposed by the biorthogonality condition $\mathbf{L}\mathbf{R} = \mathbf{I}$ once $\Psi$ is correctly identified as the internal-energy change at constant pressure. The full proof is given in Section~\ref{sec:fc_proof}.

\subsubsection{2D Conservative Eigenstructure}

The extension to two spatial dimensions introduces a shear wave associated with the tangential velocity component.  For the 2D state vector
\[
\mathbf{U} = [\alpha_1\rho_1,\; \alpha_2\rho_2,\; \rho u,\; \rho v,\; \rho E,\; \alpha_1]^T,
\]
the flux Jacobian in the $\mathbf{n}$-direction has eigenvalues
\begin{equation}\label{eq:fc2d_evals}
\lambda_1 = u_n - c,\quad
\lambda_2 = \lambda_3 = \lambda_4 = \lambda_5 = u_n,\quad
\lambda_6 = u_n + c.
\end{equation}
The four degenerate waves at $u_n$ are two entropy waves, one shear
wave, and one volume-fraction wave.

\paragraph{Right eigenvector matrix $\mathbf{R}$.}

The columns are ordered as $[\lambda_1,\,\mathrm{entr}_1,\,\mathrm{entr}_2,\,\mathrm{shear},\,\mathrm{VF},\,\lambda_6]$.

\begin{equation}\label{eq:fc2d_R}
\mathbf{R} =
\begin{bmatrix}
Y_1 & 1 & 0 & 0 & 0 & Y_1 \\
Y_2 & 0 & 1 & 0 & 0 & Y_2 \\
u-cn_x & u & u & \ell_x & 0 & u+cn_x \\
v-cn_y & v & v & \ell_y & 0 & v+cn_y \\
H-cu_n & \dfrac{q^2}{2} & \dfrac{q^2}{2} & u_t & \Psi & H+cu_n \\
0 & 0 & 0 & 0 & 1 & 0
\end{bmatrix},
\end{equation}
where $u_n = u n_x + v n_y$, $u_t = -u n_y + v n_x$, $q^2 = u^2+v^2$, and $(\ell_x,\ell_y) = (-n_y,n_x)$. 
\paragraph{Left eigenvector matrix $\mathbf{L}$.}

\begin{equation}\label{eq:fc2d_L}
\mathbf{L} =
\begin{bmatrix}
\dfrac{1}{2}\!\left(\chi\dfrac{q^2}{2}+\dfrac{u_n}{c}\right) &
\dfrac{1}{2}\!\left(\chi\dfrac{q^2}{2}+\dfrac{u_n}{c}\right) &
-\dfrac{1}{2}\!\left(\chi u+\dfrac{n_x}{c}\right) &
-\dfrac{1}{2}\!\left(\chi v+\dfrac{n_y}{c}\right) &
\dfrac{\chi}{2} &
-\dfrac{\chi}{2}\Psi \\[3pt]
1-\chi Y_1\dfrac{q^2}{2} &
-\chi Y_1\dfrac{q^2}{2} &
\chi Y_1 u &
\chi Y_1 v &
-\chi Y_1 &
\chi Y_1\Psi \\[3pt]
-\chi Y_2\dfrac{q^2}{2} &
1-\chi Y_2\dfrac{q^2}{2} &
\chi Y_2 u &
\chi Y_2 v &
-\chi Y_2 &
\chi Y_2\Psi \\[3pt]
-u_t & -u_t & \ell_x & \ell_y & 0 & 0 \\[3pt]
0 & 0 & 0 & 0 & 0 & 1 \\[3pt]
\dfrac{1}{2}\!\left(\chi\dfrac{q^2}{2}-\dfrac{u_n}{c}\right) &
\dfrac{1}{2}\!\left(\chi\dfrac{q^2}{2}-\dfrac{u_n}{c}\right) &
-\dfrac{1}{2}\!\left(\chi u-\dfrac{n_x}{c}\right) &
-\dfrac{1}{2}\!\left(\chi v-\dfrac{n_y}{c}\right) &
\dfrac{\chi}{2} &
-\dfrac{\chi}{2}\Psi
\end{bmatrix}.
\end{equation}
Again one verifies $\mathbf{L}\mathbf{R} = \mathbf{I}$. The 2D shear right eigenvector (column 4 of Eq.~\eqref{eq:fc2d_R}) is $[0,\,0,\,\ell_x,\,\ell_y,\,u_t,\,0]^T$. The first two entries (phasic densities) and the last entry (volume fraction) are zero. The energy entry $u_t$ is non-zero but, as shown in Section~\ref{sec:comparison}, a perturbation along this eigenvector does not change the pressure because the momentum change is purely tangential. The $\Psi$ entry again appears only in the volume-fraction column (column 5), in the energy row. The 2D left eigenvector matrix preserves the $-\chi\Psi/2$ entry in the volume-fraction column of each acoustic row, ensuring the same cancellation mechanism as in 1D.

The reason to present 2D eigenvectors in detail, separately (for both FC and SC), is to show that the system has six wave families in each spatial direction: two acoustic waves, two entropy waves, one shear wave, and one volume-fraction wave. The shear wave has eigenvalue $u_n$ and its right eigenvector has zero entries in all density, pressure, and volume-fraction rows. \Rfour{\textbf{This structural sparsity applies to the local characteristic projection and reconstruction step: entropy-wave and volume-fraction amplitudes do not enter the shear component, and the shear amplitude does not reconstruct density, pressure, or volume fraction. A central reconstruction applied only to the shear characteristic therefore introduces no reconstruction-level density or pressure perturbation. The subsequent HLLC flux evaluation and Runge--Kutta update remain fully nonlinear and may couple the evolved fields through the governing equations \cite{hoffmann2024centralized,chamarthi2026wave}.}} Assuming the characteristic variables are defined as, $\mathbf{W}$, where  $\mathbf{W} = \mathbf{L} \mathbf{U}$, it can be seen from the right eigenvectors that 

\begin{equation}\label{eq:density_decomp}
  \boxed{\alpha_1\rho_1 = W_1 + W_2 + W_6.}
\end{equation}
 Further details are presented in Ref.~\cite{chamarthi2026wave}. The Abgrall condition~\cite{abgrall1996prevent} requires that a uniform pressure $p$ and velocity $\mathbf{u}$ be preserved exactly across a moving material interface at the discrete level. We now prove that the conservative eigensystem satisfies this condition. The proof is self-contained: $\Psi$ is derived where it is needed rather than introduced separately.

\paragraph{Step 1: Admissible contact-wave jump.}\label{sec:fc_proof}

Consider a material interface moving at speed $u_n$ across which the partial
densities $m_k = \alpha_k\rho_k$ and the volume fraction $\alpha_1$ change
arbitrarily, while $p$ and $\mathbf{u}$ remain constant. The corresponding
jump in the conservative state vector $\mathbf{U}$ is determined component
by component as follows. The partial-density components $\Delta m_1$ and
$\Delta m_2$ are unconstrained. The momentum component follows from $du = 0$:
$\Delta(\rho u) = u\,\Delta\rho$. The volume-fraction component is
unconstrained: $\Delta\alpha_1$.

The energy component requires separate treatment. Decomposing
$\rho E = \rho e + \tfrac{1}{2}\rho u^2$ and noting that $u$ is constant,
the kinetic contribution is $\Delta(\tfrac{1}{2}\rho u^2) =
\tfrac{1}{2}u^2\Delta\rho$. To determine $\Delta(\rho e)$ at constant $p$,
we write the stiffened-gas mixture internal energy as
\begin{equation}\label{eq:rhoe_form}
\rho e = p\,\Gamma_m(\alpha_1) + \mathcal{C}_m(\alpha_1),
\end{equation}
where
\begin{equation}\label{eq:GammaC}
\Gamma_m(\alpha_1) = \frac{\alpha_1}{\gamma_1-1}
                     + \frac{1-\alpha_1}{\gamma_2-1},\qquad
\mathcal{C}_m(\alpha_1) =
\frac{\alpha_1\gamma_1\pi_{\infty,1}}{\gamma_1-1}
+ \frac{(1-\alpha_1)\gamma_2\pi_{\infty,2}}{\gamma_2-1}.
\end{equation}
Differentiating Eq.~\eqref{eq:rhoe_form} at constant $p$ gives
\[
d(\rho e)\big|_p =
\underbrace{\left[p\,\frac{d\Gamma_m}{d\alpha_1}
+ \frac{d\mathcal{C}_m}{d\alpha_1}\right]}_{\displaystyle\equiv\;\Psi}
d\alpha_1,
\]
which defines the \emph{thermodynamic jump term}
\begin{equation}\label{eq:Psi}
\Psi \;\equiv\;
p\left(\frac{1}{\gamma_1-1} - \frac{1}{\gamma_2-1}\right)
+ \left(\frac{\gamma_1\pi_{\infty,1}}{\gamma_1-1}
  - \frac{\gamma_2\pi_{\infty,2}}{\gamma_2-1}\right).
\end{equation}
Thus $d(\rho e) = \Psi\,d\alpha_1$: the term $\Psi$ measures the rate of
change of mixture internal energy with respect to volume fraction at
constant pressure. For ideal gases ($\pi_{\infty,k}=0$) it reduces to
$\Psi = p\bigl[1/(\gamma_1-1) - 1/(\gamma_2-1)\bigr]$, and it vanishes
identically in single-phase regions where $\gamma_1 = \gamma_2$.

Collecting all contributions, the admissible interface jump is
\begin{equation}\label{eq:fc_jump1d}
\Delta\mathbf{U}_{\mathrm{int}} =
\begin{pmatrix}
\Delta m_1 \\[2pt]
\Delta m_2 \\[2pt]
u\,\Delta\rho \\[2pt]
\tfrac{1}{2}u^2\Delta\rho + \Psi\,\Delta\alpha_1 \\[2pt]
\Delta\alpha_1
\end{pmatrix}, \qquad \Delta\rho = \Delta m_1 + \Delta m_2.
\end{equation}
The energy component comprises two contributions of distinct origin: the
kinematic term $\tfrac{1}{2}u^2\Delta\rho$, which reflects the change in
mixture kinetic energy at constant velocity, and the thermodynamic term
$\Psi\,\Delta\alpha_1$, which reflects the change in mixture internal energy
at constant pressure. These two contributions cancel against separate parts
of the acoustic left eigenvector in Step~4. The 2D generalisation replaces
the momentum component $u\,\Delta\rho$ with $(u\,\Delta\rho,\,v\,\Delta\rho)$
and the kinetic term with $\tfrac{1}{2}(u^2+v^2)\Delta\rho$; the
thermodynamic term $\Psi\,\Delta\alpha_1$ is unchanged.

\paragraph{Step 2: Velocity equilibrium.}

The volume-fraction right eigenvector (column 4 of $\mathbf{R}$) is
$\mathbf{r}_\alpha = [0,\,0,\,0,\,\Psi,\,1]^T$. A volume-fraction wave of
amplitude $\delta\alpha_1$ induces $dm_1 = dm_2 = 0$, hence $d\rho = 0$.
Expanding $d(\rho u) = \rho\,du + u\,d\rho = 0$ with $d\rho = 0$ yields
$du = 0$. 

\paragraph{Step 3: Pressure equilibrium.}

Since $d\rho = du = 0$, the kinetic energy is unchanged and
$d(\rho e) = d(\rho E) = \Psi\,\delta\alpha_1$. Differentiating
Eq.~\eqref{eq:rhoe_form} at constant $\rho$ and $\rho_k$,
\[
d(\rho e) = \Gamma_m\,dp + \Psi\,d\alpha_1.
\]
Substituting $d(\rho e) = \Psi\,\delta\alpha_1$ and rearranging,
\[
\Gamma_m\,dp = \Psi\,\delta\alpha_1 - \Psi\,\delta\alpha_1 = 0.
\]
Since $\Gamma_m > 0$, it follows that $dp = 0$. 

\paragraph{Step 4: Zero acoustic projection.}

The acoustic wave amplitude is $w_1 = \boldsymbol{\ell}_1 \cdot
\Delta\mathbf{U}_\mathrm{int}$, where $\boldsymbol{\ell}_1$ denotes the
first row of $\mathbf{L}$ in Eq.~\eqref{eq:fc1d_L}. The dot product
separates into two independent cancellations.

\smallskip
\noindent\emph{Kinematic cancellation.} Collecting all terms proportional
to $\Delta\rho$ and using $\Delta m_1 + \Delta m_2 = \Delta\rho$,
\[
\left(\frac{\chi u^2}{4}+\frac{u}{2c}
  -\frac{\chi u^2}{2}-\frac{u}{2c}
  +\frac{\chi u^2}{4}\right)\Delta\rho = 0.
\]

\smallskip
\noindent\emph{Thermodynamic cancellation.} The energy component of
$\boldsymbol{\ell}_1$ is $+\tfrac{\chi}{2}$, which acting on the
thermodynamic contribution $\Psi\,\Delta\alpha_1$ to the energy component
of $\Delta\mathbf{U}_\mathrm{int}$ yields $+\tfrac{\chi\Psi}{2}\Delta\alpha_1$.
The volume-fraction component of $\boldsymbol{\ell}_1$ is
$-\tfrac{\chi\Psi}{2}$, acting on $\Delta\alpha_1$, yielding
$-\tfrac{\chi\Psi}{2}\Delta\alpha_1$. The sum is identically zero:
\[
+\frac{\chi\Psi}{2}\Delta\alpha_1 - \frac{\chi\Psi}{2}\Delta\alpha_1 = 0.
\]
This cancellation is not coincidental: $\Psi$ enters the energy component
of $\Delta\mathbf{U}_\mathrm{int}$ through the thermodynamic identity
derived in Step~1, and it enters the volume-fraction component of
$\boldsymbol{\ell}_1$ with opposite sign through the biorthogonality
condition $\mathbf{L}\mathbf{R} = \mathbf{I}$. The two are a matched pair
by construction.

Hence $w_1 = 0$. By the antisymmetry of $\boldsymbol{\ell}_1$ and
$\boldsymbol{\ell}_5$ in the $u/c$ coefficient, $w_5 = 0$ as well.

\paragraph{Step 5: Decomposition onto the degenerate subspace.}

The contact jump Eq.~\eqref{eq:fc_jump1d} admits the exact decomposition
\[
\Delta\mathbf{U}_{\mathrm{int}}
= \Delta m_1\,\mathbf{r}_2
+ \Delta m_2\,\mathbf{r}_3
+ \Delta\alpha_1\,\mathbf{r}_4,
\]
where $\mathbf{r}_2 = [1,\,0,\,u,\,u^2/2,\,0]^T$ and
$\mathbf{r}_3 = [0,\,1,\,u,\,u^2/2,\,0]^T$ are the entropy wave eigenvectors
and $\mathbf{r}_4 = [0,\,0,\,0,\,\Psi,\,1]^T$ is the volume-fraction
eigenvector. The acoustic eigenvectors carry zero amplitude.

%% ============================================================
%% ============================================================
%% SECTION 4 -- SEMI-CONSERVATIVE EIGENSTRUCTURE + EQUILIBRIUM PROOF
%% ============================================================
\subsection{Semi-Conservative Eigenstructure}
\label{sec:sc}

In the semi-conservative (SC) formulation, pressure $p$ replaces total energy $\rho E$ as the fifth state variable:
\begin{equation}\label{eq:SC_vec}
\mathbf{V} = \bigl[\alpha_1\rho_1,\; \alpha_2\rho_2,\; \rho u,\; \rho v,\; p,\; \alpha_1\bigr]^T.
\end{equation}
This choice makes the mechanical equilibrium conditions structurally explicit: any eigenvector with a zero in its pressure slot automatically preserves pressure equilibrium without the need for an algebraic correction term. The volume-fraction eigenvector in this formulation has a strictly zero pressure component; no thermodynamic jump term $\Psi$ is required. The retention of the conserved momentum components $\rho u$ and $\rho v$ in the state vector is the key structural feature that distinguishes this formulation from a fully primitive one. Because the state vector contains momentum directly, the Jacobian separates cleanly into contributions from normal-velocity changes (acoustic waves), tangential-velocity changes (shear wave), density changes (entropy waves), and volume-fraction changes. This separation is more transparent here than in the FC case because no enthalpy-pressure cross-coupling appears in the shear eigenvector. Specifically, the shear right eigenvector has zero entries in all density, pressure, and volume-fraction rows, identical in character to the single-species Euler shear eigenvector. This is the structural property invoked by Hoffmann, Chamarthi and Frankel~\cite{hoffmann2024centralized} to justify central reconstruction of the shear wave in single-fluid flows, and it carries over directly to the multiphase SC system.

\subsubsection{1D Semi-Conservative Eigenstructure}

For the 1D state vector
\[
\mathbf{V} = [m_1,\, m_2,\, \rho u,\, p,\, \alpha_1]^T,
\]
with $m_k = \alpha_k\rho_k$ and $\rho = m_1+m_2$, the eigenvalues are identical to the FC case:
\[
\lambda_1 = u - c,\qquad
\lambda_{2,3,4} = u,\qquad
\lambda_5 = u + c.
\]
The eigenvalues are the same because the wave speeds of the five-equation system are a property of the governing equations, not of the variable set used to represent them. What differs between FC and SC is the eigenvector structure, not the characteristic directions.

\paragraph{Right eigenvectors.}

Comparing the SC right eigenvectors to the FC counterparts in Eq.~\eqref{eq:fc1d_R}, the most important difference is in the volume-fraction column. In FC, the energy row of $\mathbf{r}_4$ carries $\Psi$. In SC, the pressure row of $\mathbf{r}_4$ is identically zero. There is no thermodynamic correction to be made because pressure is stored directly and a pure volume-fraction wave leaves $p$ unchanged by construction. The entropy wave columns $\mathbf{r}_2$ and $\mathbf{r}_3$ carry a zero pressure entry, reflecting that a change in phasic density at constant velocity and volume fraction does not directly perturb the pressure variable in the SC state vector. The acoustic columns $\mathbf{r}_1$ and $\mathbf{r}_5$ carry $c^2$ in the pressure row, which is the standard acoustic pressure-density coupling scaled by the mixture sound speed squared.

\begin{equation}\label{eq:sc1d_R}
\mathbf{r}_1 =
\begin{pmatrix}
Y_1 \\[2pt] Y_2 \\[2pt] u-c \\[2pt] c^2 \\[2pt] 0
\end{pmatrix},\quad
\mathbf{r}_2 =
\begin{pmatrix}
1 \\[2pt] 0 \\[2pt] u \\[2pt] 0 \\[2pt] 0
\end{pmatrix},\quad
\mathbf{r}_3 =
\begin{pmatrix}
0 \\[2pt] 1 \\[2pt] u \\[2pt] 0 \\[2pt] 0
\end{pmatrix},\quad
\mathbf{r}_4 =
\begin{pmatrix}
0 \\[2pt] 0 \\[2pt] 0 \\[2pt] 0 \\[2pt] 1
\end{pmatrix},\quad
\mathbf{r}_5 =
\begin{pmatrix}
Y_1 \\[2pt] Y_2 \\[2pt] u+c \\[2pt] c^2 \\[2pt] 0
\end{pmatrix}.
\end{equation}
The volume-fraction eigenvector $\mathbf{r}_4$ has zeros in all mechanical components, including the pressure slot.

\paragraph{Left eigenvectors.}

A corresponding left eigenvector matrix $\mathbf{L}$ satisfying $\mathbf{L}\mathbf{R} = \mathbf{I}$ can be written as
\begin{equation}\label{eq:sc1d_L}
\begin{aligned}
\boldsymbol{\ell}_1 &= \bigl[\tfrac{u}{2c},\; \tfrac{u}{2c},\; -\tfrac{1}{2c},\; \tfrac{1}{2c^2},\; 0\bigr], \\
\boldsymbol{\ell}_2 &= \bigl[1,\, 0,\, 0,\, -Y_1/c^2,\, 0\bigr], \\
\boldsymbol{\ell}_3 &= \bigl[0,\, 1,\, 0,\, -Y_2/c^2,\, 0\bigr], \\
\boldsymbol{\ell}_4 &= \bigl[0,\, 0,\, 0,\, 0,\, 1\bigr], \\
\boldsymbol{\ell}_5 &= \bigl[-\tfrac{u}{2c},\; -\tfrac{u}{2c},\; \tfrac{1}{2c},\; \tfrac{1}{2c^2},\; 0\bigr].
\end{aligned}
\end{equation}
One verifies that $\mathbf{L}\mathbf{R} = \mathbf{I}$ and $\mathbf{L}\mathbf{A} = \boldsymbol{\Lambda}\mathbf{L}$ with $\boldsymbol{\Lambda} = \mathrm{diag}(u-c,\,u,\,u,\,u,\,u+c)$. The SC left eigenvectors are noticeably simpler than their FC counterparts in Eq.~\eqref{eq:fc1d_L}. The acoustic rows contain no $\Psi$ terms, only the standard acoustic pressure coupling $1/(2c^2)$ in the pressure column. The entropy wave rows carry only $-Y_k/c^2$ in the pressure slot. The volume-fraction row is simply $[0,\,0,\,0,\,0,\,1]$: the volume fraction is fully decoupled from all mechanical variables. This simplicity is a direct consequence of storing pressure directly. The nonlinear EOS coupling that generates $\Psi$ and its sign-matched counterpart $-\chi\Psi/2$ in the FC eigensystem is absorbed into the variable choice and disappears from the equations entirely.

\subsubsection{2D Semi-Conservative Eigenstructure}

The 2D SC system adds the shear wave, just as in the FC case. The critical difference is the structure of the shear right eigenvector. In the SC system, column 4 of the right eigenvector matrix has non-zero entries only in the two momentum rows ($-n_y$ and $n_x$, the tangential direction components). All density rows, the pressure row, and the volume-fraction row are exactly zero. This is structurally identical to the shear eigenvector of the single-species Euler equations in 2D. In the FC case, the shear eigenvector carries a non-zero entry $u_t$ in the energy row; while this does not produce a pressure perturbation (as shown in Section~\ref{sec:comparison}), the structural argument is less direct. This is one of the advantages of the SC formulation. For the 2D state vector
\[
\mathbf{V} = [m_1,\, m_2,\, \rho u,\, \rho v,\, p,\, \alpha_1]^T,
\]
the flux Jacobian in the $\mathbf{n}$-direction has eigenvalues
\[
\lambda_1 = u_n - c,\qquad
\lambda_{2,3,4,5} = u_n,\qquad
\lambda_6 = u_n + c,
\]
with $u_n = u n_x + v n_y$ as before.

\paragraph{Right eigenvector matrix.}

A convenient choice of right eigenvectors is
\begin{equation}\label{eq:sc2d_R}
\mathbf{R} =
\begin{bmatrix}
Y_1 & 1 & 0 & 0 & 0 & Y_1 \\
Y_2 & 0 & 1 & 0 & 0 & Y_2 \\
u-cn_x & u & u & -n_y & 0 & u+cn_x \\
v-cn_y & v & v & n_x & 0 & v+cn_y \\
c^2 & 0 & 0 & 0 & 0 & c^2 \\
0 & 0 & 0 & 0 & 1 & 0
\end{bmatrix},
\end{equation}
with columns ordered as $[u_n-c,\ \mathrm{comp}_1,\ \mathrm{comp}_2,\ \mathrm{shear},\ \mathrm{VF},\ u_n+c]$. The shear column has non-zero entries only in the momentum rows, and the VF column has a zero pressure component.

\paragraph{Left eigenvector matrix.}

The associated left eigenvector matrix can be written as
\begin{equation}\label{eq:sc2d_L}
\mathbf{L} =
\begin{bmatrix}
\dfrac{u_n}{2c} & \dfrac{u_n}{2c} & -\dfrac{n_x}{2c} & -\dfrac{n_y}{2c} & \dfrac{1}{2c^2} & 0 \\[3pt]
1 & 0 & 0 & 0 & -\dfrac{Y_1}{c^2} & 0 \\[3pt]
0 & 1 & 0 & 0 & -\dfrac{Y_2}{c^2} & 0 \\[3pt]
-u_t & -u_t & -n_y & n_x & 0 & 0 \\[3pt]
0 & 0 & 0 & 0 & 0 & 1 \\[3pt]
-\dfrac{u_n}{2c} & -\dfrac{u_n}{2c} & \dfrac{n_x}{2c} & \dfrac{n_y}{2c} & \dfrac{1}{2c^2} & 0
\end{bmatrix},
\end{equation}
with $u_t = -u n_y + v n_x$. One verifies $\mathbf{L}\mathbf{R} = \mathbf{I}$. The last column (volume-fraction column) of both acoustic rows is zero, reflecting that volume-fraction waves carry no acoustic content and produce no pressure perturbation. The last row is $[0,\,0,\,0,\,0,\,0,\,1]$, confirming that $\alpha_1$ is entirely decoupled from all other wave families in the SC system. The shear row (row 4) contains $-u_t$ in the phasic density columns and the tangential normal components in the momentum columns, with zeros in the pressure and volume-fraction columns. Thus, it annihilates a pure material-contact jump with constant velocity, while its finite-amplitude form is the density-weighted tangential-velocity departure derived below.
\label{sec:sc_proof}

\Rthree{\paragraph{Relation to the primitive shear characteristic.}
The distinction is clearest at an $x$-face, for which the shear field is the transverse velocity $v$. Consider the primitive state vector $\mathbf{P}=[m_1,\,m_2,\,u,\,v,\,p,\,\alpha_1]^T$ used in earlier primitive-characteristic reconstruction~\cite{chamarthi2025physics}. The primitive shear characteristic and its right-eigenvector update are
\begin{equation}\label{eq:primitive_shear}
w_s^{P}=v,
\qquad
\delta\mathbf{P}_s=[0,\,0,\,0,\,\delta w_s,\,0,\,0]^T.
\end{equation}
Thus, central reconstruction of $w_s^P$ is simply central reconstruction of the transverse velocity itself. For the SC state vector $\mathbf{V}=[m_1,\,m_2,\,\rho u,\,\rho v,\,p,\,\alpha_1]^T$, reconstruction uses the fourth row of Eq.~\eqref{eq:sc2d_L} frozen at an interface reference velocity $\bar v$. Its shear projection of a stencil state is
\begin{equation}\label{eq:sc_shear}
\widetilde w_s^{\mathrm{SC}}(\bar v;\mathbf V)
=[-\bar v,\,-\bar v,\,0,\,1,\,0,\,0]\mathbf V
=-\bar v(m_1+m_2)+\rho v
=\rho(v-\bar v),
\qquad
\delta\mathbf{V}_s=[0,\,0,\,0,\,\delta w_s,\,0,\,0]^T.
\end{equation}
If the shear row is evaluated at the same local state, so that $\bar v=v$, its projection vanishes exactly: $-v(m_1+m_2)+\rho v=-\rho v+\rho v=0$. When the row is frozen at an interface reference state, it instead measures the density-weighted departure $\rho(v-\bar v)$ in Eq.~\eqref{eq:sc_shear}. The volume-fraction component remains exactly zero in both cases. The vector entries in Eqs.~\eqref{eq:primitive_shear} and~\eqref{eq:sc_shear} look identical, but their meaning is different: the fourth primitive entry is $v$, whereas the fourth SC entry is $\rho v$. Consequently, the same shear-amplitude update gives $\delta v=\delta w_s$ in the primitive basis, but $\delta v=\delta w_s/\rho$ in the SC basis because the phasic masses are unchanged. Central reconstruction therefore acts on $v$ in the primitive basis and on the tangential-momentum perturbation $\rho(v-\bar v)$ in the SC basis. For a variable-density field, a reconstruction operator $\mathcal{C}$ does not commute with this change of variable,
\begin{equation}\label{eq:shear_noncommute}
\mathcal{C}\!\left[\rho(v-\bar v)\right]
\neq
\rho^*\!\left(\mathcal{C}[v]-\bar v\right).
\end{equation}
The two central reconstructions are therefore generally different; they reduce to the same operation only for constant density, up to a scalar normalization. The corresponding single-species observation, that characteristic centralization at $x$- and $y$-faces acts through $\rho v$ and $\rho u$, respectively, was identified in~\cite{chamarthi2024generalized}. Primitive-variable and primitive-characteristic reconstructions are standard in compressible multiphase-flow methods, and were also used in our earlier formulations. Accordingly, centralizing the primitive shear characteristic initially appeared to be the natural implementation. In the variable-density shear problems relevant here, however, this choice centralizes $v$ rather than the transverse momentum perturbation and did not provide the required robust inviscid treatment. This observation motivated reconstruction in the FC/SC characteristic variables themselves, for which centralization acts on $\rho(v-\bar v)$ and the update is applied to $\rho v$. The resulting formulation is the one validated in the present work. In the earlier primitive formulations, central treatment of the tangential velocity was consequently restricted by continuity, viscosity, artificial-viscosity, or sensor conditions~\cite{chamarthi2025wave,chamarthi2025physics}. The $y$-face argument follows identically, with $u$ as the transverse velocity.}

\paragraph{Step 1: Admissible contact-wave jump.}

With $p$ and $\mathbf{u}$ constant across a material interface, the admissible jump in the SC variables is
\begin{equation}\label{eq:sc_jump}
\Delta\mathbf{V}_{\mathrm{int}} =
\begin{bmatrix}
\Delta m_1 \\[2pt]
\Delta m_2 \\[2pt]
u\,\Delta\rho \\[2pt]
0 \\[2pt]
\Delta\alpha_1
\end{bmatrix}
\quad\text{(1D)},\qquad
\Delta\mathbf{V}_{\mathrm{int}} =
\begin{bmatrix}
\Delta m_1 \\[2pt]
\Delta m_2 \\[2pt]
u\,\Delta\rho \\[2pt]
v\,\Delta\rho \\[2pt]
0 \\[2pt]
\Delta\alpha_1
\end{bmatrix}
\quad\text{(2D)},
\end{equation}
with $\Delta\rho = \Delta m_1 + \Delta m_2$. The pressure component is identically zero; no $\Psi$ term appears.

\paragraph{Step 2: Velocity equilibrium.}

From the VF eigenvector $\mathbf{r}_4 = [0,\,0,\,0,\,0,\,1]^T$ in 1D, a pure VF wave of amplitude $\delta\alpha_1$ yields
\[
dm_1 = 0,\quad dm_2 = 0,\quad d(\rho u) = 0,\quad dp = 0,\quad d\alpha_1 = \delta\alpha_1.
\]
Thus $d\rho = 0$, and $d(\rho u) = \rho\,du + u\,d\rho = 0$ implies $du = 0$. The same reasoning applies in 2D for both components of $\mathbf{u}$. 

\paragraph{Step 3: Structural pressure equilibrium.}

Because the pressure slot of $\mathbf{r}_4$ is zero by construction, a pure VF wave never perturbs $p$, regardless of the EOS. Storing $p$ directly therefore ensures $dp = 0$ across any eigenvector whose pressure component vanishes. No EOS-based correction such as $\Psi$ is required. 

\paragraph{Step 4: Zero acoustic projection (1D).}

Using Eq.~\eqref{eq:sc1d_L} and the jump~\eqref{eq:sc_jump},
\begin{equation}\label{eq:sc_acoustic_proj}
\boldsymbol{\ell}_1\cdot\Delta\mathbf{V}_{\mathrm{int}}
= \frac{u}{2c}\Delta m_1
 + \frac{u}{2c}\Delta m_2
 - \frac{1}{2\rho c} \cdot \rho u\,\Delta\rho
 + \frac{1}{2\rho c^2}\cdot 0
 = \frac{u}{2c}\Delta\rho - \frac{u}{2c}\Delta\rho = 0.
\end{equation}
The cancellation is purely kinematic: the momentum jump $u\,\Delta\rho$ matches the density jump at constant $u$. By symmetry, $\boldsymbol{\ell}_5\cdot\Delta\mathbf{V}_{\mathrm{int}} = 0$ as well. 

\paragraph{Step 5: Decomposition onto the degenerate subspace.}

The SC contact jump decomposes into the three degenerate $u$-waves as
\[
\Delta\mathbf{V}_{\mathrm{int}} =
\Delta m_1\,\mathbf{r}_2 + \Delta m_2\,\mathbf{r}_3 + \Delta\alpha_1\,\mathbf{r}_4,
\]
with
\[
\mathbf{r}_2 = [1,\,0,\, u,\,0,\,0]^T,\quad
\mathbf{r}_3 = [0,\,1,\, u,\,0,\,0]^T,\quad
\mathbf{r}_4 = [0,\,0,\,0,\,0,\,1]^T.
\]
Thus the density jump is carried by the two entropy waves at constant velocity and pressure, and the VF wave carries $\Delta\alpha_1$ without any mechanical perturbation. No acoustic wave is excited. 

\paragraph{Extension to 2D.}

All steps extend directly to 2D by replacing $u$ with $u_n = u n_x + v n_y$ and $\Delta(\rho u)$, $\Delta(\rho v)$ with their normal and tangential counterparts. The shear wave is not excited by the admissible contact jump because $\Delta\mathbf{V}_{\mathrm{int}}$ has no component in the tangential momentum direction when $\Delta u = \Delta v = 0$.

%% ============================================================
%% SECTION 5 -- COMPARISON
%% ============================================================
\subsection{Comparison of Formulations}
\label{sec:comparison}

The FC and SC formulations share the same eigenvalues, the same physical wave families, and the same admissible contact jumps. Their differences lie in the choice of stored energy variable and the way mechanical equilibrium is enforced at material interfaces. Table~\ref{tab:comparison} summarizes the main structural distinctions.

\begin{table}[htb]
\centering
\caption{Structural differences between the FC and SC eigenstructures.}
\label{tab:comparison}
\begin{tabular}{lll}
\toprule
Feature & Fully conservative (FC) & Semi-conservative (SC) \\
\midrule
Stored energy variable & $\rho E$ & $p$ \\
Volume-fraction eigenvector & Energy component $\Psi$ & Zero pressure component \\
Pressure equilibrium & Algebraic cancellation via $\Psi$ & Structural zero in pressure slot \\
Shear energy row & $u_t$ (kinematic only) & No explicit energy row \\
Pressure recovery & From $\rho E$ and EOS & Direct from $p$ \\
EOS generality & Requires $\Psi$ derivation per EOS & Structural; any EOS \\
\bottomrule
\end{tabular}
\end{table}

In the FC formulation, the volume-fraction eigenvector includes the thermodynamic jump term $\Psi$ in its energy component. This term encodes the internal-energy change required when $\alpha_1$ varies at constant $p$ and $\mathbf{u}$ and appears with opposite sign in the left eigenvectors, ensuring exact cancellation of EOS-induced perturbations in the acoustic fields. This multi-phase coupling is absent in single-species Euler eigensystems and must be derived explicitly for the five-equation model.

In the SC formulation, pressure is stored directly, and the pressure slot of the VF eigenvector is identically zero. Mechanical equilibrium at material interfaces is then enforced by the structure of the eigensystem: any pure VF wave leaves $p$ and $\mathbf{u}$ unchanged for any EOS, and no auxiliary jump term is required. The SC eigenvectors are correspondingly simpler, and the biorthogonality of the characteristic matrices is more transparent.

The shear wave has the same physical role in both formulations: it carries tangential-velocity perturbations only and leaves densities, pressure, and volume fraction unchanged. In FC this appears as a non-zero energy row proportional to $u_t$ but with zero induced pressure; in SC it appears purely in the momentum components. In either case the shear amplitude is structurally decoupled from the thermodynamic and interface fields, justifying the use of a central (non-dissipative) reconstruction scheme for this wave family.

Finally, the entropy waves differ algebraically between FC and SC because the stored energy variable differs, but they represent the same contact physics: density and composition changes at constant pressure and velocity. In both formulations, admissible material-interface jumps project entirely onto the entropy wave and VF eigenvectors, with zero projection onto the acoustic eigenvectors, so Abgrall’s equilibrium condition is satisfied by design.

%% ============================================================
\section{Numerical Method}
\label{sec:reconstruction}

\subsection{Finite-Volume Discretisation}
\label{sec:fv}

The five-equation model is discretised on a uniform Cartesian grid using a conservative finite-volume method \cite{godunov1959}.  The semi-discrete update for a generic conserved state vector $\mathbf{Q}_{ij}$ reads 

\begin{equation}\label{eq:fv_update}
\frac{\mathrm{d}\mathbf{Q}_{ij}}{\mathrm{d}t}
=
-\frac{1}{\Delta x}\bigl(\mathbf{F}_{i+1/2,j}-\mathbf{F}_{i-1/2,j}\bigr)
-\frac{1}{\Delta y}\bigl(\mathbf{G}_{i,j+1/2}-\mathbf{G}_{i,j-1/2}\bigr),
\end{equation}
where $\mathbf{F}$ and $\mathbf{G}$ are numerical fluxes evaluated by an HLLC Riemann solver \cite{toro1994restoration}, unless otherwise stated, applied to left/right interface states reconstructed in characteristic space (Section~\ref{sec:grab}).  Time integration uses third-order strong-stability-preserving Runge--Kutta scheme \cite{Jiang1995}; the time step satisfies a CFL condition based on the maximum eigenvalue of the flux Jacobian.

At each face, the reconstructed left and right states are expressed in conservative form before they are supplied to the HLLC solver. \Rthree{For these conservative face states, the HLLC solver uses the left- and right-going signal-speed estimates
\begin{equation}\label{eq:hllc_speeds}
S_L=\min\!\left(u_{n,L}-c_L,\;u_n^{a}-c^{a}\right),\qquad
S_R=\max\!\left(u_{n,R}+c_R,\;u_n^{a}+c^{a}\right).
\end{equation}
For the one-dimensional calculations, $(u_n^{a},c^{a})$ is obtained from the Roe-averaged state. For the multidimensional calculations, $u_n^{a}=(u_{n,L}+u_{n,R})/2$ and $c^{a}=(c_L+c_R)/2$. This signal-speed average is distinct from the state used to evaluate the characteristic matrices. The contact speed is
\begin{equation}\label{eq:hllc_contact_speed}
S_*=
\frac{
p_R-p_L+\rho_L u_{n,L}(S_L-u_{n,L})
-\rho_R u_{n,R}(S_R-u_{n,R})
}{
\rho_L(S_L-u_{n,L})-\rho_R(S_R-u_{n,R})
}.
\end{equation}
The same definitions are used for FC and SC reconstructions; only the characteristic basis used to obtain the interface states differs.} \Rone{For SC characteristic reconstruction, the conservative face state is obtained by recovering total energy from the reconstructed pressure and mixture EOS:
\begin{equation}\label{eq:sc_energy_recovery}
(\rho E)^{\rm rec}
=
\frac{p+\gamma\pi_\infty}{\gamma-1}
+\frac{1}{2}\rho\left(u^2+v^2\right).
\end{equation}
This recovery is not specific to SC: whenever primitive variables, or characteristic variables formed from primitive-variable increments, are reconstructed, the corresponding conservative face state, including $\rho E$, must likewise be recovered before evaluating a conservative HLLC flux. This reconstruction-to-conservative-state sequence is used in the five-equation formulation of Allaire et al.~\cite{allaire2002five} and in subsequent high-order multicomponent methods of Johnsen and Colonius~\cite{johnsen2006implementation}, Nonomura et al.~\cite{Nonomura2012}, Coralic and Colonius~\cite{coralic2014finite}, Wong et al.~\cite{wong2021positivity}, Zhang et al.~\cite{zhang2024hybrid}, Li and Fu~\cite{li2024family}, and Collis et al.~\cite{collis2026robust}.}

\subsection{Characteristic Decomposition and Wave-Appropriate Reconstruction}
\label{sec:grab}

The reconstruction procedure follows the wave-appropriate philosophy established in the papers~\cite{chamarthi2025wave,chamarthi2025physics,hoffmann2024centralized}: variables are projected into characteristic space, each wave family is reconstructed with the scheme most consistent with its physics, and the result is projected back to physical space. The per-interface procedure at face $i+\tfrac{1}{2}$ in one spatial direction is as follows. The same procedure applies in both coordinate directions using the appropriate face normal $\mathbf{n} = (n_x, n_y)$ and tangent $\boldsymbol{\ell} = (-n_y, n_x)$.

\subsubsection{Discontinuity Sensors}
\label{sec:sensors}
Reconstruction procedure requires evaluation of two discontinuity sensors (interface and shocks):

\paragraph{Interface sensor $\psi_i$.}
A smoothness-based contact and material-interface sensor \cite{chamarthi2025wave,chamarthi2025physics} is constructed from the entropy like variable $s_i = p_i/\rho_i^{\hat{\gamma}}$:
\begin{equation}\label{eq:sensor_ab}
a_i = \tfrac{13}{12}\lvert s_{i-2}-2s_{i-1}+s_i\rvert
    + \tfrac{1}{4}\lvert s_{i-2}-4s_{i-1}+3s_i\rvert,\quad
b_i = \tfrac{13}{12}\lvert s_i-2s_{i+1}+s_{i+2}\rvert
    + \tfrac{1}{4}\lvert 3s_i-4s_{i+1}+s_{i+2}\rvert,
\end{equation}
\begin{equation}\label{eq:sensor_psi}
\psi_i = \frac{2a_i b_i + \varepsilon}{a_i^2 + b_i^2 + \varepsilon},\qquad
\varepsilon = \frac{0.9\psi_c}{1-0.9\psi_c}\,\xi,\quad
\xi = 10^{-2},\quad \psi_c = 0.35.
\end{equation}
THINC is activated at interface $i+\tfrac{1}{2}$ when
$\min(\psi_{i-1},\psi_i,\psi_{i+1},\psi_{i+2}) < \psi_c$.
\Rfour{In two dimensions, the same sensor is applied dimension by dimension on the directional stencil associated with each face. Thus, for an $x$-face the stencil is $(i-1,j),\ldots,(i+2,j)$, whereas for a $y$-face it is $(i,j-1),\ldots,(i,j+2)$. The threshold used in all computations is $\psi_c=0.35$.}

\paragraph{Shock sensor $\sigma_i$.}
Ducros shock sensor~\cite{ducros1999large} prevents the use of the central scheme for the shear wave near regions with shock:
\begin{equation}\label{eq:sigma_sensor}
\sigma_i = \frac{\left| -p_{i-2} + 16 p_{i-1} - 30 p_{i} + 16 p_{i+1} - p_{i+2} \right|}{\left| p_{i-2} + 16 p_{i-1} + 30 p_{i} + 16 p_{i+1} + p_{i+2} \right|}
\frac{ \left( \nabla \cdot \mathbf{u} \right)^2}
{ \left( \nabla \cdot \mathbf{u} \right)^2 + \left| \nabla \times \mathbf{u} \right|^2+\epsilon_D},
\end{equation}
\Rfour{with $\epsilon_D=10^{-30}$ in all calculations.}

\subsubsection{Acoustic waves ($k = 1, 6$)}

Acoustic waves carry pressure perturbations and are responsible for shocks. MUSCL scheme \cite{van1977towards} is applied unconditionally:
\begin{equation}\label{eq:muscl}
W_L^{(k)} = W_i^{(k)}
  + \tfrac{1}{4}\bigl[(1-\kappa)\,\mathrm{mm}(\Delta^-,2\Delta^0)
                    +(1+\kappa)\,\mathrm{mm}(\Delta^0,2\Delta^-)\bigr],
\end{equation}
and symmetrically for $W_R^{(k)}$, where $\Delta^- = W_i^{(k)}-W_{i-1}^{(k)}$, $\Delta^0 = W_{i+1}^{(k)}-W_i^{(k)}$, $\kappa = 1/3$, and $\mathrm{mm}(a,b) = \tfrac{1}{2}(\mathrm{sgn}(a)+\mathrm{sgn}(b)) \min(\lvert a\rvert,\lvert b\rvert)$.

\subsubsection{Entropy waves ($k = 2, 3$)}

These waves carry density and species-composition changes at constant pressure and velocity. MUSCL (Eq.~\eqref{eq:muscl}) is the default. When
the interface sensor detects a material interface ($\min(\psi_{i-1},\psi_i,\psi_{i+1},\psi_{i+2}) < \psi_c$), the symmetry-preserving algebraic THINC interface-value formula~\cite{Xiao2005,wakimura2022symmetry} replaces MUSCL for the entropy fields. \Rone{The THINC expression below is the algebraic interface-value form of Wakimura, Takagi, and Xiao~\cite{wakimura2022symmetry}, obtained from a hyperbolic-tangent profile after enforcing the cell-average constraint. For interface $i+\tfrac{1}{2}$, $W_{L,i+1/2}^{(k)}$ denotes the right boundary value reconstructed from cell $i$, and $W_{R,i+1/2}^{(k)}$ denotes the left boundary value reconstructed from cell $i+1$; the upper and lower signs in Eq.~\eqref{eq:thinc_LR} give these two interface states, respectively.}

\begin{equation}\label{eq:thinc_LR}
W_{L,R}^{(k)} = q_a \pm q_d\,\frac{T_1 \pm T_2/T_1}{1\pm T_2}
\quad\text{if }
(W_{i+1}^{(k)}-W_i^{(k)})(W_i^{(k)}-W_{i-1}^{(k)}) > 0,
\end{equation}
otherwise $W_L^{(k)} = W_i^{(k)}$, $W_R^{(k)} = W_{i+1}^{(k)}$, with

\begin{equation}\label{eq:thinc_params}
q_a = \tfrac{1}{2}(W_{i+1}^{(k)}+W_{i-1}^{(k)}),\quad
q_d = \tfrac{1}{2}(W_{i+1}^{(k)}-W_{i-1}^{(k)}),\quad
\xi_i = \frac{W_i^{(k)}-q_a}{q_d},
\end{equation}

\begin{equation}\label{eq:thinc_T}
T_1 = \tanh\!\bigl(\tfrac{\beta}{2}\bigr),\quad
T_2 = \tanh\!\bigl(\tfrac{\xi_i\beta}{2}\bigr),\quad
\beta = 1.8.
\end{equation}
\Rthree{Unlike the component-wise TENOA--THINC selection of Li and Fu~\cite{li2024family}, which permits THINC selection independently for any variable identified as discontinuous, THINC is applied here only to the entropy-wave fields. This distinction is essential at a material interface: pressure and velocity are continuous and must not be sharpened. Applying THINC to those fields creates nonphysical jumps and can generate pressure oscillations, as demonstrated in the wave-appropriate analyses of~\cite{chamarthi2025wave,chamarthi2025physics}.}

\subsubsection{Shear wave ($k = 4$, two dimensions only)}

The shear right eigenvector has zero entries in all phasic density, pressure, and volume-fraction rows. The shear wave carries only tangential velocity perturbations and is structurally decoupled from the interface and acoustic fields. When $\max(\sigma_{i-1},\sigma_i,\sigma_{i+1},\sigma_{i+2}) < 0.01$, a fourth-order central interpolation is used:

\begin{equation}\label{eq:central_shear}
W_{i+1/2}^{(4)} = -\tfrac{1}{12}W_{i-1}^{(4)} + \tfrac{7}{12}W_{i}^{(4)} + \tfrac{7}{12}W_{i+1}^{(4)} - \tfrac{1}{12}W_{i+2}^{(4)},
\end{equation}
with $W_L^{(4)} = W_R^{(4)} = W_{i+1/2}^{(4)}$. MUSCL (Eq.~\eqref{eq:muscl}) is used in shocked regions ($\max(\sigma_{i-1},\sigma_i,\sigma_{i+1},\sigma_{i+2}) \geq 0.01$).

\subsubsection{Volume-fraction wave ($k = 5$)}

Volume fraction is discontinuous across every material interface. THINC (Eqs.~\eqref{eq:thinc_LR} to \eqref{eq:thinc_T}) is applied whenever the
interface sensor fires; MUSCL otherwise.

\subsection{Complete Algorithm}
\label{sec:algorithm}

The per-interface reconstruction for a face at $i+\tfrac{1}{2}$ in one spatial direction proceeds as follows. The same procedure applies in both
coordinate directions using the appropriate face normal.

\begin{algorithm}[H]
\caption{Wave-appropriate characteristic reconstruction at interface $i+\tfrac{1}{2}$.}
\label{alg:grab}
\begin{enumerate}
  \item[\textbf{Step~1.}] \textbf{Sensors.}  Evaluate the interface sensor
        $\psi_i$ (Eq.~\eqref{eq:sensor_psi}) and the Ducros-weighted shock
        sensor $\sigma_i$ (if necessary) (Eq.~\eqref{eq:sigma_sensor}).
	  \item[\textbf{Step~2.}] \textbf{Roe-average state.}  Compute the
	        density-weighted averages $\hat{\rho}$, $\hat{u}$, $\hat{v}$,
	        $\hat{p}$, $\hat{\alpha}_1$, $\hat{Y}_k$, $\hat{\gamma}$,
	        $\hat{\pi}_\infty$, and $\hat{c}$.
\Rfour{The Roe-averaged sound speed is not computed by directly averaging the left and right sound speeds. It is evaluated from the Roe-averaged thermodynamic state,
\begin{equation}\label{eq:roe_sound_speed}
\hat{c}^2=
\frac{\hat{\gamma}\left(\hat{p}+\hat{\pi}_\infty\right)}
{\hat{\rho}},
\end{equation}
with $\hat{\gamma}$ and $\hat{\pi}_\infty$ obtained from the same Roe-averaged mixture composition used in the eigenvectors.}
	  \item[\textbf{Step~3.}] \textbf{Eigenvectors.}  Using the Roe-average
	        state and the face normal $\mathbf{n}$, assemble the right
        eigenvector matrix $\mathbf{R}$ and the
        left eigenvector matrix $\mathbf{L}$.
  \item[\textbf{Step~4.}] \textbf{Projection to characteristic space.}
        For each stencil cell $\ell \in \{i-1,\ldots,i+2\}$:
        \begin{equation}\label{eq:projection}
        \mathbf{W}_\ell = \mathbf{L}\,\mathbf{V}_\ell,
        \end{equation}
        where $\mathbf{V}_\ell$ is the state vector (FC or SC) at cell
        $\ell$.
  \item[\textbf{Step~5.}] \textbf{Wave-appropriate reconstruction.}  For
        each characteristic field $k$ independently:
    \begin{equation}\label{eq:wa_assign}
    W_{L,R}^{(k)} = \begin{cases}
      \text{MUSCL (Eq.~\eqref{eq:muscl})} & k = 1,6 \;\text{(acoustic)}, \\[4pt]
      \begin{cases}
        \text{THINC (Eq.~\eqref{eq:thinc_LR})} & \text{if } \min(\psi_{i-1},\psi_i,\psi_{i+1},\psi_{i+2}) < \psi_c, \\
        \text{MUSCL} & \text{otherwise},
      \end{cases} & k = 2,3 \;\text{(entropy; \textbf{option-1})}, \\[4pt]
      \begin{cases}
        W_L^{(4)} = W_R^{(4)} = W_{i+1/2}^{(4)} \;\text{(Eq.~\eqref{eq:central_shear})} & \text{if } \max(\sigma) < 0.01, \\
        \text{MUSCL} & \text{otherwise},
      \end{cases} & k = 4 \;\text{(shear; \textbf{option-2})}, \\[4pt]
      \begin{cases}
        \text{THINC} & \text{if } \min(\psi_{i-1},\psi_i,\psi_{i+1},\psi_{i+2}) < \psi_c, \\
        \text{MUSCL} & \text{otherwise},
      \end{cases} & k = 5 \;\text{(volume fraction; \textbf{option-1})}.
    \end{cases}
    \end{equation}
  \item[\textbf{Step~6.}] \textbf{Back-projection.}  Recover the
        left and right interface states in physical space:
        \begin{equation}\label{eq:backprojection}
        \mathbf{V}_{L}\big|_{i+1/2} = \mathbf{R}\,\mathbf{W}_{L},
        \qquad
        \mathbf{V}_{R}\big|_{i+1/2} = \mathbf{R}\,\mathbf{W}_{R}.
        \end{equation}
  \item[\textbf{Step~7.}] \textbf{Positivity fix.}  If any reconstructed
        phasic density $\hat{m}_k < 0$ or volume fraction
        $\hat{\alpha}_1 \notin [0,1]$, reset to the first-order upwind value which are also reconstructed from the characteristic variables.
	  \item[\textbf{Step~8.}] \textbf{Riemann solve.}  Pass
	        $(\mathbf{V}_L,\mathbf{V}_R)\big|_{i+1/2}$ to the HLLC solver.
	        \Rone{For the SC formulation, recover $\rho E$ from $(\rho,\mathbf{u},p)$
	        via the stiffened-gas EOS before flux evaluation, and use the recovered total energy in the energy flux and conservation diagnostic.}
\end{enumerate}
\end{algorithm}

Several aspects of the algorithm must be noted:

\begin{itemize}
\item Algorithm~\ref{alg:grab} is applied without modification to  gas-gas simulations, combining interface sharpening for the entropy wave with central reconstruction for the shear wave. For gas-liquid configurations, only one of these options is used at a time, as noted explicitly for each test case;  combining both is found to be unreliable in the majority of cases and is therefore avoided.
	\item An additional critical aspect is implementing the positivity fix within the algorithm. When negative densities and pressures are detected in the flow, the values at the relevant cell interfaces are typically computed using first-order upwind values. However, direct computation with conservative variables can introduce pressure and velocity oscillations that may not be immediately apparent but can ultimately result in simulation failure. Consequently, in these situations, first-order upwind values derived from characteristic variables are employed to ensure positivity. Furthermore, the robustness of the positivity can be improved by considering the work of \cite{baumgart2024ensuring} but is not considered here.
	\item In contrast to the algorithm described in \cite{chamarthi2025wave}, which utilizes primitive variables with a low-resolution scheme in liquid regions and characteristic variables with a high-resolution scheme in gas regions, such an approach is unnecessary and can induce pressure oscillations if non primitive variables are reconstructed. In the present study, all simulations employ characteristic-variable reconstruction exclusively. If a high-resolution scheme is required in gas regions, the algorithm from \cite{chamarthi2025wave} may be adopted; however, liquid regions should also be reconstructed using characteristic variables (either FC or SC) once identified by the stiffened gas parameter.
	\item Although the present study primarily reports results obtained with the MUSCL scheme, other nonlinear schemes may also be applicable for these simulations. Where feasible, results using a high-resolution scheme, specifically a sixth-order central method combined with the monotonicity-preserving scheme of Suresh and Huynh \cite{suresh1997accurate} on fine grids are included. The WENOIS scheme of Wong et  al.~\cite{wong2021positivity} is found to be more robust than MUSCL in certain scenarios but is computationally intensive and results are not included.
\end{itemize}

\section{Numerical Results}
\label{sec:results}
The FC and SC eigensystems, along with the wave-appropriate characteristic reconstruction described in Section~\ref{sec:grab}, are evaluated using a suite of one- and two-dimensional benchmark problems. Throughout this section, the scheme that combines wave-appropriate characteristic reconstruction with MUSCL limiting in either the FC or SC eigensystem is referred to as Wave-MUSCL-FC or Wave-MUSCL-SC, respectively. As detailed in the earlier section, shear waves are computed using a central scheme if the Ducros sensor detects no shocks; this applies to both gas-gas and gas-liquid configurations when FC or SC variables are employed. The selected test cases are designed to isolate specific theoretical claims. For instance, the isolated material-interface advection test (Example~\ref{ex:isol}) directly examines the equilibrium condition and the necessity of characteristic-space reconstruction. The triple-point problem (Example~\ref{ex:triplet}) simultaneously evaluates mechanical equilibrium preservation and the central-scheme treatment of the shear wave, enabling direct comparison between schemes with and without the structural shear-wave argument. In all cases, the FC and SC formulations employ identical reconstruction parameters; thus, any differences in results are attributable solely to the choice of eigensystem. \Rone{Unless explicitly stated otherwise, dimensional one-dimensional tests use SI units: density in kg\,m$^{-3}$, velocity in m\,s$^{-1}$, pressure and stiffened-gas pressure constants in Pa, length in m, and time in s. Nondimensional tests retain the units of the cited reference problem.}

\begin{example}\label{ex:isol}{Material Interface Advection (Water in Air)}
\end{example}

\begin{figure}[H]
\centering
\subfigure[Density profile, Example \ref{ex:isol}]{\includegraphics[width=0.45\textwidth]{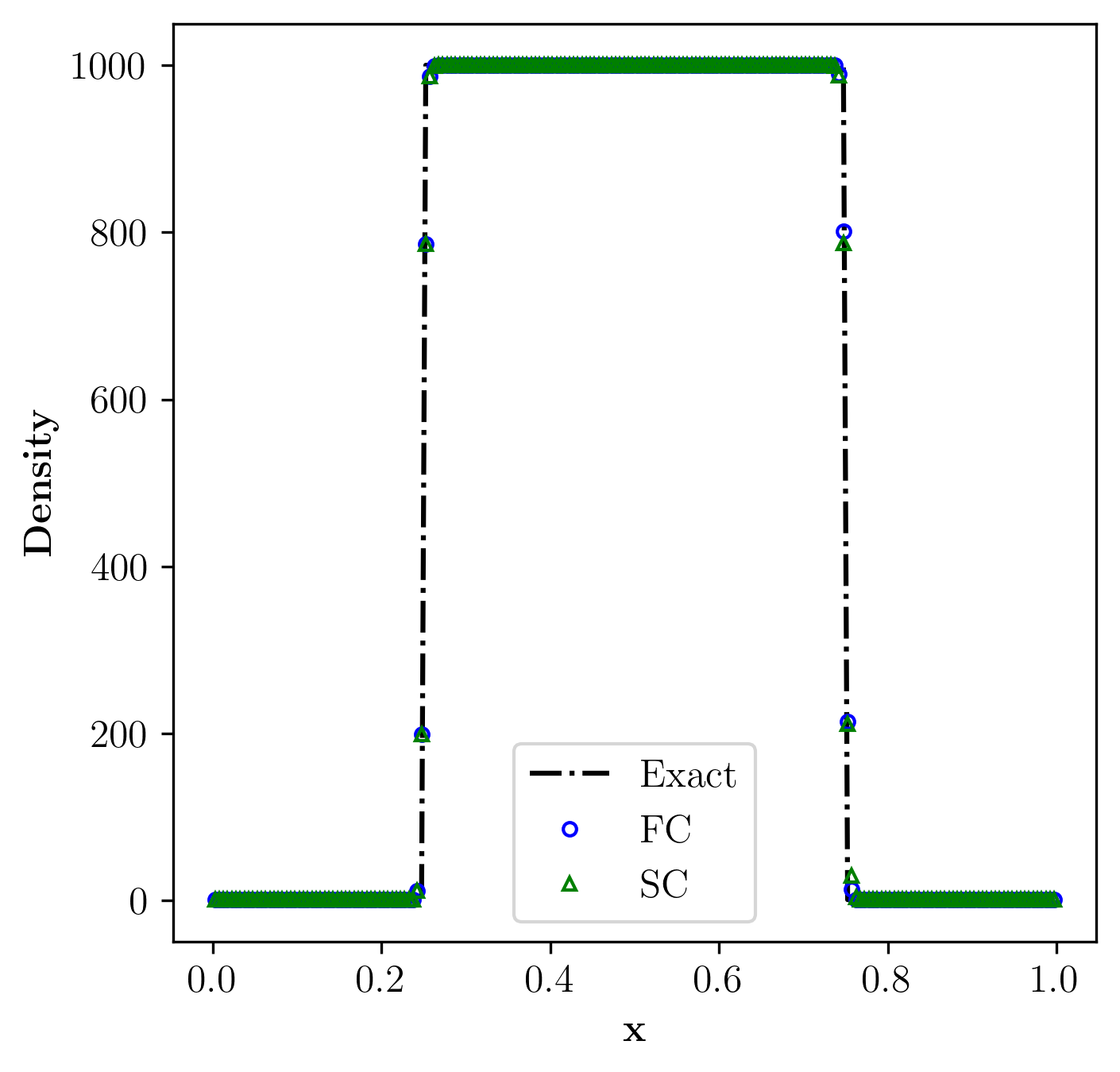}
\label{fig:iso-met_l}}
\subfigure[Velocity profile, Example \ref{ex:isol}]{\includegraphics[width=0.45\textwidth]{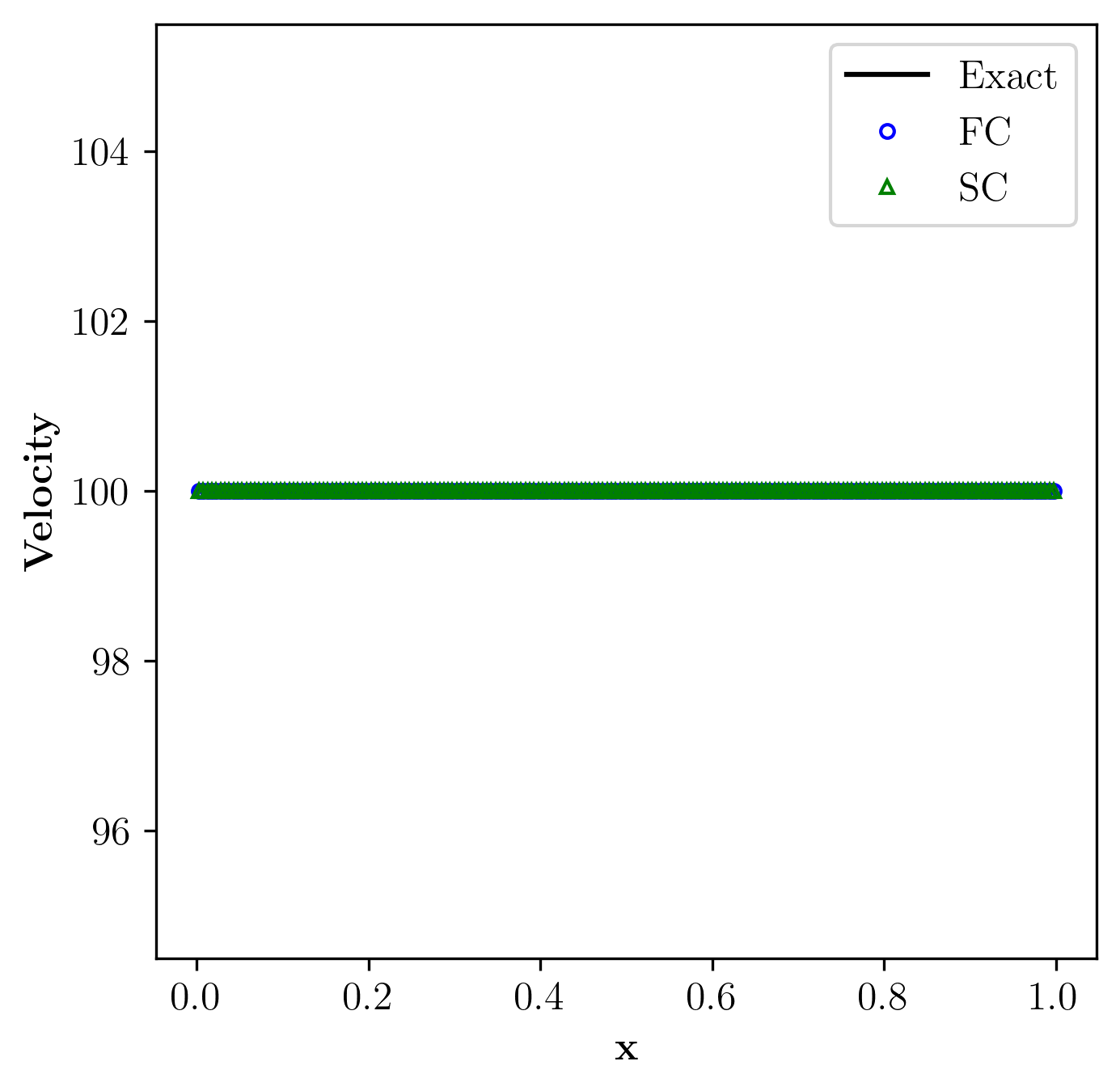}
\label{fig:iso-met2_l2}}
\subfigure[Pressure profile, Example \ref{ex:isol}]{\includegraphics[width=0.45\textwidth]{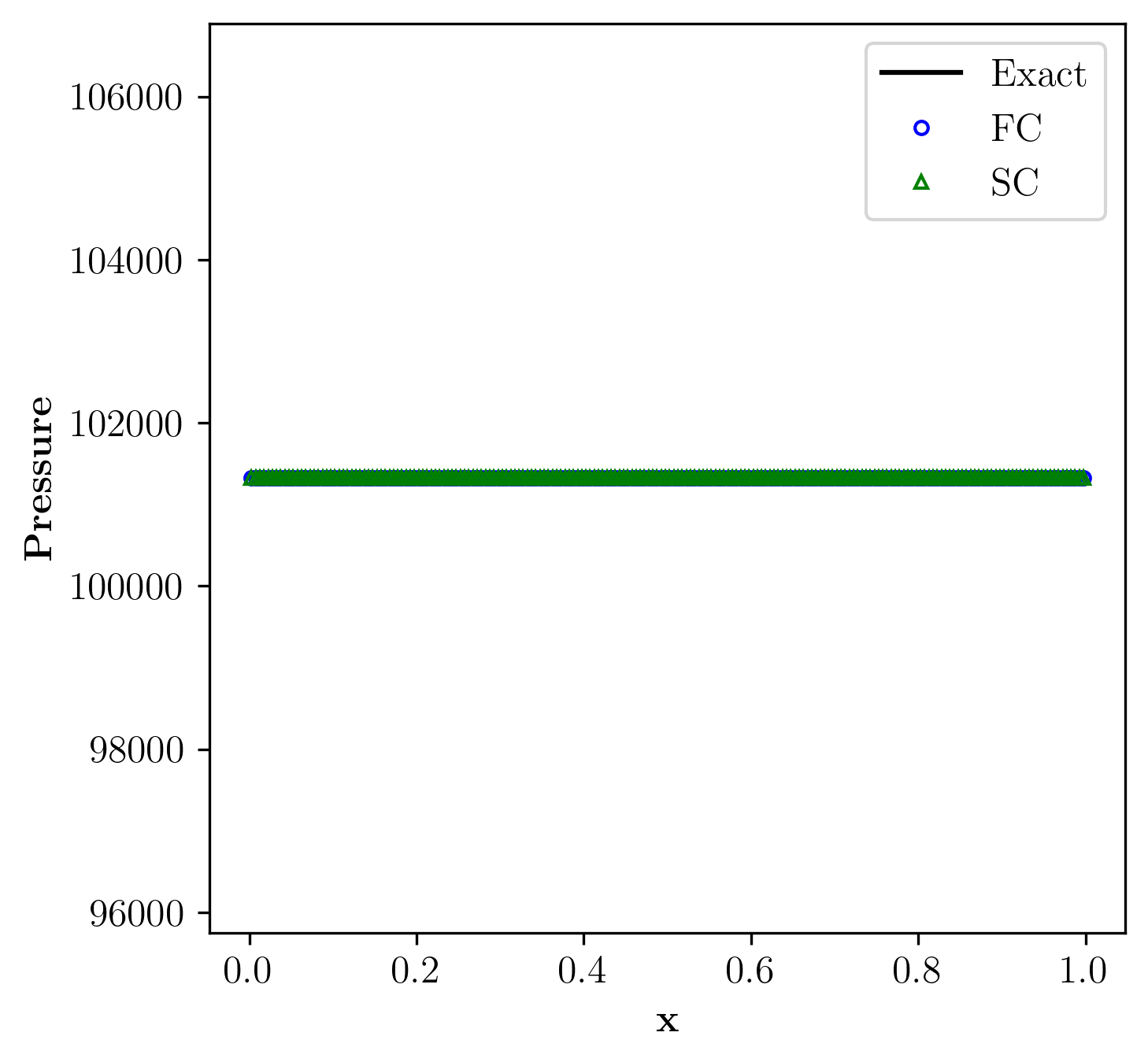}
\label{fig:iso-met_p}}
\subfigure[Volume fraction profile, Example \ref{ex:isol}]{\includegraphics[width=0.45\textwidth]{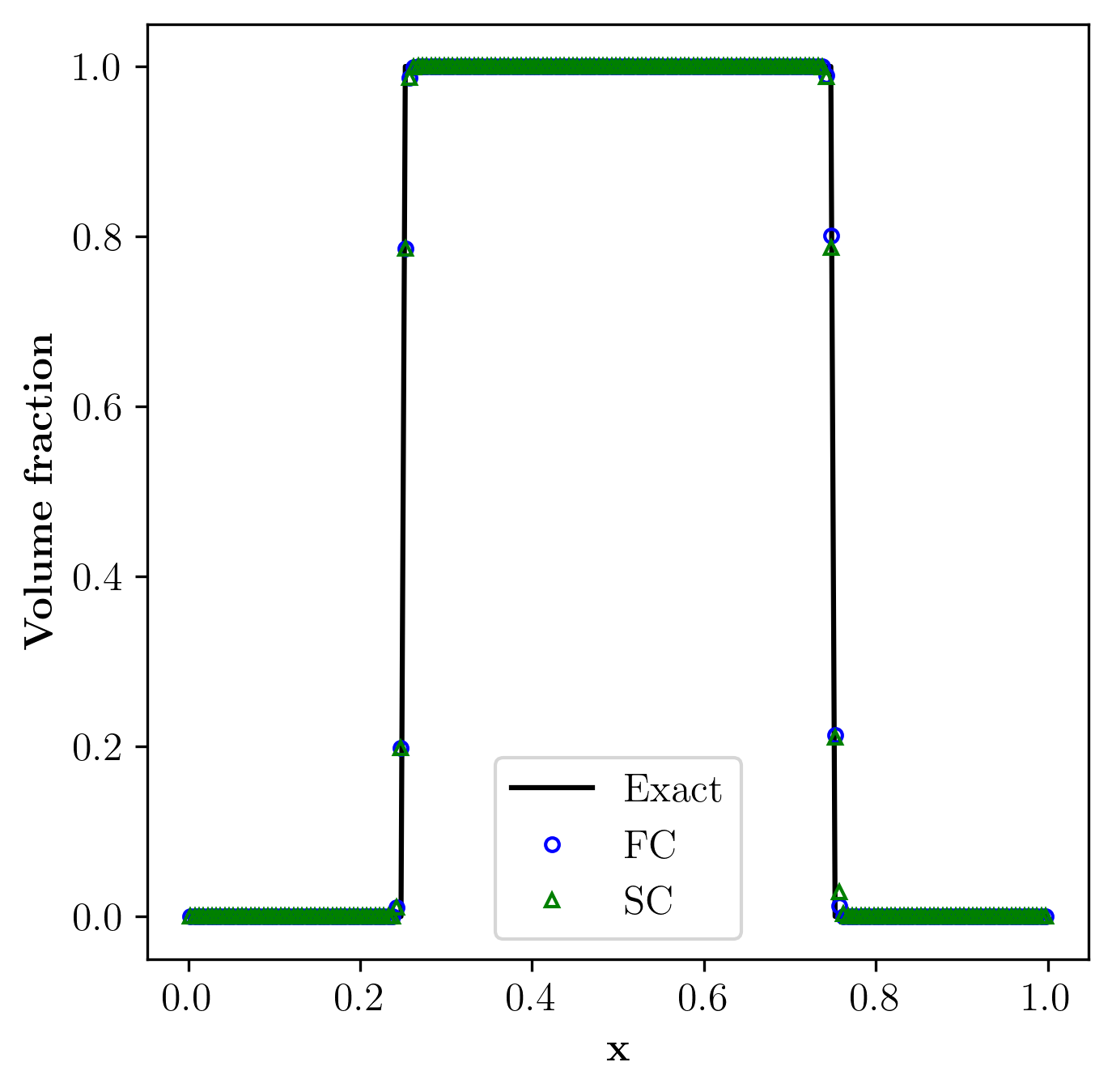}
\label{fig:iso-met2_vf}}
    \caption{Numerical solution for isolated contact test case using $N = 200$ grid points at $t = 0.1$,  Example \ref{ex:isol}, where Solid or Dashed line: reference solution; blue circles: FC; green triangles: SC for density, velocity, volume fraction, and pressure.}
    \label{fig_iso-ml-p}
\end{figure}

\begin{figure}[H]
\centering
\subfigure[Pressure profile (conservative), Example \ref{ex:isol}]{\includegraphics[width=0.45\textwidth]{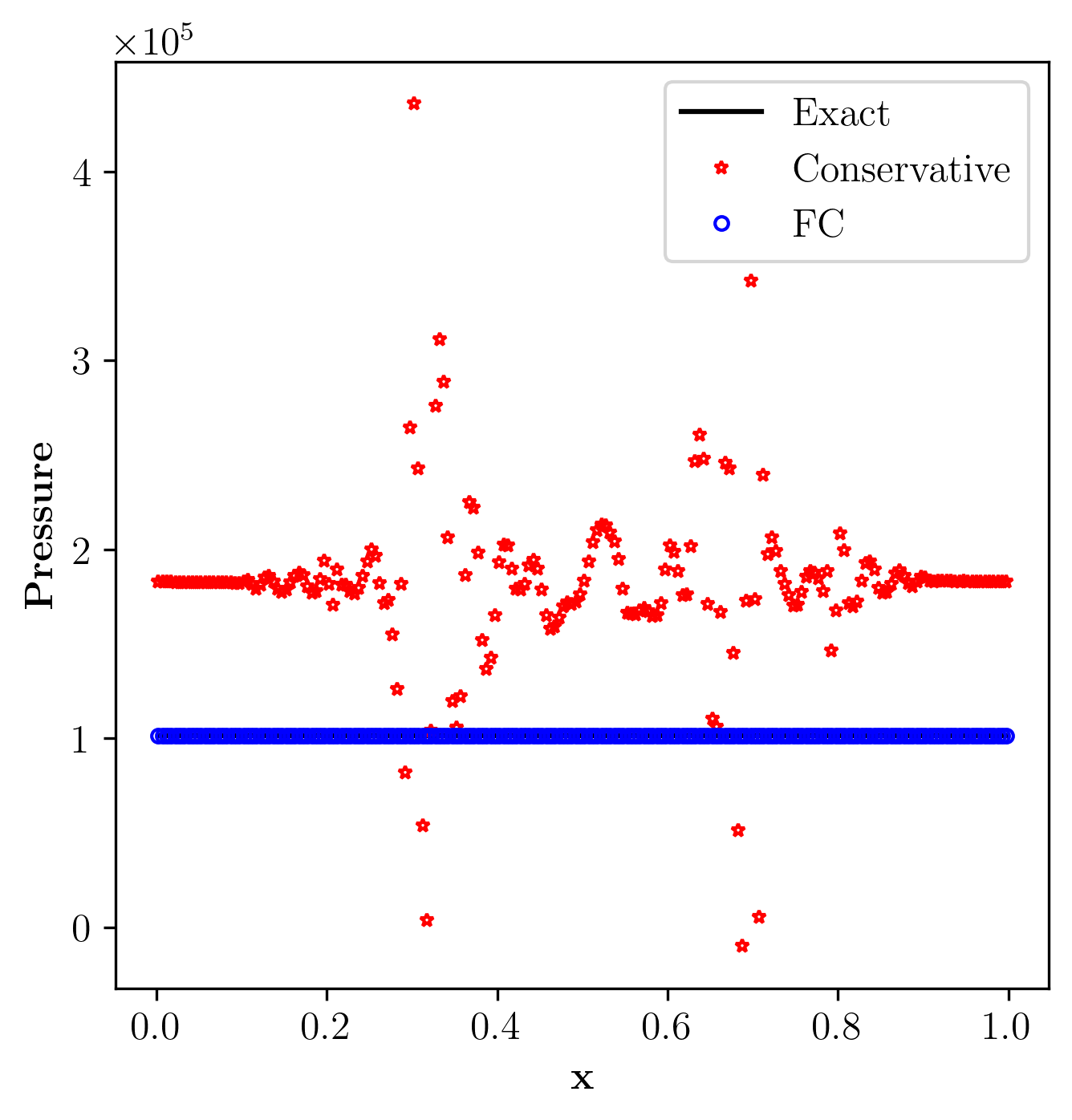}
\label{fig:iso-bad_p}}
\subfigure[Velocity profile (conservative), Example \ref{ex:isol}]{\includegraphics[width=0.45\textwidth]{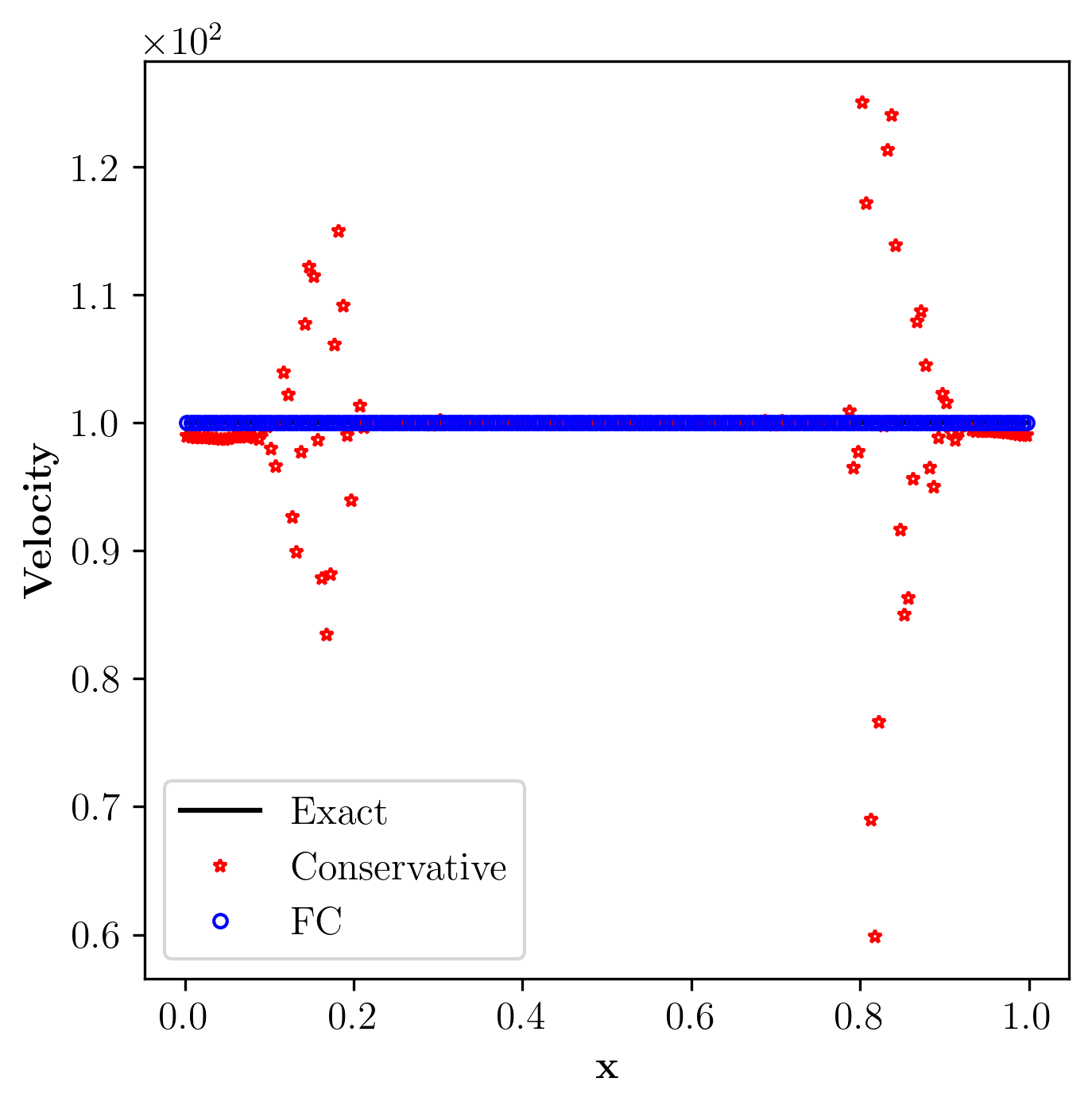}
\label{fig:iso-bad_vf}}
\subfigure[Pressure profile (hybrid), Example \ref{ex:isol}]{\includegraphics[width=0.45\textwidth]{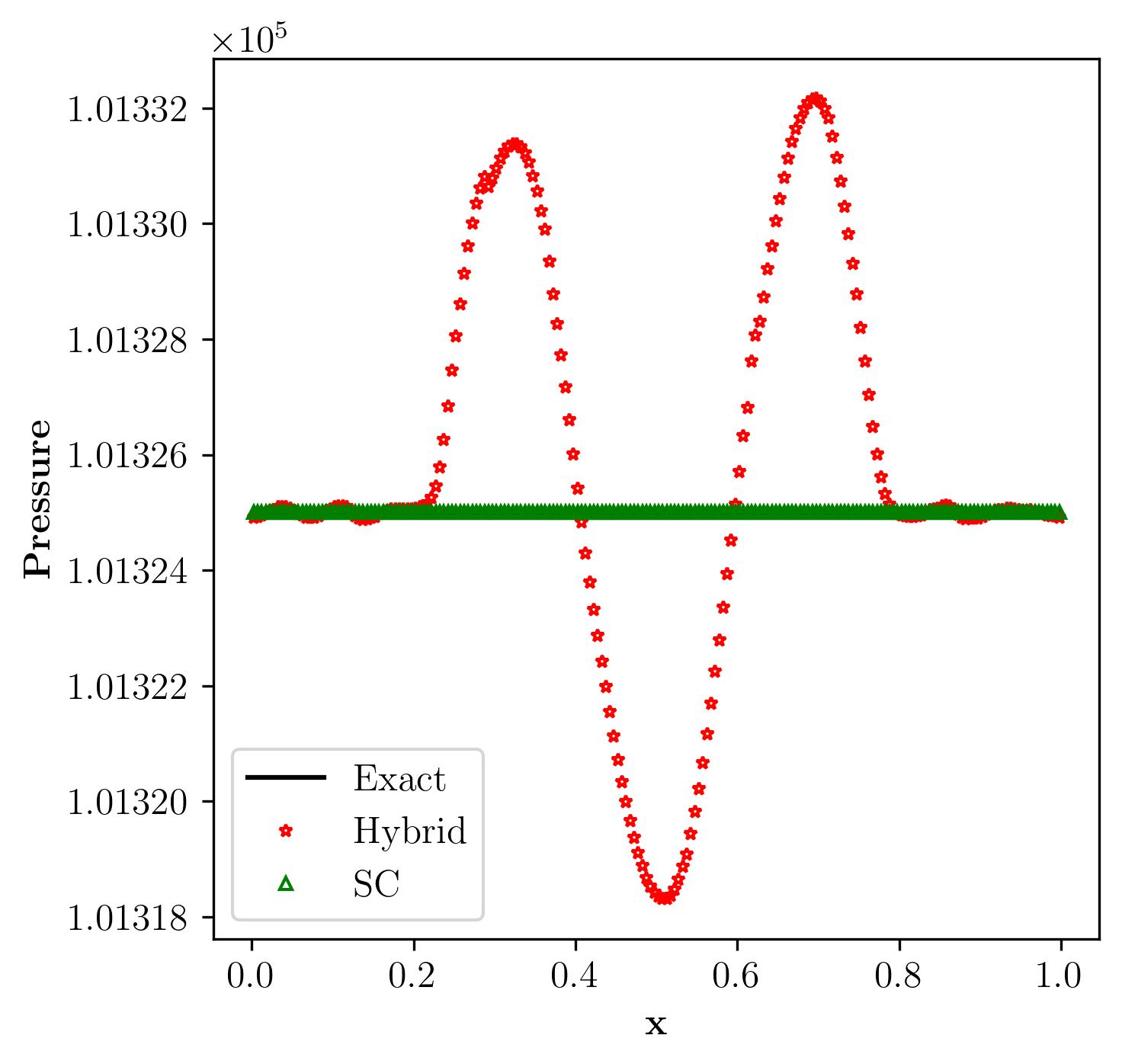}
\label{fig:iso-bad_l}}
\subfigure[Velocity profile (hybrid), Example \ref{ex:isol}]{\includegraphics[width=0.45\textwidth]{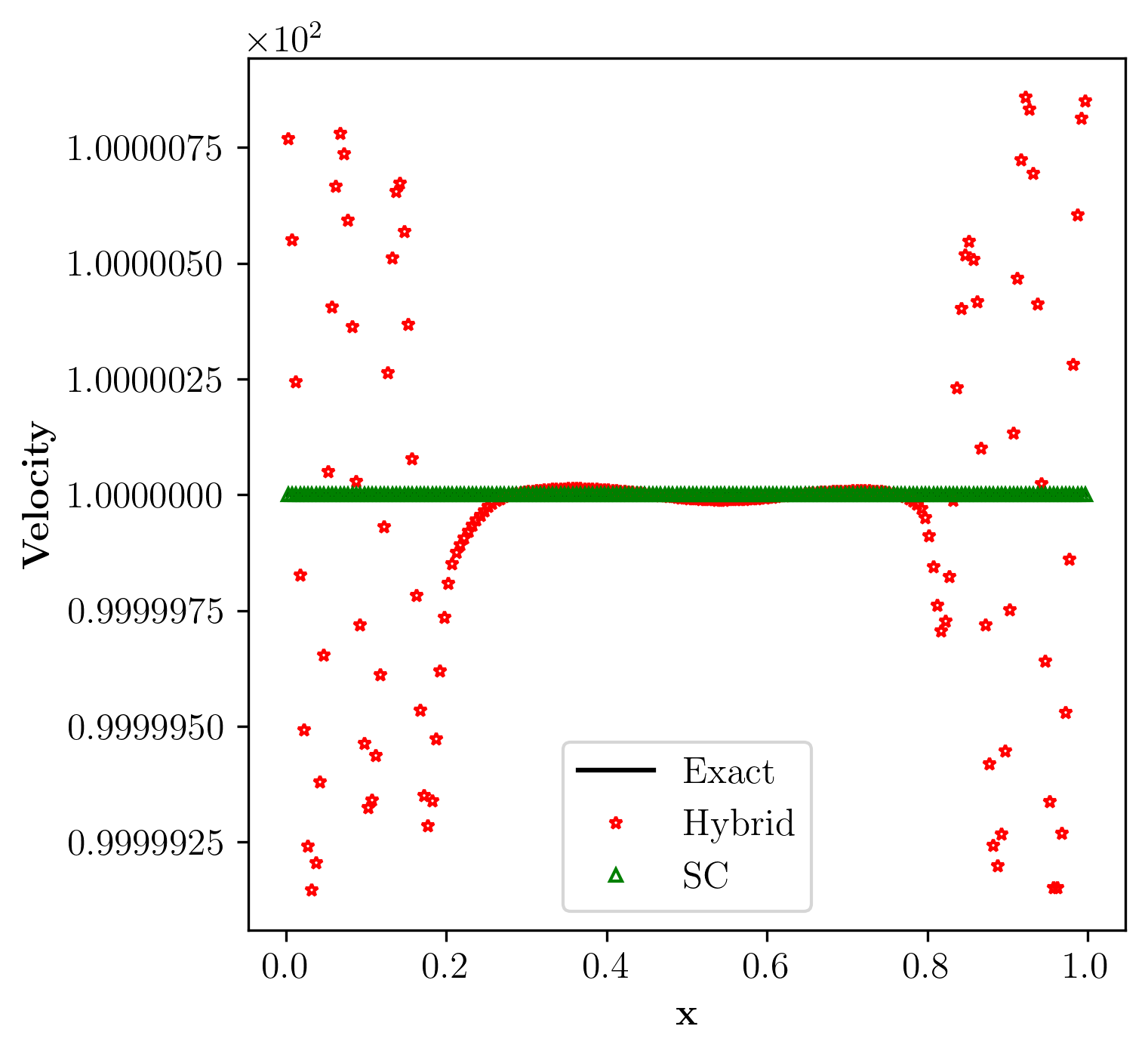}
\label{fig:iso-bad_l2}}
    \caption{Numerical solution for isolated contact test case using $N = 200$ grid points at $t = 0.1$,  Example \ref{ex:isol}, where Solid line: reference solution; blue circles: FC; green triangles: SC; red star: conservative or hybrid variables for both pressure and velocity.}
    \label{fig_iso-ml-c}
\end{figure}

The isolated material interface advection test provides the most direct assessment of pressure-equilibrium preservation: a water block is carried at a constant velocity through an air domain with periodic boundary conditions, so that no shocks or rarefactions are generated, and any spurious oscillations in pressure or velocity arise solely from the reconstruction of the interface itself. This test was taken from~\cite{coralic2014finite,wong2021positivity} and is a standard benchmark for diffuse-interface methods. The domain is $x \in [0,1]$, discretised on $N=200$ cells, and the simulation is advanced to $t_{\mathrm{end}} = 0.1$:
\begin{equation}\label{eq:tc5}
(\alpha_1\rho_1,\;\alpha_2\rho_2,\;u,\;p,\;\alpha_1) = \begin{cases}
(1000,\; 0,\; 100,\; 101325,\; 1) & 0.25 \leq x < 0.75,\\
(0,\; 1.2,\; 100,\; 101325,\; 0) & \text{otherwise},
\end{cases}
\end{equation}
with fluid properties $\gamma_1 = 4.4$, $\pi_1 = 6\times10^8$, $\gamma_2 = 1.4$, $\pi_2 = 0$. \Rfour{In the numerical implementation, pure-phase values in Eq.~\eqref{eq:tc5} are initialized as $\alpha_1=1-\epsilon_\alpha$ and $\alpha_1=\epsilon_\alpha$, with $\epsilon_\alpha=10^{-8}$, to avoid singular EOS evaluations in nominally absent phases.} Figure~\ref{fig_iso-ml-p} shows density, velocity, pressure, and volume fraction profiles for the FC and SC formulations with characteristic-space THINC reconstruction. Both schemes transport the water block without generating spurious pressure or velocity oscillations: the pressure remains constant at $p = 101325$~Pa, and the velocity remains uniform at $u = 100$~ms$^{-1}$ throughout the domain. The volume fraction profile is captured within a few cells and retains its initial profile with minimal diffusion.
Figure~\ref{fig_iso-ml-c} illustrates the consequences of bypassing characteristic decomposition and directly reconstructing conservative or hybrid (non-primitive) variables. Independent reconstruction of phasic densities or mixture momentum in physical space causes the nonlinearity of the MUSCL scheme to violate the relation $(\rho u)^* = \rho^* u$ at the interface, resulting in $O(1)$ oscillations in pressure and velocity that persist throughout the simulation. These findings support the central practical implication of the theoretical analysis in Section~\ref{sec:reconstruction} that the reconstruction must be performed in characteristic space to preserve equilibrium at material interfaces, regardless of whether the full-conservative (FC) or semi-conservative (SC) eigensystem is used. Direct reconstruction of conservative variables produces significantly more oscillations than reconstruction of semi-conservative variables.

\begin{example}\label{ex:multi-sod}{Two-Material Sod Shock Tube}
\end{example}

This case is a two-material extension of the classical Sod shock tube, designed to test the preservation of mechanical equilibrium across a material interface under Sod-type initial data \cite{Karni1994,abgrall2001computations}. The domain is $x \in [-5,5]$, discretised on $N=200$ cells, and the simulation is run to $t_{\mathrm{end}} = 2.0$. In the left half, the flow is dominated by fluid~1, and in the right half by fluid~2, each with a distinct adiabatic exponent:
\begin{equation}\label{eq:tc3}
(\alpha_1\rho_1,\;\alpha_2\rho_2,\;u,\;p,\;\alpha_1) = \begin{cases}
(1-10^{-6},\; 0.125\times10^{-6},\; 0,\; 1,\; 1-10^{-6}) & x < 0,\\
(10^{-6},\; 0.125\times(1-10^{-6}),\; 0,\; 0.1,\; 10^{-6}) & x \geq 0,
\end{cases}
\end{equation}
with fluid properties $\gamma_1 = 1.4$, $\pi_1 = 0$, $\gamma_2 = 1.6$, $\pi_2 = 0$.

\begin{figure}[H]
\centering 
\subfigure[Density]{\includegraphics[width=0.4\textwidth]{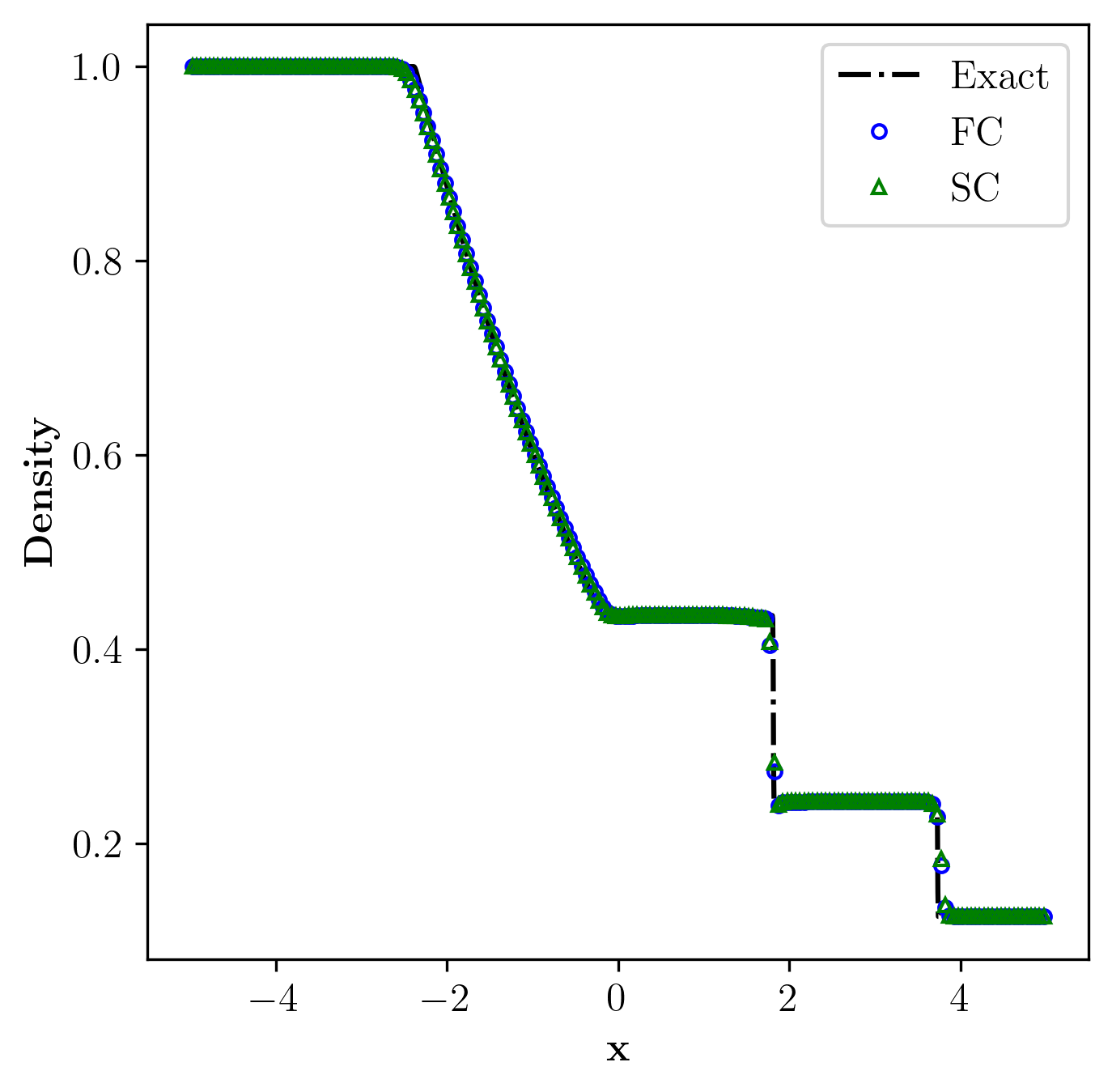}
\label{fig:multi_sod-den}}
\subfigure[Velocity]{\includegraphics[width=0.4\textwidth]{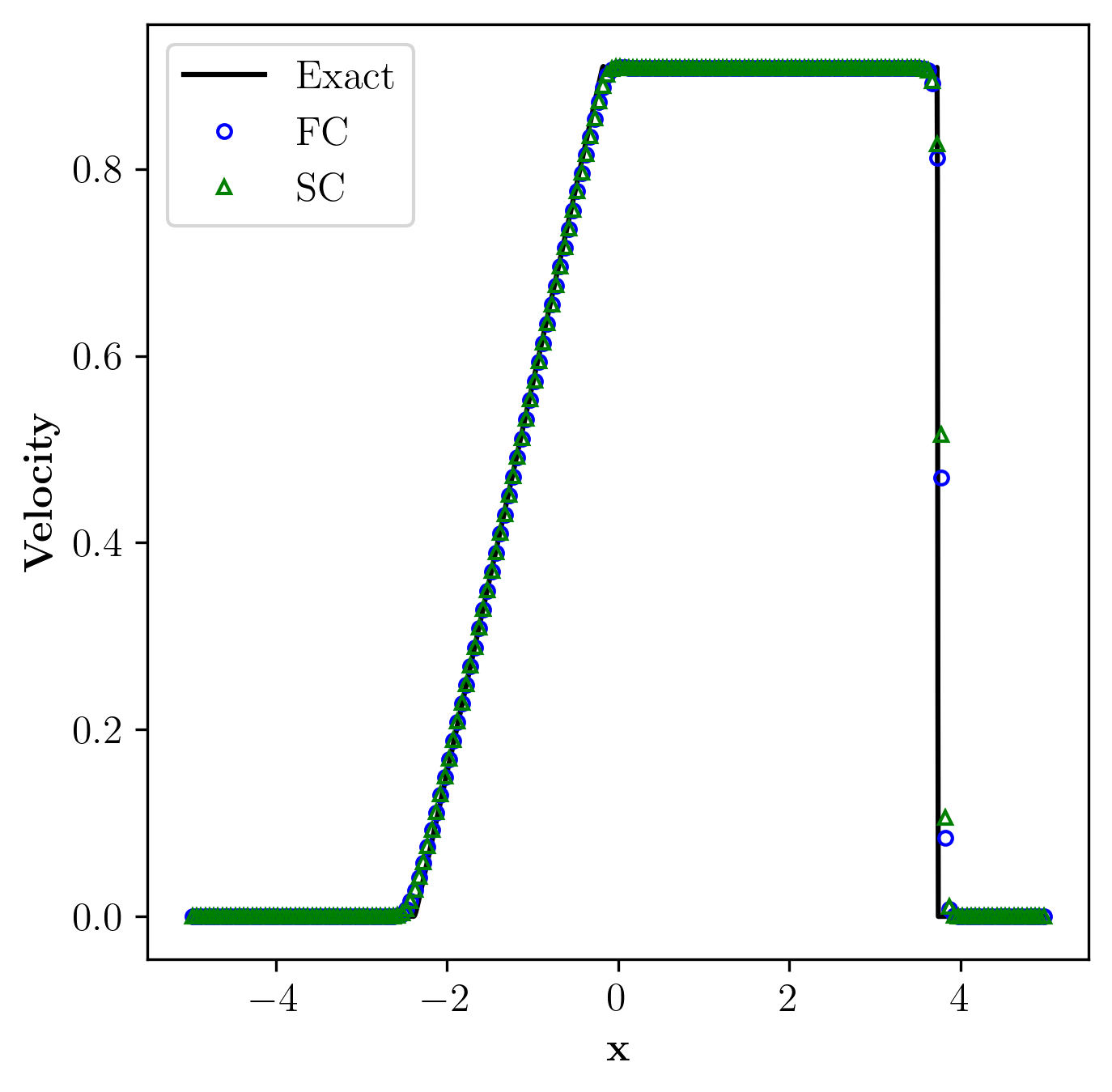}
\label{fig:multi_sod-vel}}
\subfigure[Pressure]{\includegraphics[width=0.4\textwidth]{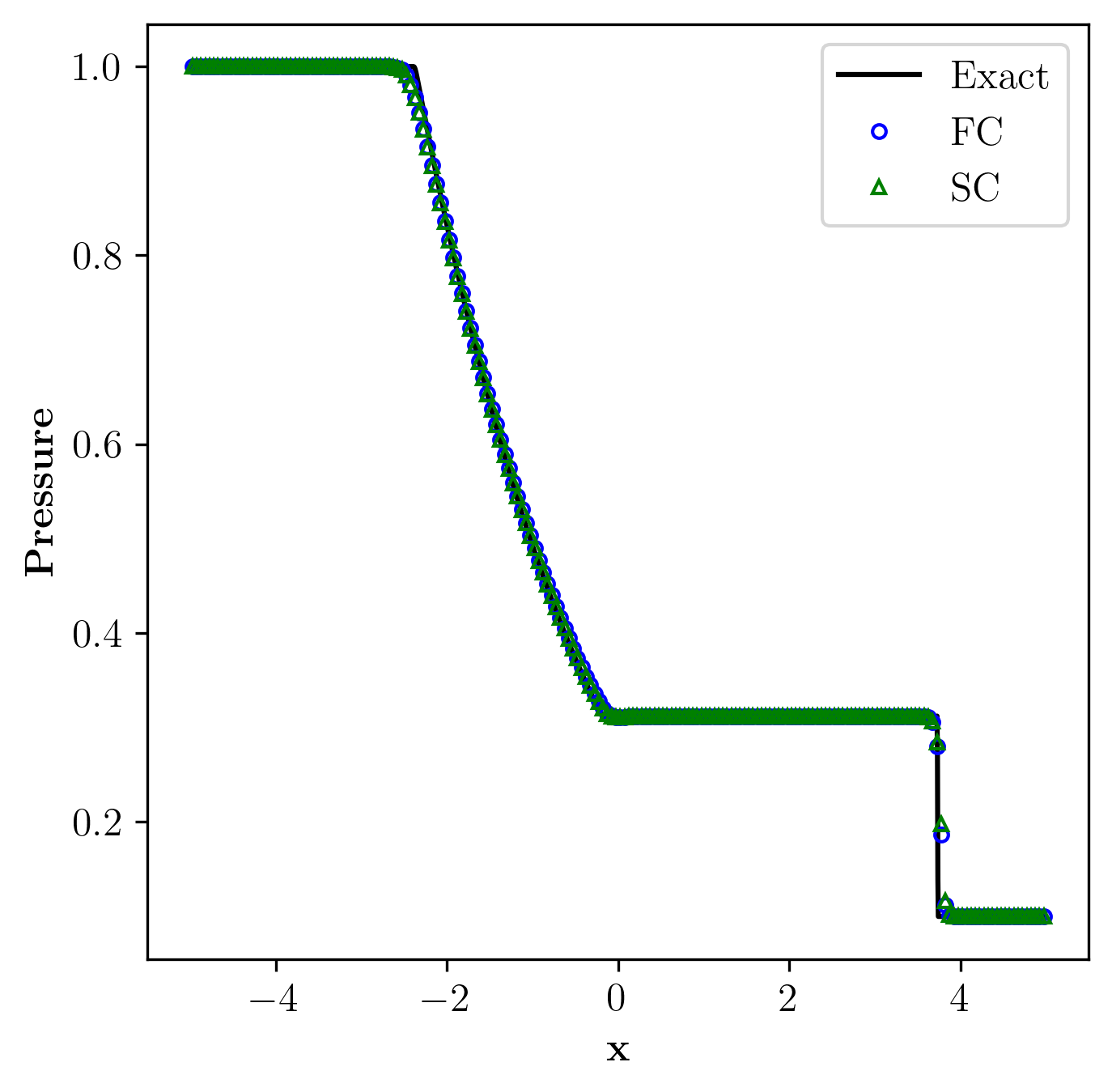}
\label{fig:multi_sod-pres}}
\subfigure[Volume fraction]{\includegraphics[width=0.4\textwidth]{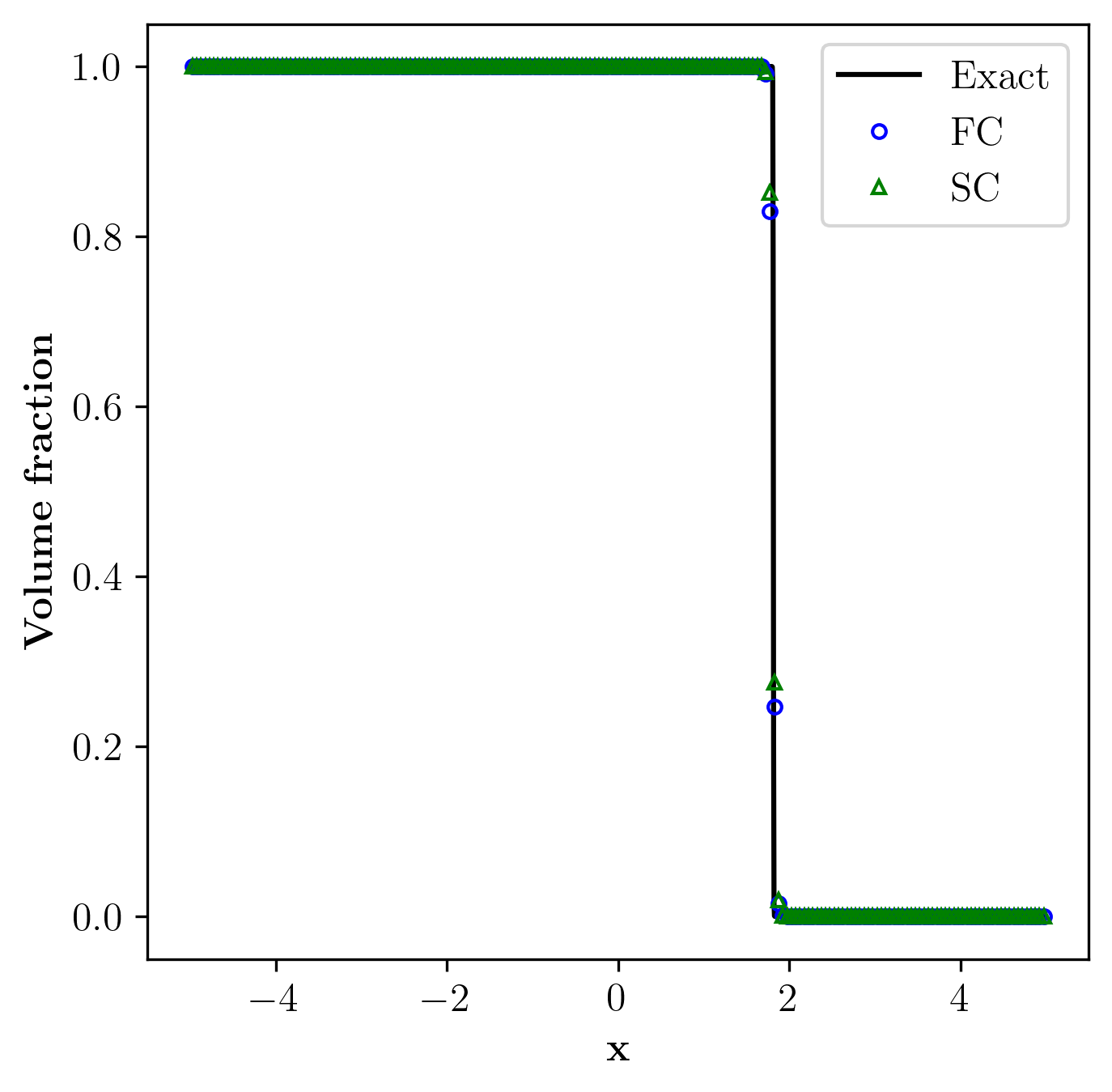}
\label{fig:multi_sod-volf}}
\caption{Numerical solution for multi-species shock tube problem in Example \ref{ex:multi-sod} on a grid size of $N=200$. Solid or Dashed line: reference solution; blue circles: FC; green triangles: SC for density, velocity, volume fraction, and pressure.}
\label{fig_multisod}
\end{figure}

Figure~\ref{fig_multisod} compares density, velocity, pressure, and volume fraction against the exact solution for both the FC and SC formulations. The solutions agree closely with the reference across the rarefaction fan, the contact discontinuity, and the right-propagating shock. Because the two fluids have different $\gamma_k$, the contact wave carries an equation-of-state mismatch that would generate spurious pressure oscillations if single-species eigenvectors were used. The clean pressure profile in Figure~\ref{fig_multisod}(c) confirms that the multi-phase $\Psi$ term (FC) and the structural equilibrium of the SC formulation both handle the $\gamma$ mismatch correctly.

\begin{figure}[H]
    \centering
    \includegraphics[width=0.4\textwidth]{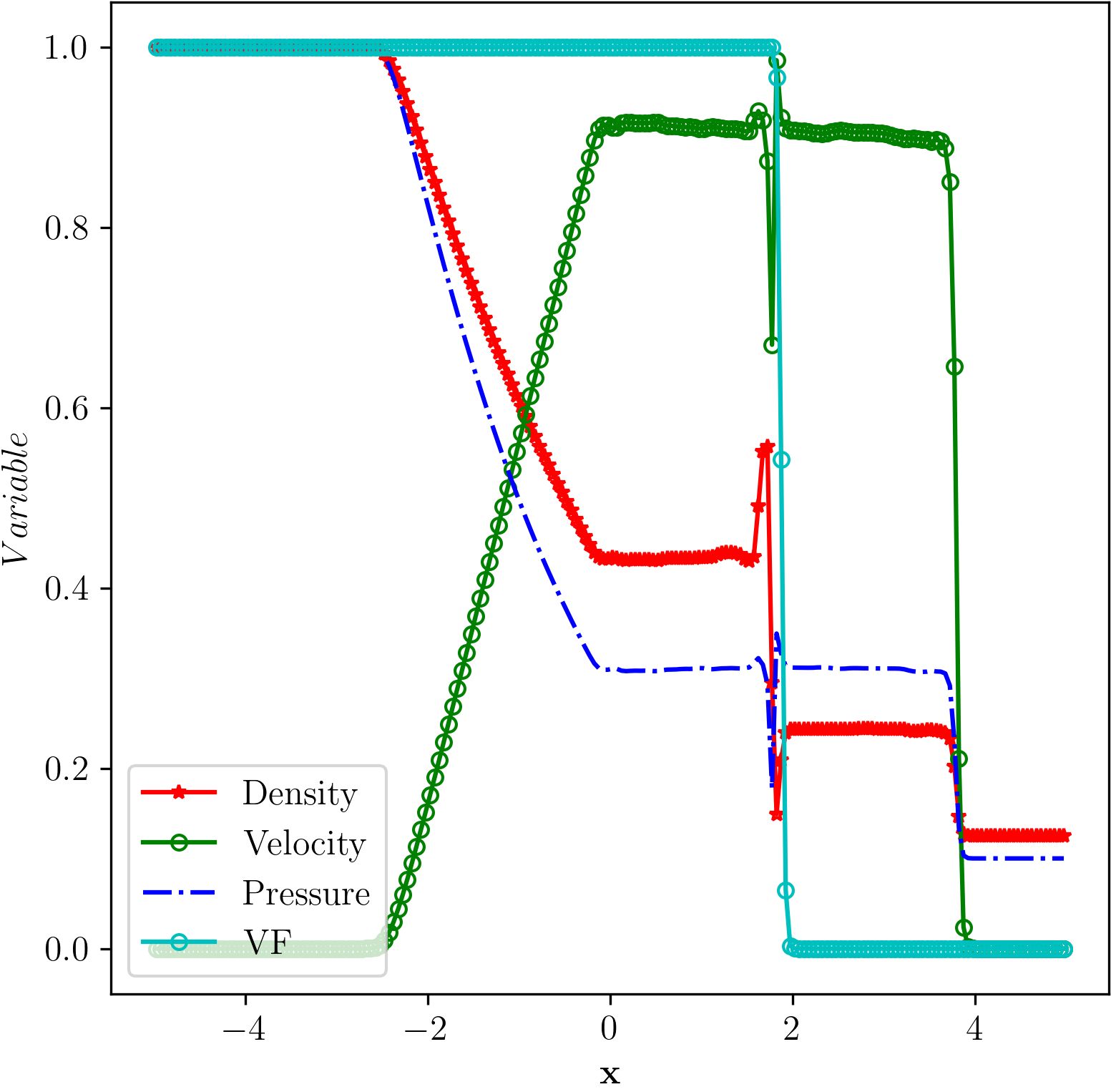}
    \caption{Numerical solution for the multi-species shock tube problem, 
    Example~\ref{ex:multi-sod}, using direct reconstruction of 
    semi-conservative variables in physical space.}
    \label{fig:bad}
\end{figure}

Furthermore, Figure~\ref{fig:bad} shows the results obtained by directly reconstructing the semi-conservative variables in physical space. Significant oscillations in density, velocity, and pressure are observed, consistent with the argument of Johnsen~\cite{Johnsen2011} that characteristic variables must be reconstructed; the present results extend this conclusion to show that characteristic-space reconstruction is necessary regardless of the variable set employed.

\begin{example}\label{ex:curtain}{Inviscid shock-curtain interaction}
\end{example}

The shock-curtain problem introduces a finite-width curtain of fluid 2 (helium-like, $\gamma_2 = 1.67$) into a domain otherwise filled with fluid 1 (air-like, $\gamma_1 = 1.4$)\cite{wong2021positivity,chamarthi2023gradient}. A right-propagating shock, initiated at the left boundary, interacts with both curtain interfaces, producing transmitted and reflected waves and two contact discontinuities. This setup tests the reconstruction method's ability to resolve two material interfaces simultaneously without cross-contamination of characteristic fields. The computational domain is $x \in [0,1]$, discretized into $N=200$ cells, and the simulation runs until $t_{\mathrm{end}} = 0.3$.
\begin{equation}\label{eq:tc4}
(\alpha_1\rho_1,\;\alpha_2\rho_2,\;u,\;p,\;\alpha_1) = \begin{cases}
(1.3765,\; 0,\; 0.3948,\; 1.57,\; 1) & 0 \leq x < 0.25,\\
(1,\; 0,\; 0,\; 1,\; 1) & 0.25 \leq x < 0.4 \text{ or } 0.6 \leq x < 1,\\
(0,\; 0.138,\; 0,\; 1,\; 0) & 0.4 \leq x < 0.6,
\end{cases}
\end{equation}
with fluid properties $\gamma_1 = 1.4$, $\pi_1 = 0$, $\gamma_2 = 1.67$, $\pi_2 = 0$.

\begin{figure}[H]
\centering
\subfigure[Density]{\includegraphics[width=0.4\textwidth]{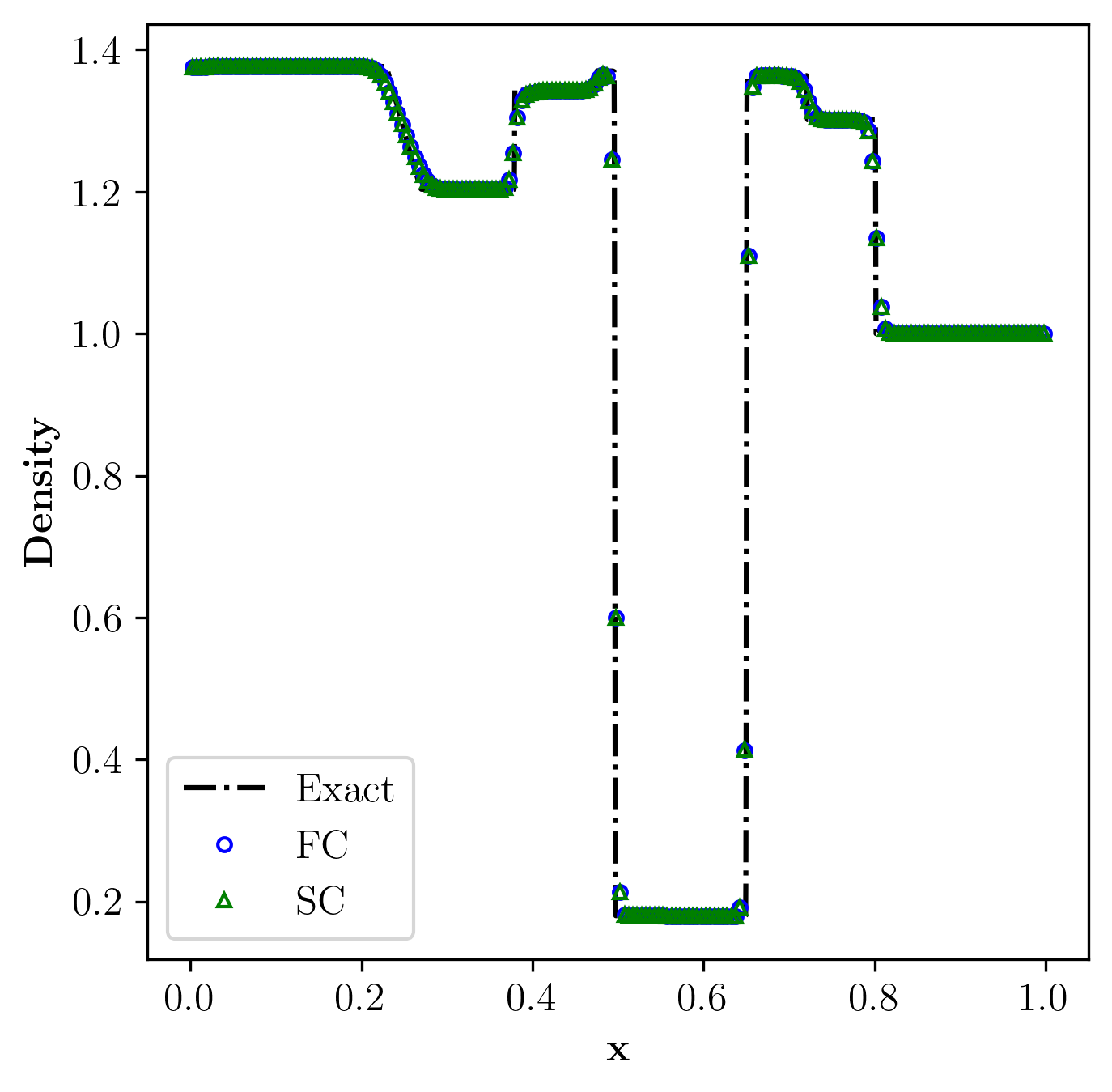}
\label{fig:multi_curtain-den}}
\subfigure[Velocity]{\includegraphics[width=0.4\textwidth]{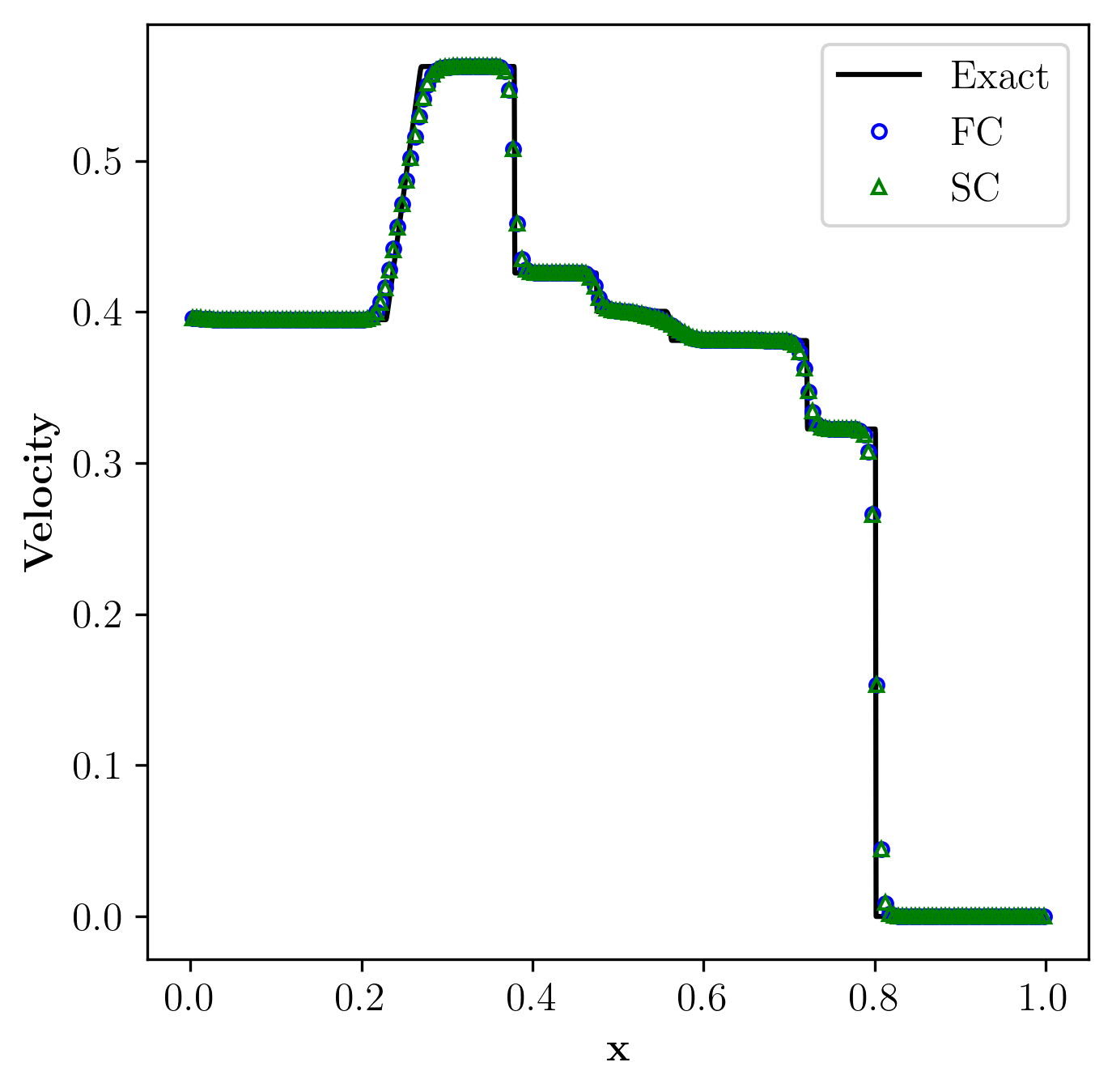}
\label{fig:multi_curtain-vel}}
\subfigure[Pressure]{\includegraphics[width=0.4\textwidth]{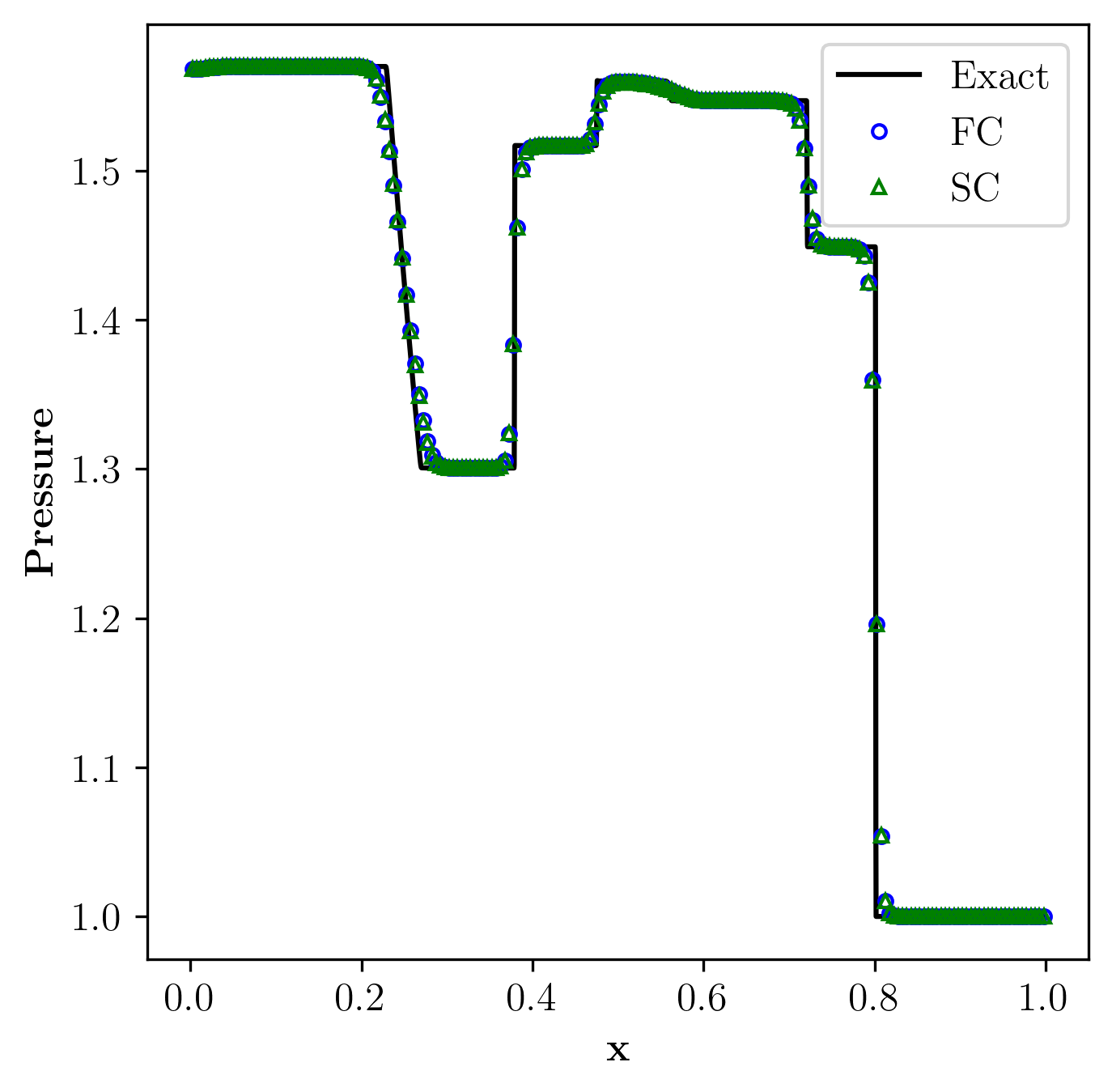}
\label{fig:multi_curtain-pres}}
\subfigure[Volume fraction, $\alpha_1$]{\includegraphics[width=0.4\textwidth]{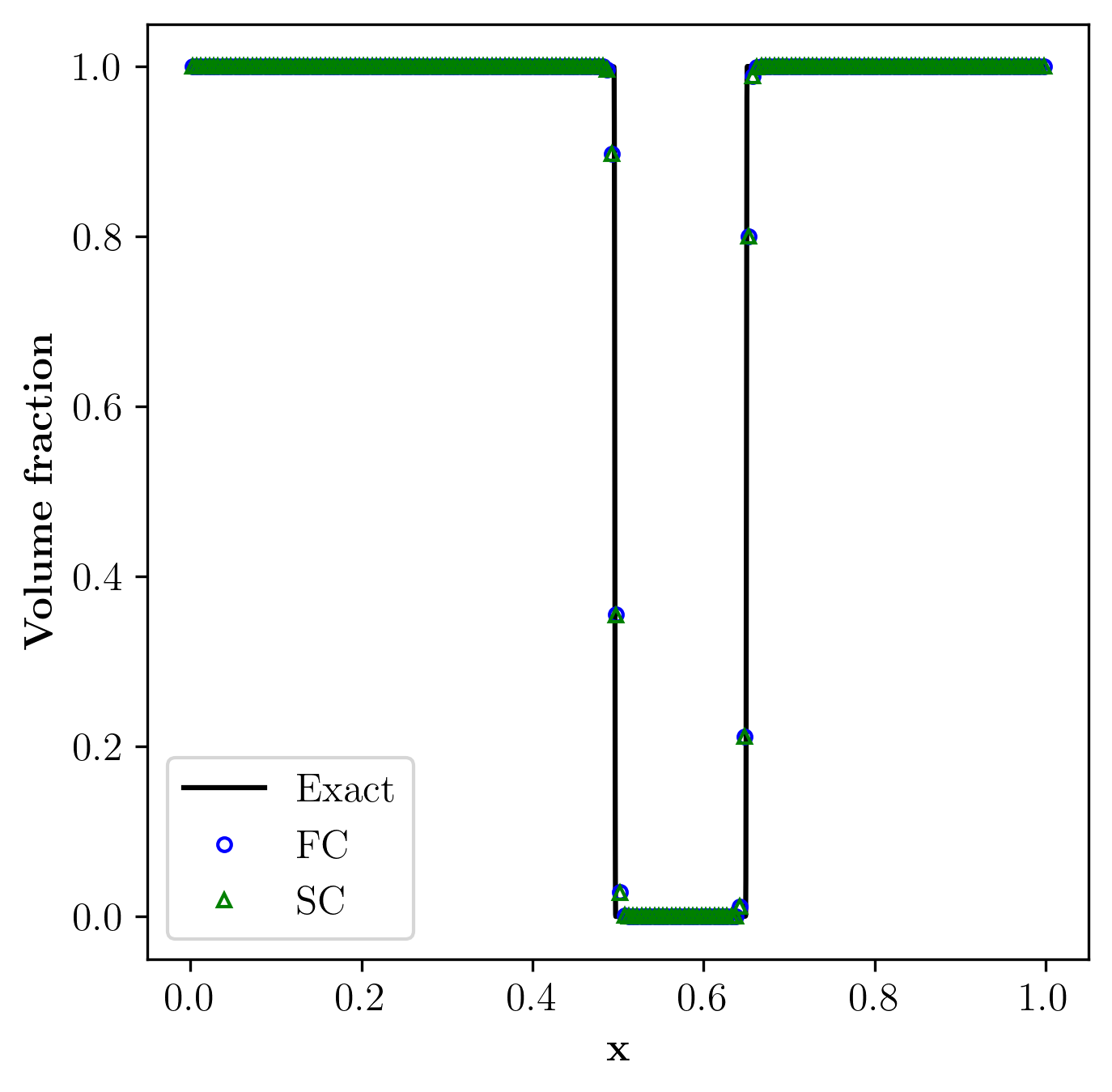}
\label{fig:multi_curtain-alpha}}
\caption{Numerical solution for the shock-curtain interaction problem in Example \ref{ex:curtain} on a grid with $N=200$ cells. Solid or dashed lines represent the reference solution; blue circles denote FC results; green triangles indicate SC results for density, velocity, volume fraction, and pressure.}
\label{fig_multi_curtain}
\end{figure}

Figure~\ref{fig_multi_curtain} shows the density, velocity, pressure, and volume fraction profiles at the final simulation time for both the FC and SC schemes, compared with the reference solution. Both numerical methods accurately capture the transmitted shock, the two contact discontinuities at the curtain boundaries, and the reflected acoustic structure. Equilibrium is maintained across both material interfaces, with no observable oscillations in the velocity or pressure profiles. The volume fraction profile sharply delineates the interface.

\begin{example}\label{ex:stiff}{Three-Region Gas--Gas Shock Tube}
\end{example}

This three-region shock-tube problem features two ideal gases ($\pi_{\infty,k} = 0$) with different adiabatic exponents, providing a stringent test of characteristic reconstruction at a gas--gas interface. The domain $x \in [-1,1]$ is divided into $N=200$ cells, and the simulation runs to $t_{\mathrm{end}} = 0.07$. The initial conditions include a high-velocity region on the left, a stationary middle region of fluid~1, and a right region filled with fluid~2.
\begin{equation}\label{eq:tc2}
(\alpha_1\rho_1,\;\alpha_2\rho_2,\;u,\;p,\;\alpha_1) = \begin{cases}
(0.386,\; 0,\; 26.59,\; 100,\; 1) & -1 < x < -0.8,\\
(0.1,\; 0,\; -0.5,\; 1,\; 1) & -0.8 \leq x < -0.2,\\
(0,\; 1,\; -0.5,\; 1,\; 0) & -0.2 \leq x < 1,
\end{cases}
\end{equation}
with fluid properties $\gamma_1 = 1.67$, $\pi_1 = 0$, $\gamma_2 = 1.4$, $\pi_2 = 0$.

\begin{figure}[H]
\centering
\subfigure[Density]{\includegraphics[width=0.4\textwidth]{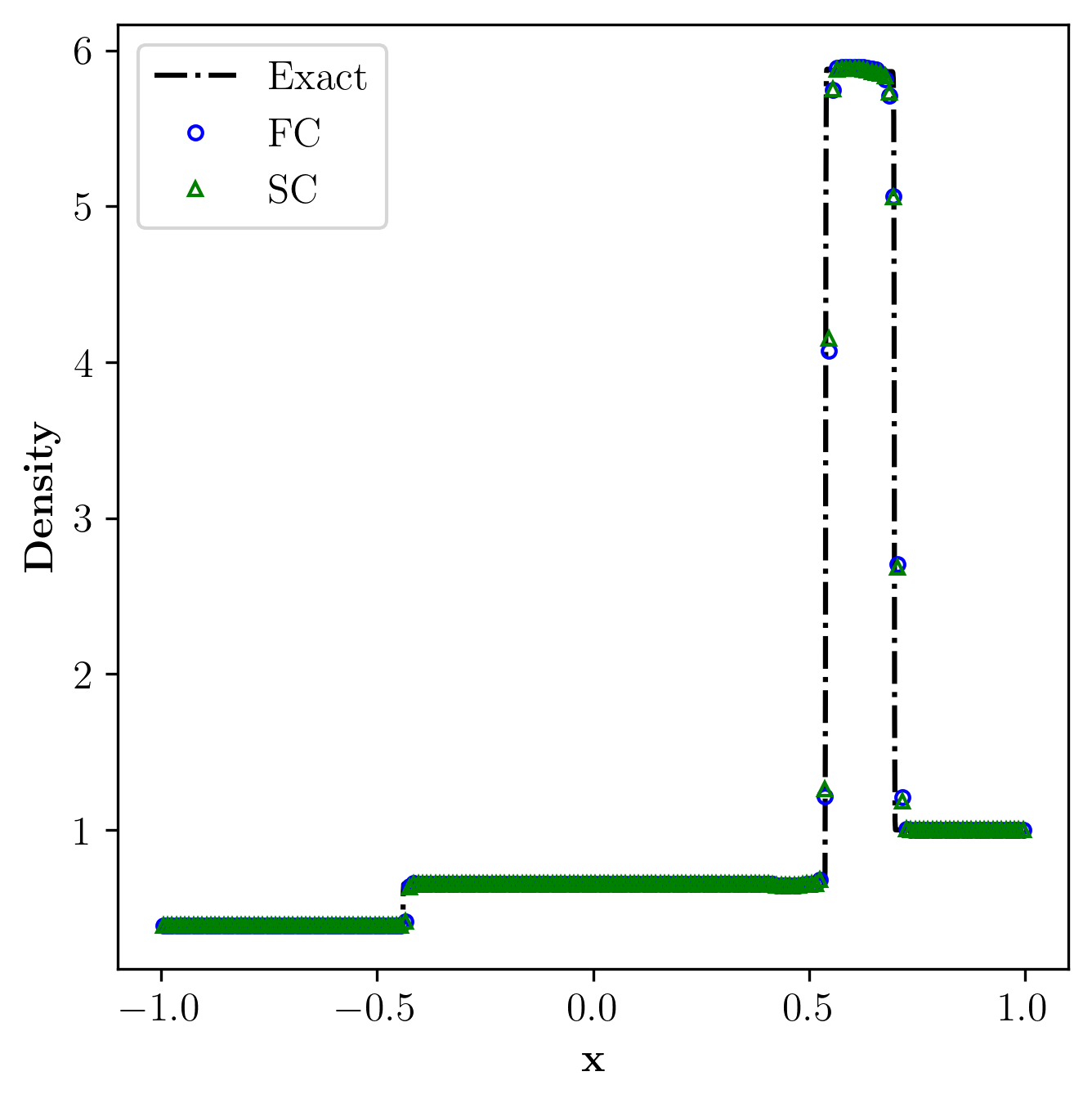}
\label{fig:multi_stiff-den}}
\subfigure[Velocity]{\includegraphics[width=0.4\textwidth]{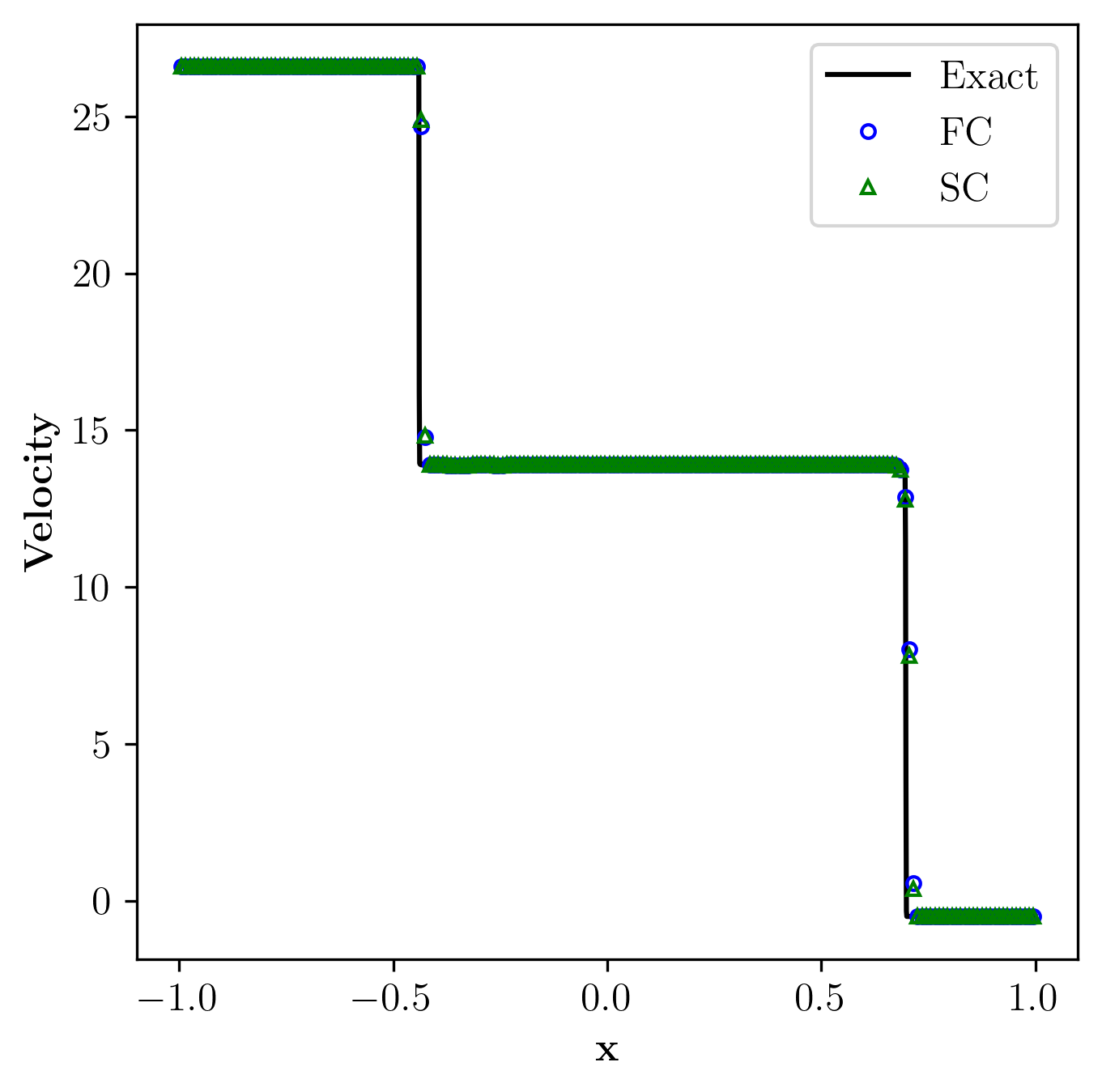}
\label{fig:multi_stiff-vel}}
\subfigure[Pressure]{\includegraphics[width=0.4\textwidth]{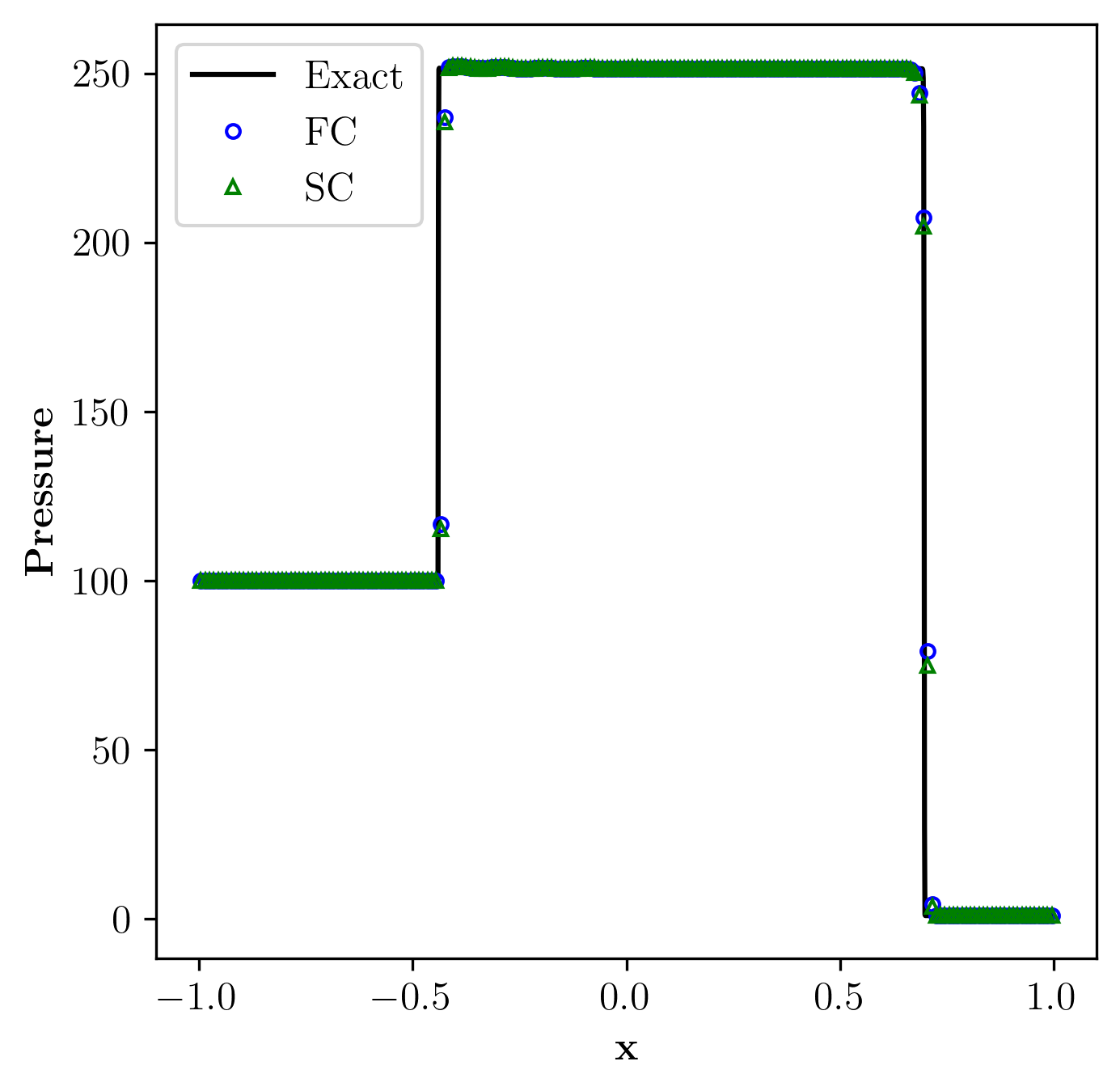}
\label{fig:multi_stiff-pres}}
\subfigure[Volume fraction, $\alpha_1$]{\includegraphics[width=0.4\textwidth]{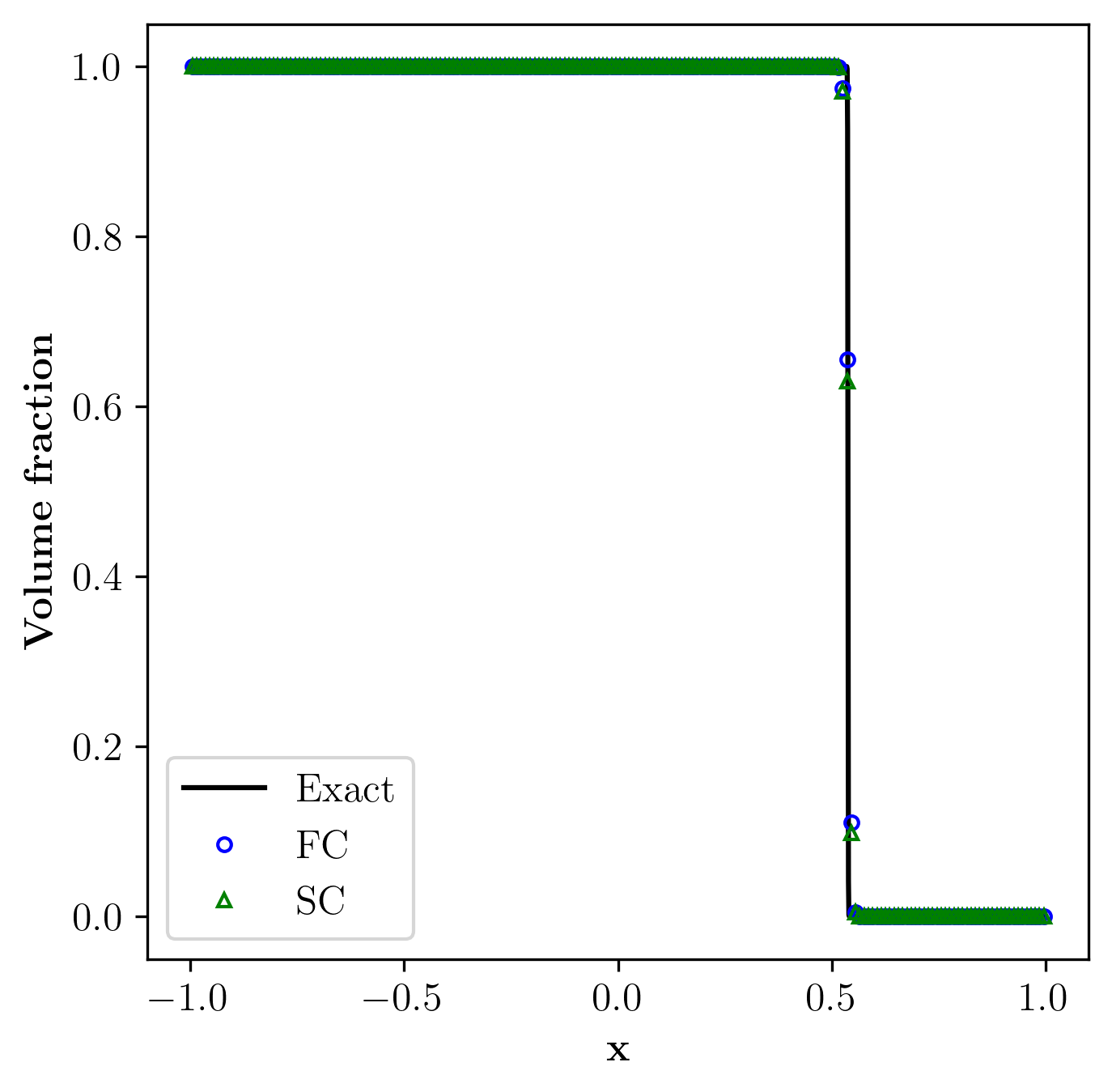}
\label{fig:multi_stiff-alpha}}
\caption{Numerical solution for the shock interface interaction problem in Example \ref{ex:stiff} on a grid with $N=200$ cells. Solid or dashed lines indicate the reference solution; blue circles represent FC; green triangles represent SC for density, velocity, volume fraction, and pressure.}
\label{fig_stiff}
\end{figure}

Figure~\ref{fig_stiff} presents density, velocity, pressure, and volume fraction profiles for the FC and SC formulations, compared with the exact reference solution. Both formulations accurately capture all wave structures: the left-propagating expansion fan, the contact discontinuity between the two gas species, and the right-propagating shock. The volume fraction transitions smoothly from unity to zero at the material interface, with no spurious acoustic signals in the pressure or velocity profiles. The material interfaces are captured within a few points using the THINC scheme for both the formulations, indicating that the contact discontinuity sensor proposed in \cite{chamarthi2025wave,chamarthi2025physics} seamlessly works with all sets of variables.

\begin{example}\label{ex:liquid}{Gas-liquid Riemann problem}
\end{example}

This test examines a one-dimensional shock-tube configuration involving a large-density-ratio gas--liquid interface, following the setup of~\cite{coralic2014finite}. The left state contains highly compressed air, while the right state is water at near-atmospheric pressure. This combination tests the characteristic reconstruction's ability to remain non-oscillatory in the presence of large thermodynamic mismatches arising from the stiffened-gas equation of state. The computational domain is $x \in [-1,1]$, discretised on $N = 200$ uniformly spaced cells, with a final time of $t_{\mathrm{end}} = 0.2$:
\begin{equation}\label{eq:tc1}
(\alpha_1\rho_1,\;\alpha_2\rho_2,\;u,\;p,\;\alpha_1) = \begin{cases}
(0,\; 1.241,\; 0,\; 2.753,\; 10^{-8}) & -1 \leq x < 0,\\
(0.991,\; 0,\; 0,\; 3.059\times10^{-4},\; 1-10^{-8}) & 0 \leq x \leq 1,
\end{cases}
\end{equation}
with fluid properties $\gamma_l = 5.5$, $\pi_l = 1.505$, $\gamma_g = 1.4$, $\pi_g = 0$.

\begin{figure}[H]
\centering
\subfigure[Density]{\includegraphics[width=0.43\textwidth]{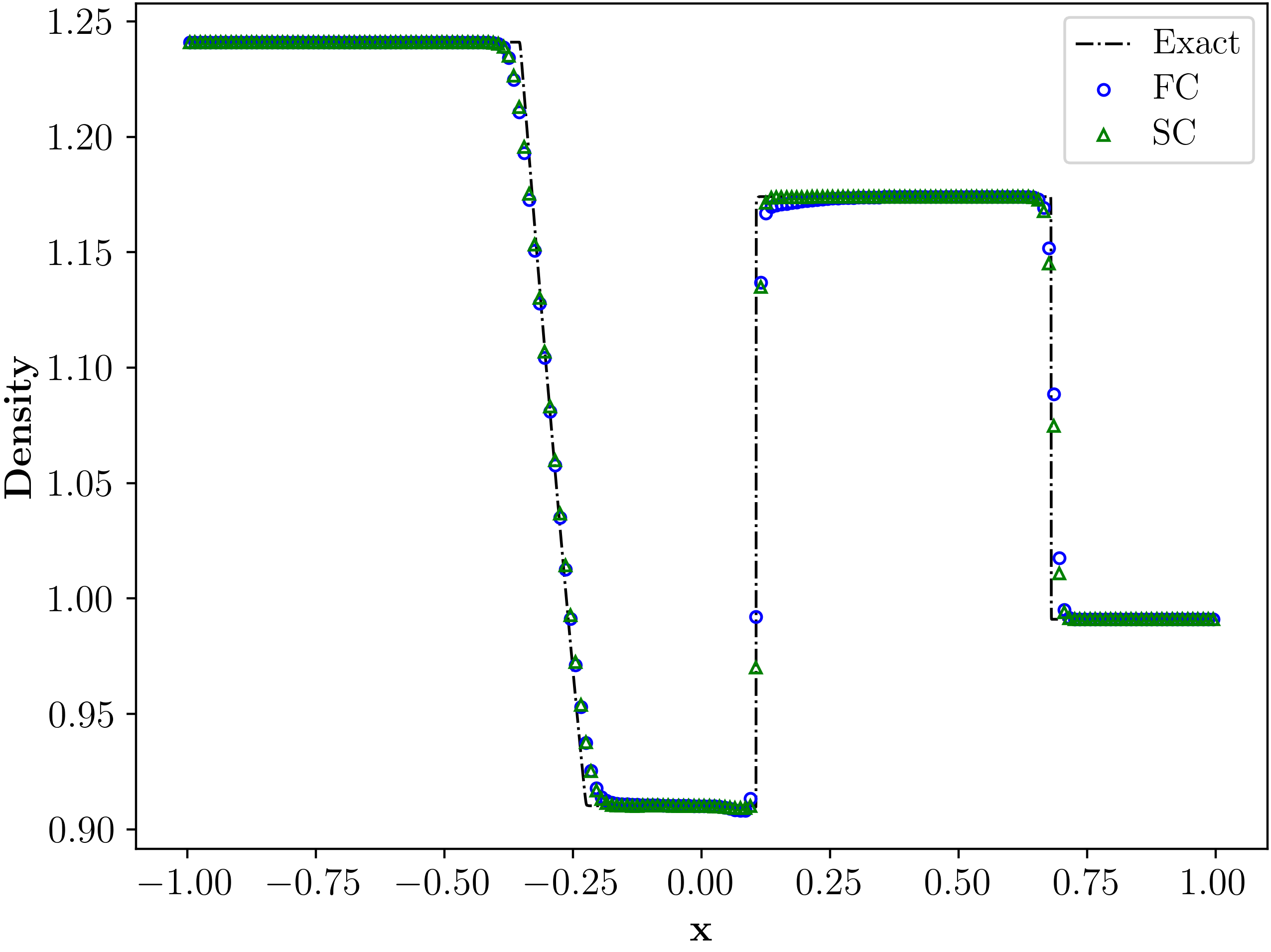}
\label{fig:multi_zhang-den}}
\subfigure[Pressure]{\includegraphics[width=0.415\textwidth]{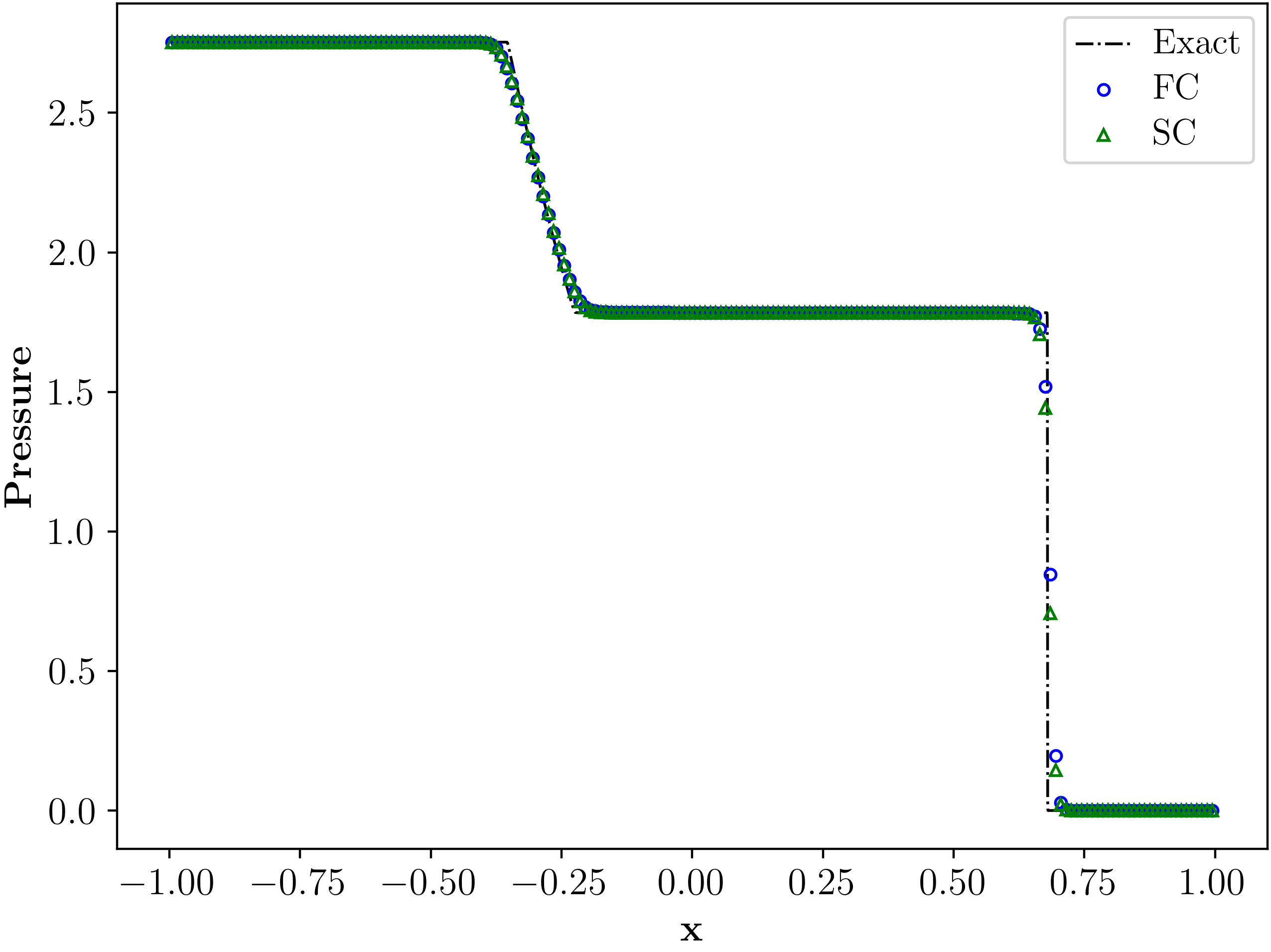}
\label{fig:multi_zhang-pres}}
\subfigure[\Rfour{Velocity}]{\includegraphics[width=0.43\textwidth]{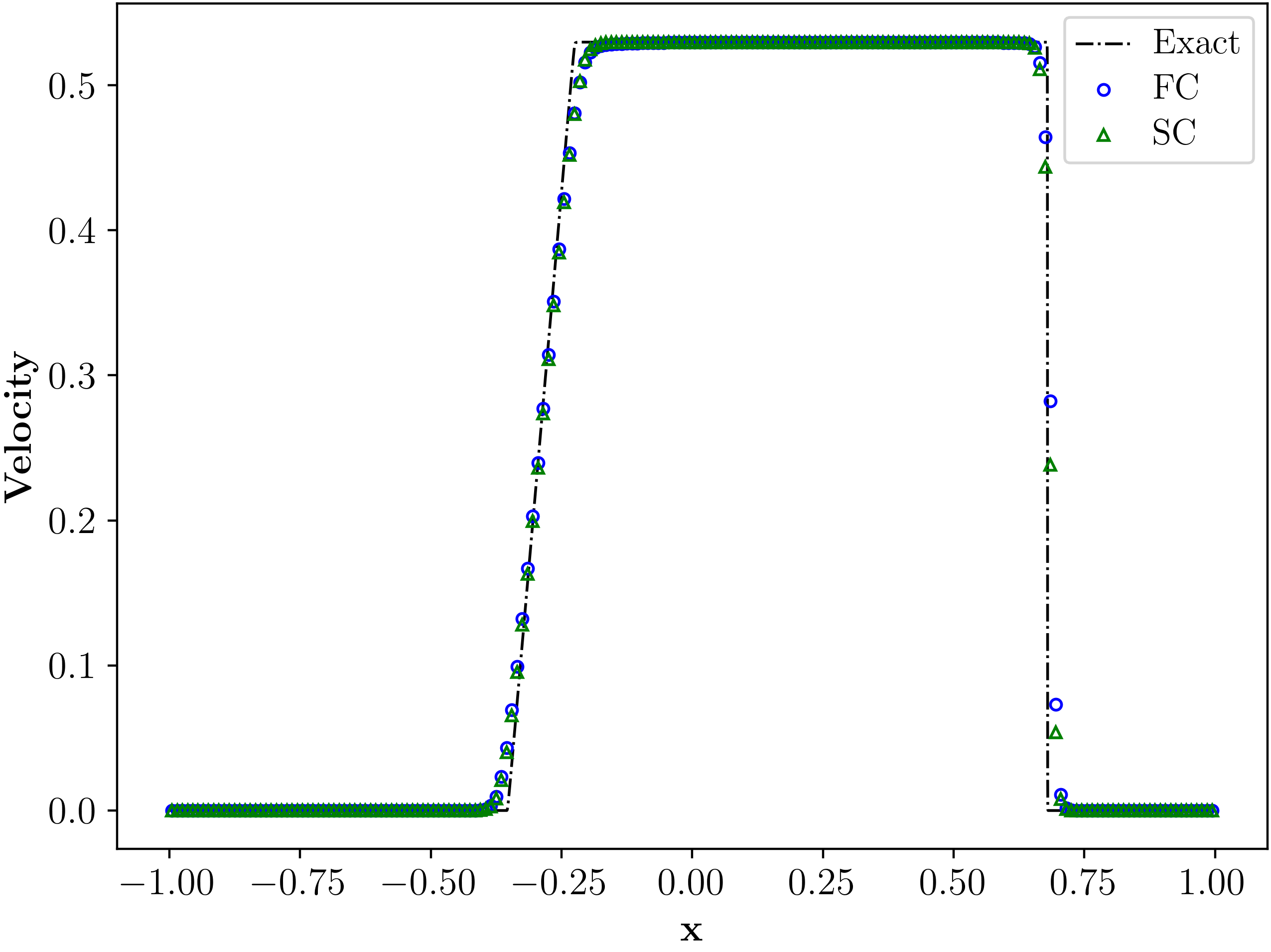}
\label{fig:multi_zhang-vel}}
\subfigure[\Rfour{Volume fraction}]{\includegraphics[width=0.43\textwidth]{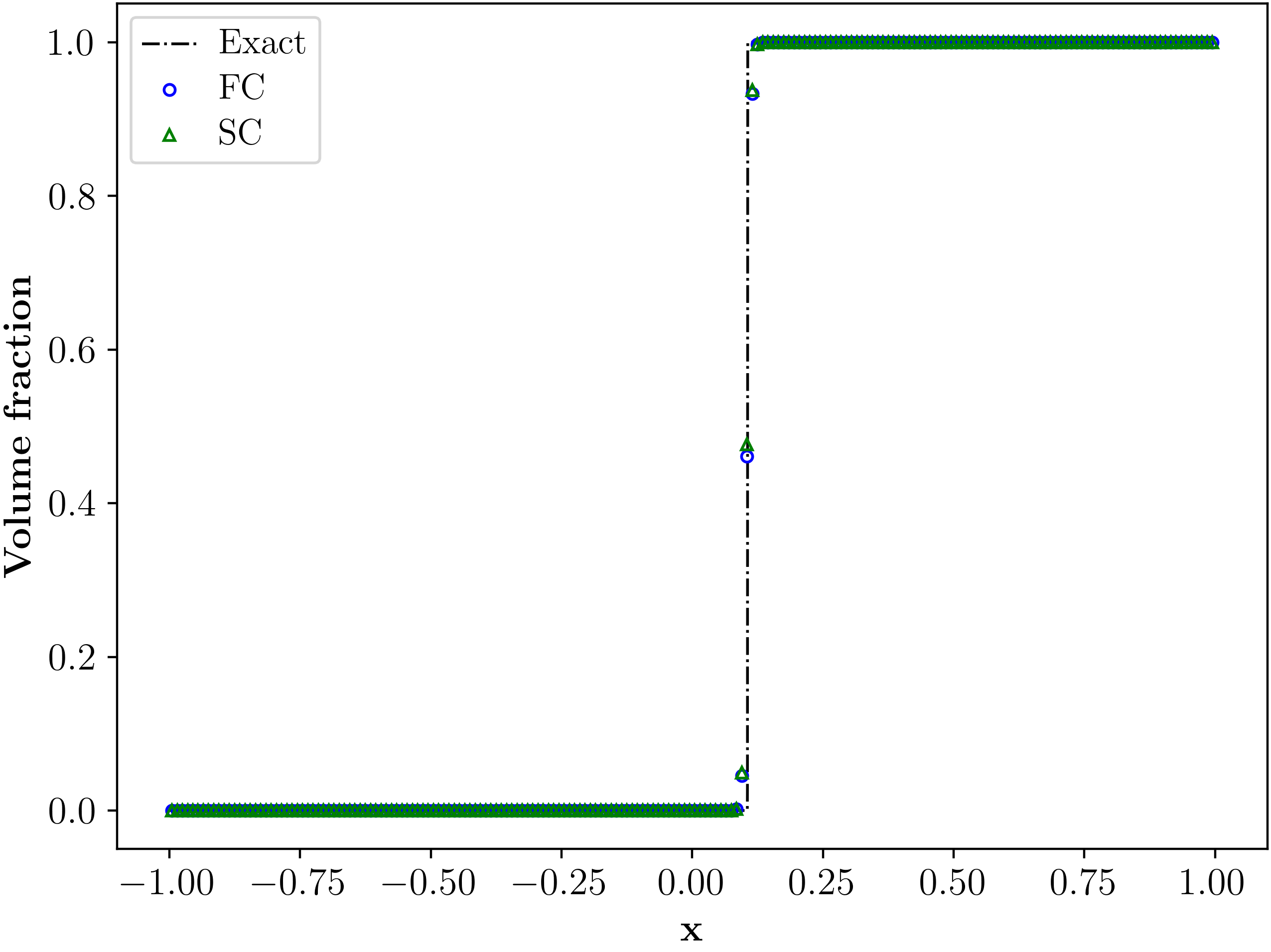}
\label{fig:multi_zhang-vfrac}}
\caption{Numerical solution for shock interface interaction problem in Example \ref{ex:liquid} on a grid size of $N=200$. Solid line: reference solution; blue circles: FC; green triangles: SC.}
\label{fig_zhang}
\end{figure}

Figure~\ref{fig_zhang} compares the numerical solutions of the FC and SC formulations against the exact reference. Both schemes capture the density profile and pressure profile without spurious oscillations, confirming that the thermodynamic coupling term~$\Psi$ derived for the FC formulation correctly cancels the equation-of-state mismatch at the interface. The SC formulation produces equivalent results through its structurally cleaner eigensystem. Both schemes accurately resolve the sharp pressure jump at the shock, and no non-physical pressure oscillations appear at the material interface, consistent with mechanical equilibrium. \Rfour{The velocity and volume-fraction profiles confirm the absence of a spurious interface impulse and identify the material-contact location relative to the shock.}

\begin{figure}[H]
\centering
\includegraphics[width=0.70\textwidth]{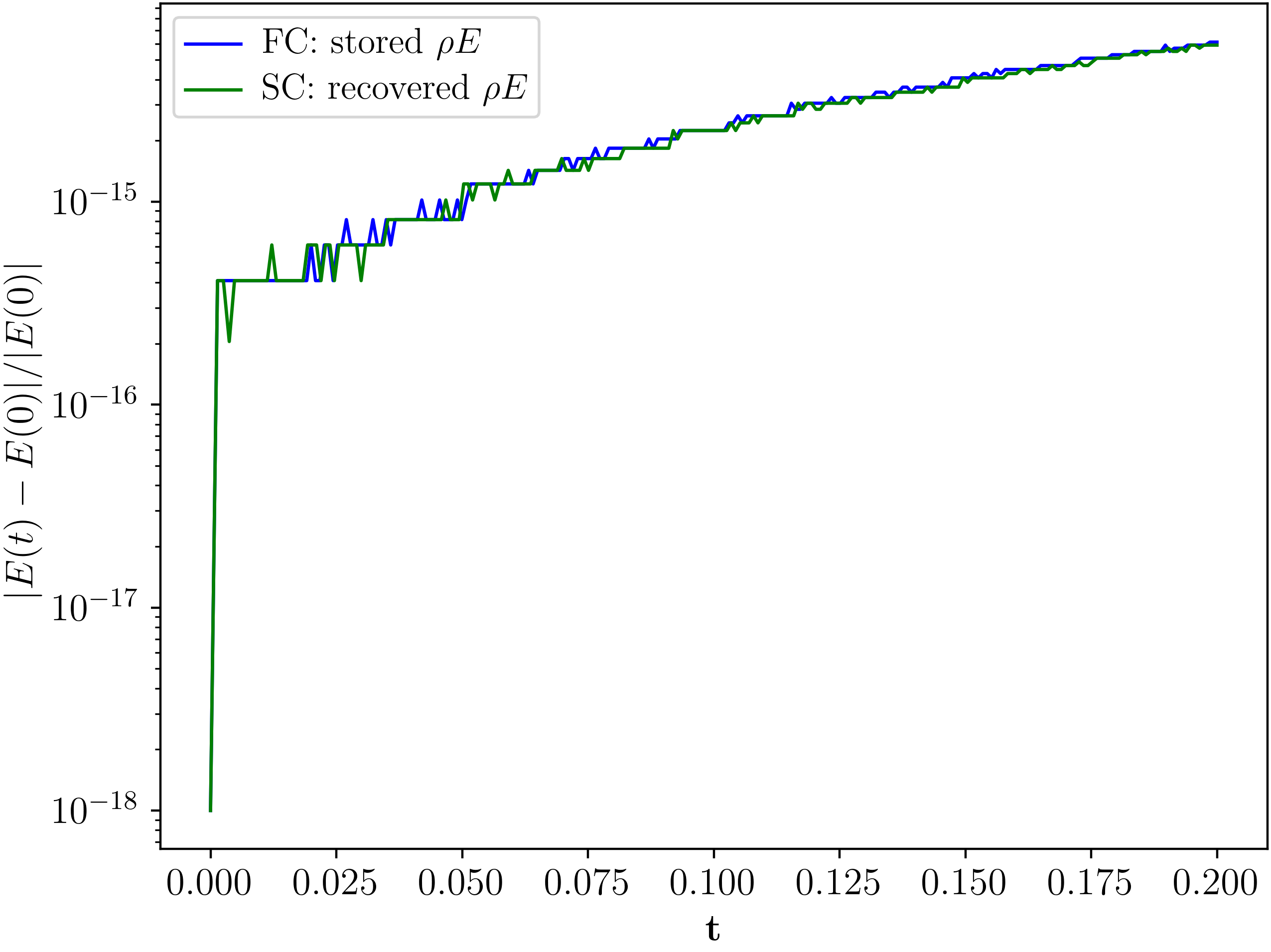}
\caption{\Rone{Relative domain-integrated total-energy error for the gas--liquid Riemann problem, Example \ref{ex:liquid}. FC reports the stored conservative energy; SC reports total energy recovered from its reconstructed pressure and the mixture EOS.}}
\label{fig:energy_history_gas_liquid}
\end{figure}

\Rone{For this gas--liquid shock tube, the SC energy monitor is computed from the pressure-recovered cell energy,
\begin{equation}\label{eq:energy_monitor}
E_\Omega^{\rm SC}(t)=\sum_i(\rho E)_i^{\rm rec}(t)\Delta x,
\end{equation}
and is compared with $E_\Omega^{\rm FC}(t)=\sum_i(\rho E)_i(t)\Delta x$. The relative error is
\begin{equation}\label{eq:energy_error_closed}
\varepsilon_E(t)=\frac{\left|E_\Omega(t)-E_\Omega(0)\right|}{\left|E_\Omega(0)\right|}.
\end{equation}
Figure~\ref{fig:energy_history_gas_liquid} shows that both histories remain at round-off level throughout the calculation.}

%\subsection{Test Case 6 -- Liquid--Gas Shock Tube (High Pressure)}
\begin{example}{Liquid-gas shock tube}\label{ex:multi-species}
\end{example}

This high-pressure liquid--gas shock tube, studied by Wong et al.~\cite{wong2021positivity}, involves a factor-of-$10^4$ pressure ratio across a water--air interface, making it one of the most demanding one-dimensional benchmarks for multiphase Riemann solvers. The significant density ratio ($\rho_{\rm water}/\rho_{\rm air} \sim 10^3$) and the large stiffness parameter $\pi_{\infty,1} = 3.43\times10^8$ for the water phase combine to create strong transmitted and reflected waves whose accurate capture requires both a robust reconstruction strategy and a correct characteristic decomposition of the stiffened-gas eigensystem. The domain is $x \in [0,1]$, discretised on $N=200$ cells, and the simulation is advanced to $t_{\mathrm{end}} = 2.4\times10^{-4}$:
\begin{equation}\label{eq:tc6}
(\alpha_1\rho_1,\;\alpha_2\rho_2,\;u,\;p,\;\alpha_1) = \begin{cases}
(1000,\; 10^{-8},\; 0,\; 10^9,\; 1-10^{-8}) & x < 0.75,\\
(10^{-8},\; 1,\; 0,\; 10^5,\; 10^{-8}) & x \geq 0.75,
\end{cases}
\end{equation}
with fluid properties $\gamma_1 = 6.12$, $\pi_1 = 3.43\times10^8$, $\gamma_2 = 1.4$, $\pi_2 = 0$.

\begin{figure}[H]
\centering
\subfigure[Density]{\includegraphics[width=0.45\textwidth]{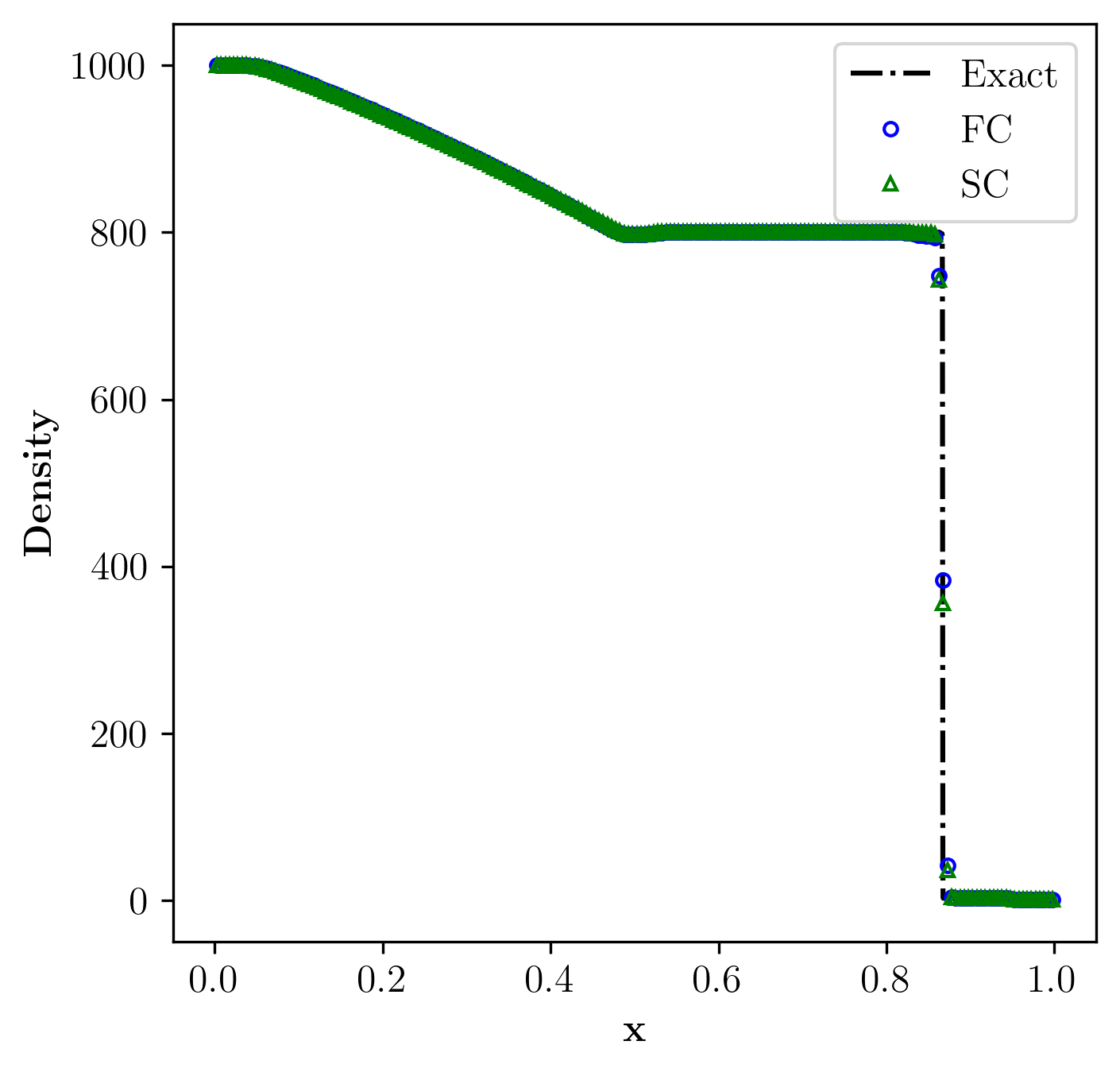}
\label{fig:multi_11-den}}
\subfigure[Velocity]{\includegraphics[width=0.45\textwidth]{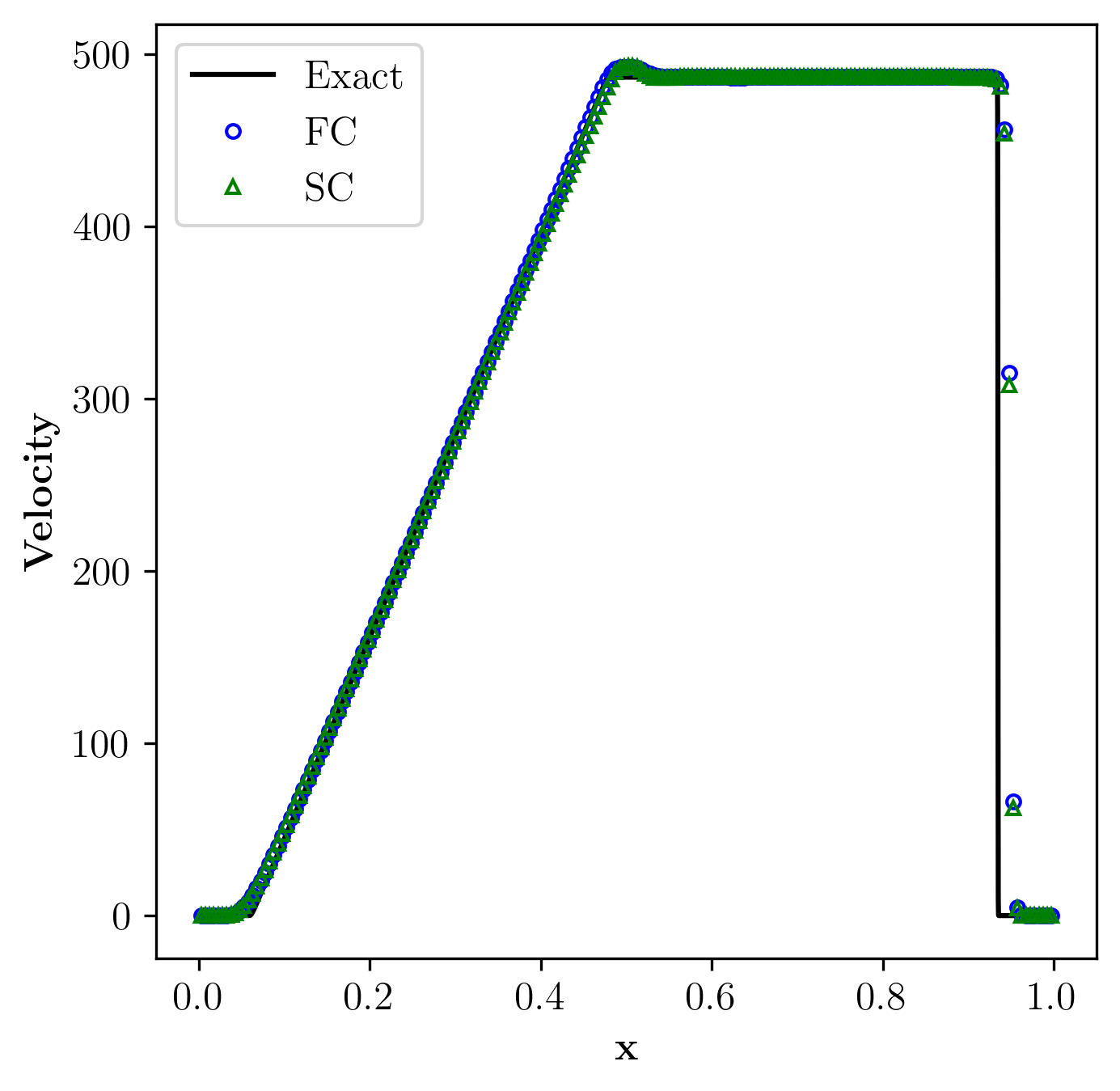}
\label{fig:multi_11-alpha}}
\subfigure[\Rthree{Pressure (semilogarithmic scale)}]{\includegraphics[width=0.45\textwidth]{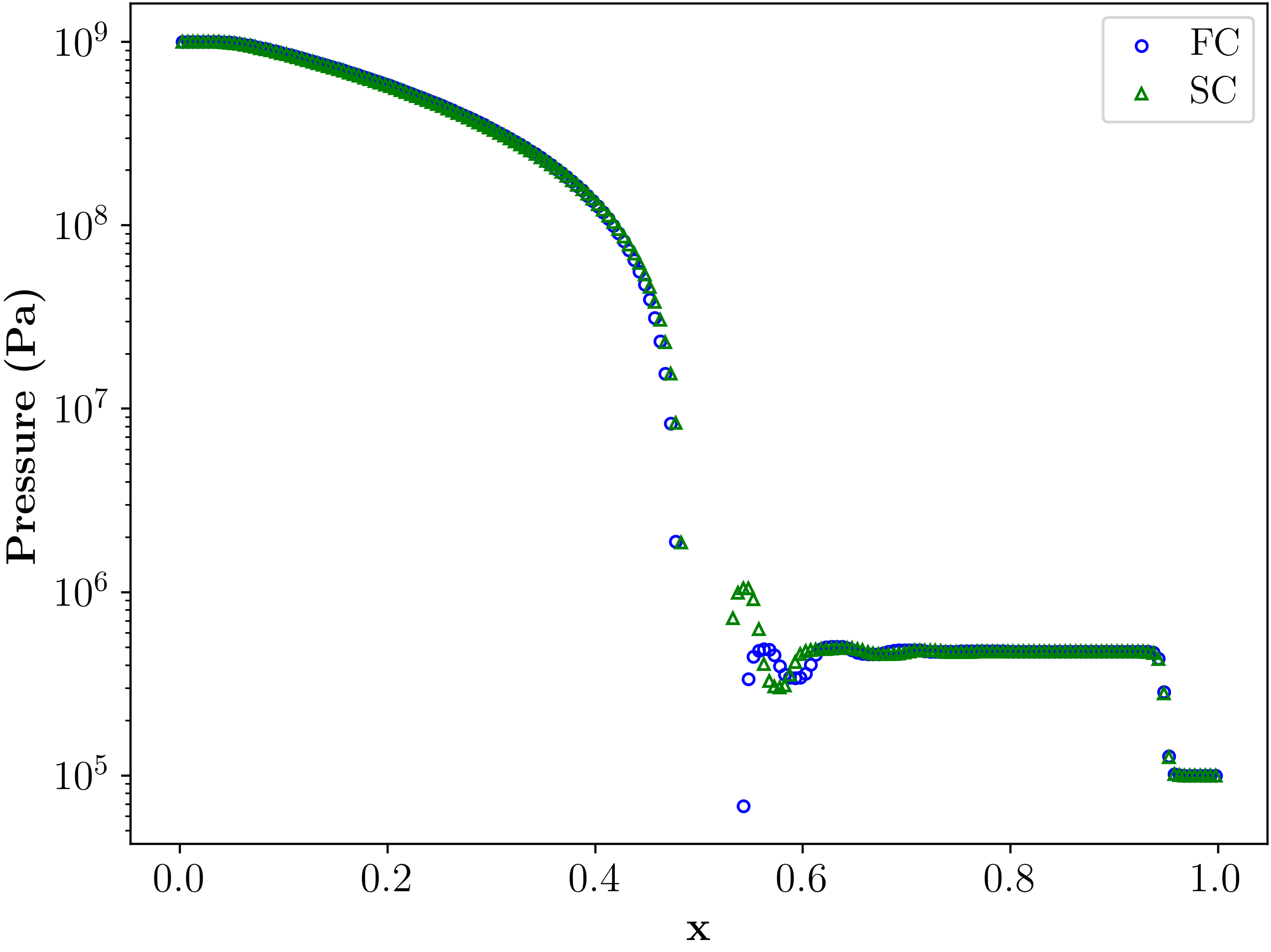}
\label{fig:multi_11-pres}}
\subfigure[\Rfour{Volume fraction}]{\includegraphics[width=0.45\textwidth]{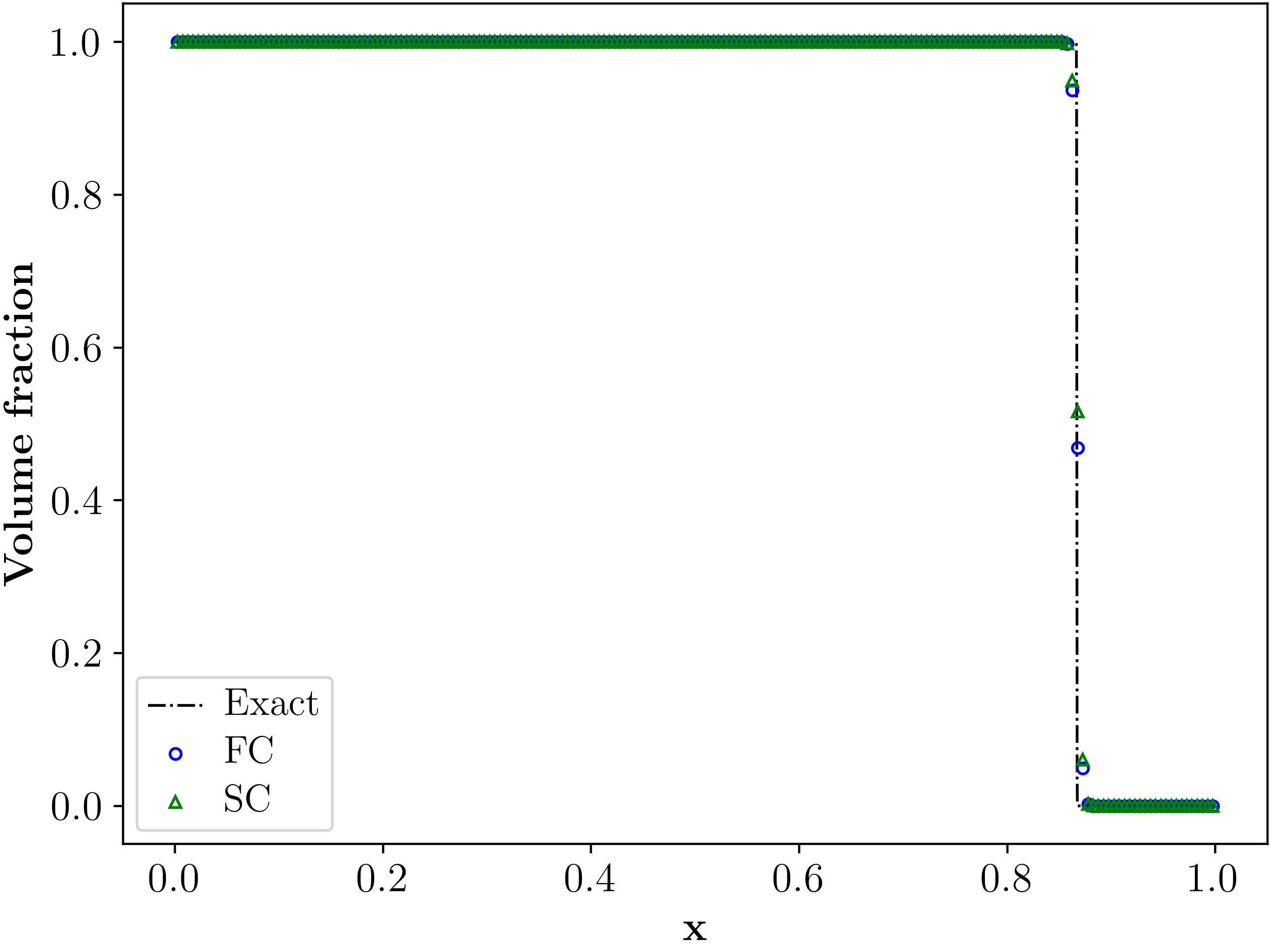}
\label{fig:multi_11-vfrac}}
\caption{Numerical solution for the liquid--gas shock tube problem in Example \ref{ex:multi-species} on a grid with $N=200$ cells. Solid line: reference solution; blue circles: FC; green triangles: SC.}
\label{fig_11}
\end{figure}

Figure~\ref{fig_11} presents a comparison of density, velocity, pressure, and volume fraction obtained from the FC and SC formulations with the reference solution on the $N=200$ grid. \Rfour{The volume fraction tracks the water--air contact, while the pressure profile verifies the transmitted-shock and post-shock states.} \Rthree{The pressure panel uses a semilogarithmic scale to expose small pressure deviations near the material interface.} Both methods accurately capture the transmitted shock propagating into the air and the rarefaction wave receding into the water. No spurious oscillations are observed in the velocity or density fields at the gas–liquid interface, demonstrating that the multi-phase eigenstructure effectively manages the large stiffness contrast.

\begin{example}{Water Column Expansion}\label{ex:column}
\end{example}

This test case, adapted from Wong et al.~\cite{wong2021positivity}, describes the expansion of a highly compressed water column into a lighter gas. \textcolor{black}{The $10^4$ initial pressure ratio, combined with the large water-phase stiffness ($\pi_{\infty,1} = 3.43\times10^8$), drives a strong rarefaction back through the water and a transmitted compression wave in the gas.} The problem is particularly sensitive to the accuracy of the characteristic decomposition: any misrepresentation of the contact-wave eigenvector generates unphysical acoustic sources that corrupt the rarefaction profile. The domain is $x \in [0,1.5]$, discretised on $N=200$ cells, and the simulation is advanced to $t_{\mathrm{end}} = 3\times10^{-4}$:
\begin{equation}\label{eq:tc7}
(\alpha_1\rho_1,\;\alpha_2\rho_2,\;u,\;p,\;\alpha_1) = \begin{cases}
(1000,\; 10^{-8},\; 0,\; 10^9,\; 1-10^{-8}) & x < 0.8,\\
(10^{-8},\; 20,\; 0,\; 10^5,\; 10^{-8}) & x \geq 0.8,
\end{cases}
\end{equation}
with fluid properties $\gamma_1 = 6.12$, $\pi_1 = 3.43\times10^8$, $\gamma_2 = 1.4$, $\pi_2 = 0$.

\begin{figure}[H]
\centering
\subfigure[Density]{\includegraphics[width=0.45\textwidth]{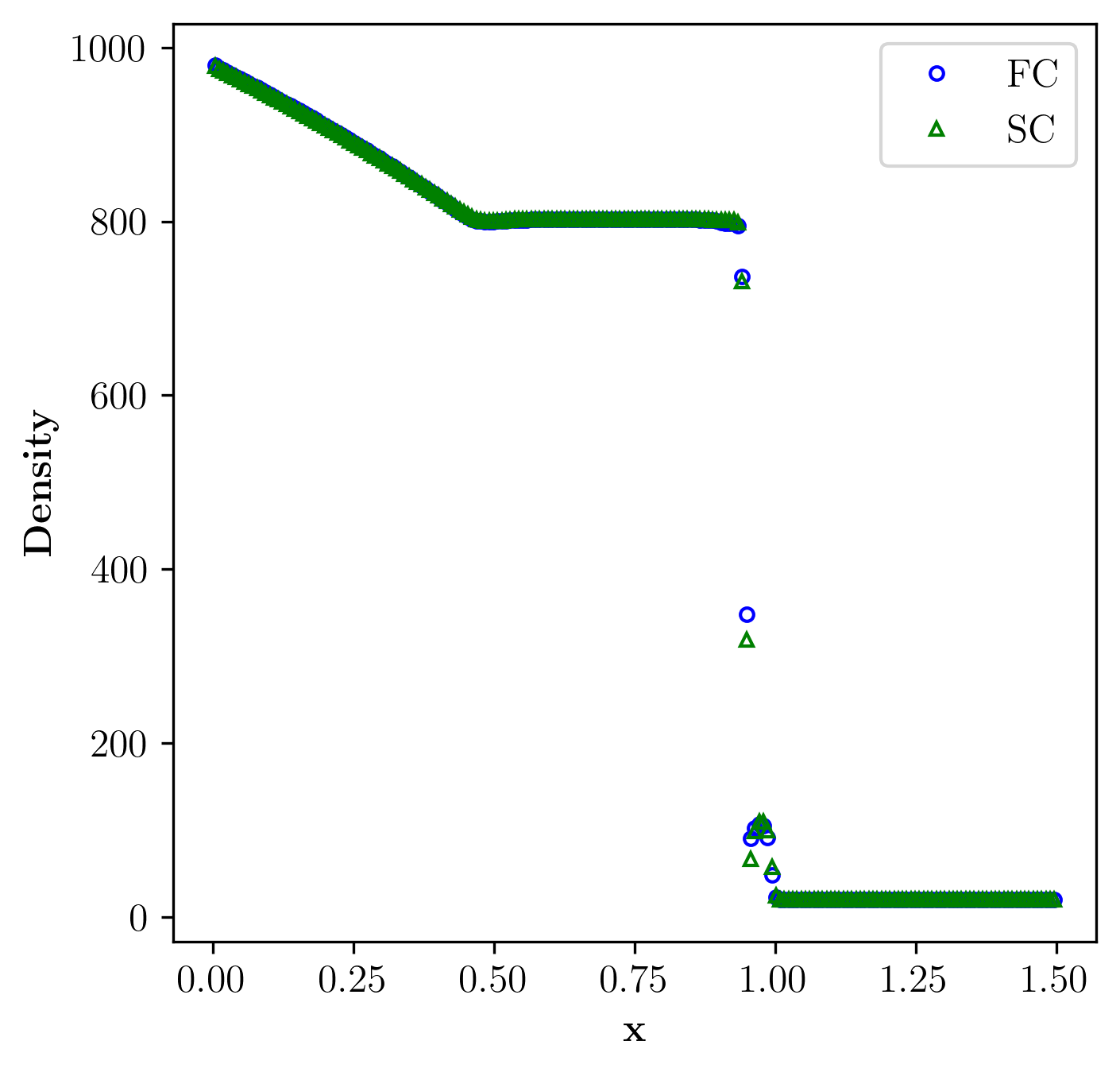}
\label{fig:multi_12-den}}
\subfigure[Velocity]{\includegraphics[width=0.45\textwidth]{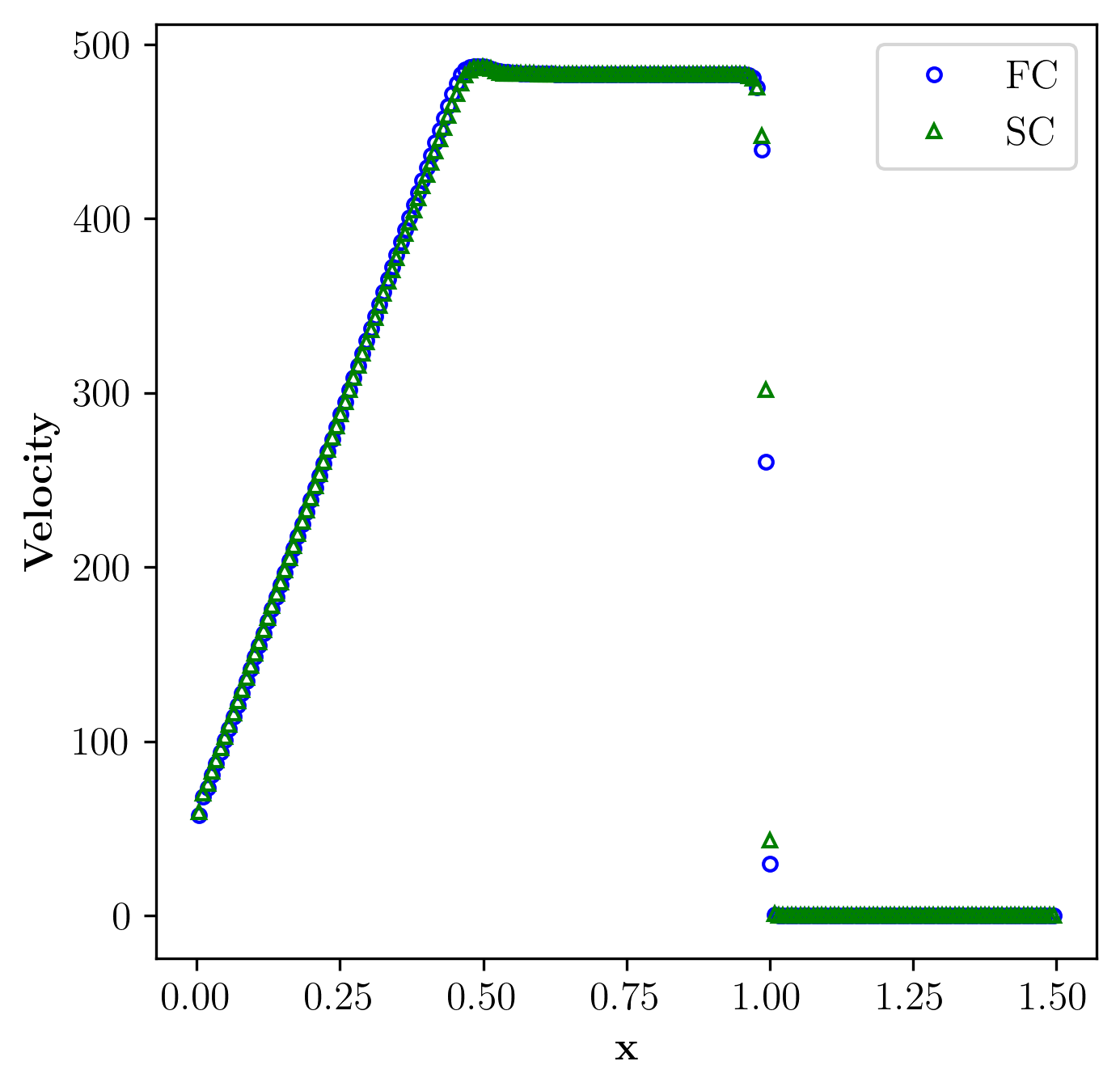}
\label{fig:multi_12-alpha}}
\subfigure[\Rfour{Pressure (with material-interface inset)}]{\includegraphics[width=0.45\textwidth]{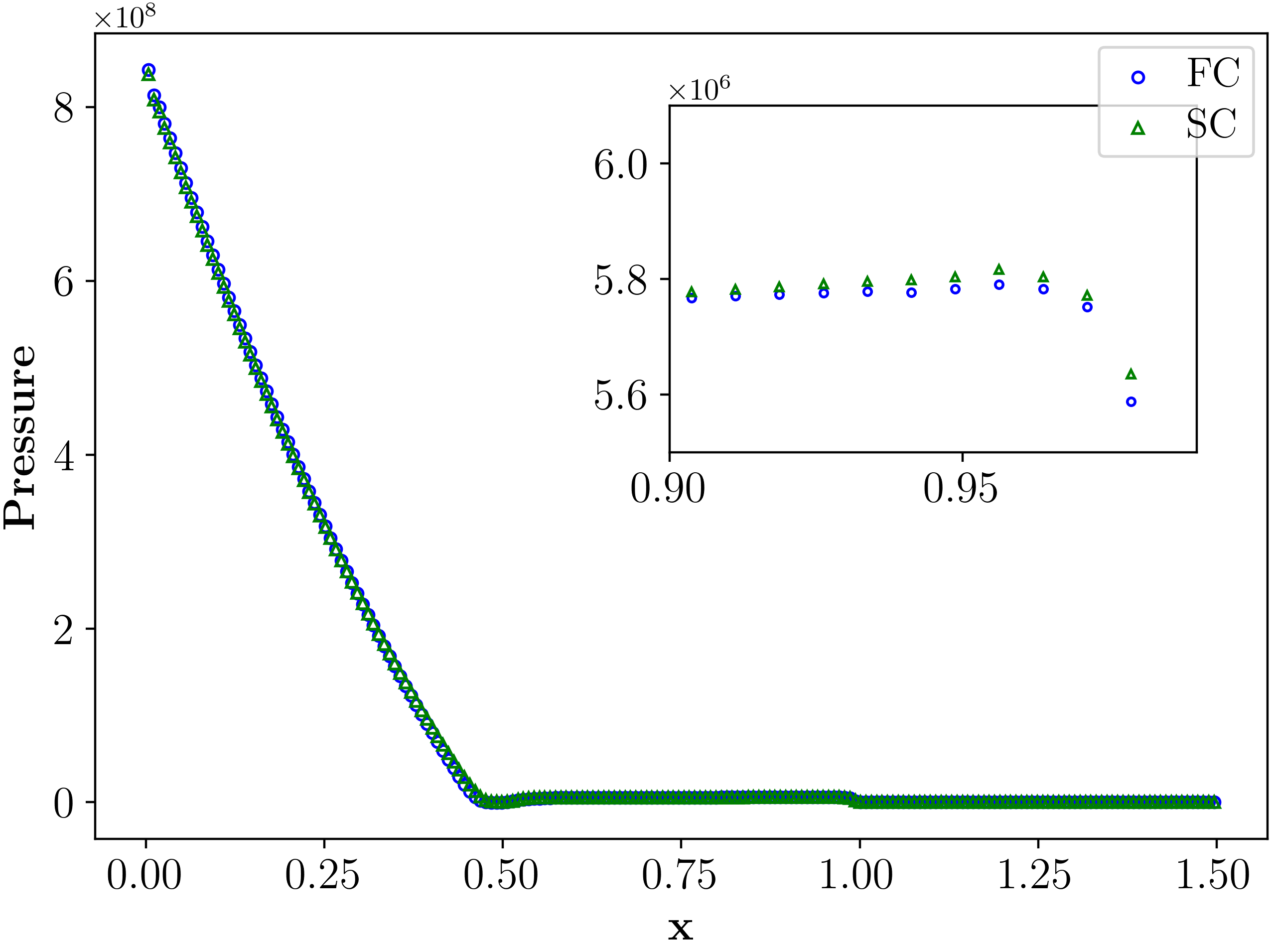}
\label{fig:multi_12-pres}}
\subfigure[Volume fraction]{\includegraphics[width=0.45\textwidth]{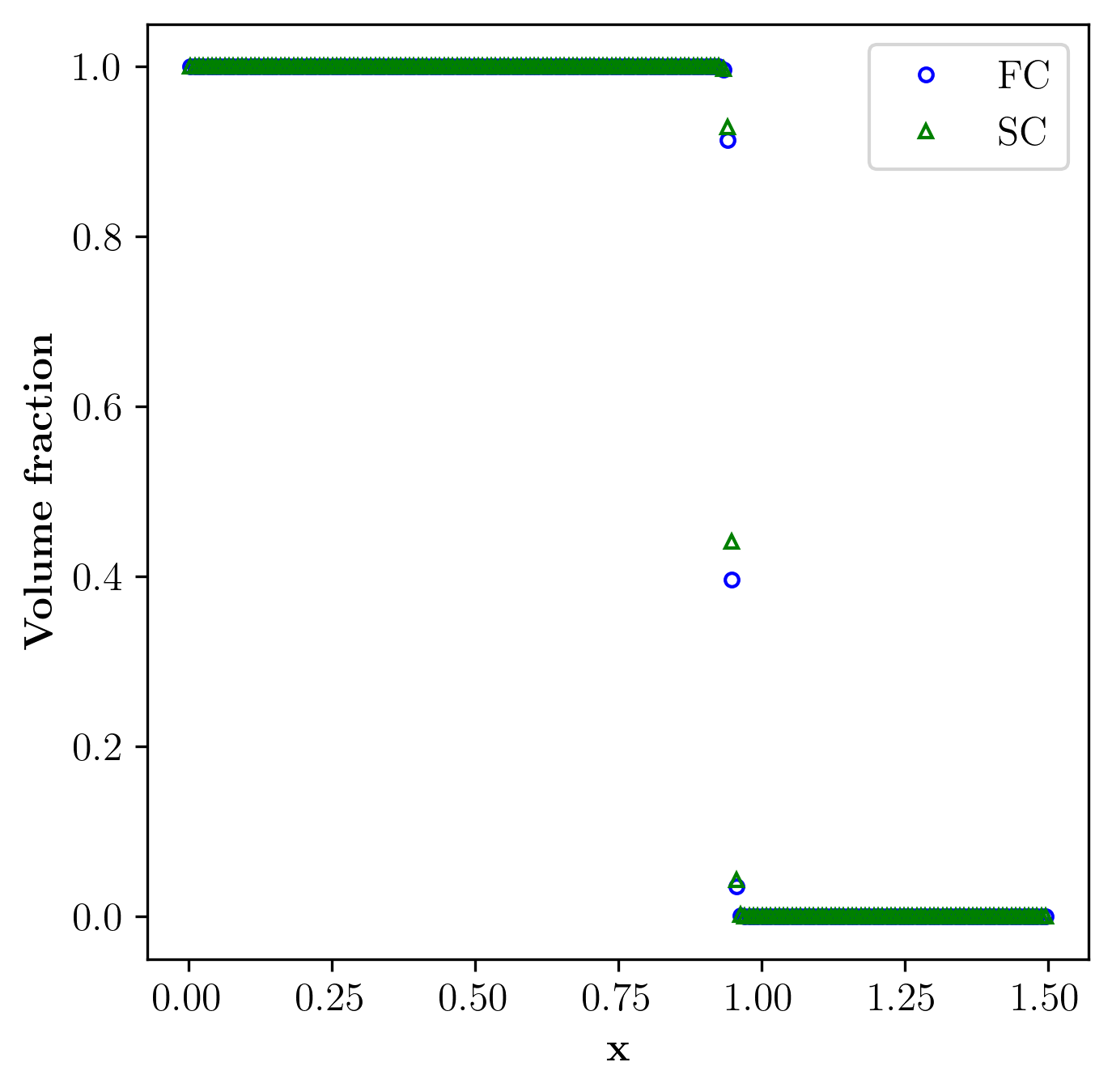}
\label{fig:multi_12-prim}}
\caption{Numerical solution for the water-column expansion problem in Example \ref{ex:column} on a grid size of $N=200$. Blue circles: FC; green triangles: SC. \Rfour{The inset in the pressure panel shows the final material-interface neighborhood near $x\simeq0.95$.}}
\label{fig_8}
\end{figure}

Figure~\ref{fig_8} shows density, velocity, pressure, and volume fraction for the FC and SC formulations on the $N=200$ grid. The velocity profile exhibits the self-similar expansion-fan structure characteristic of a centered rarefaction, and the pressure decreases monotonically across the fan region without non-physical kinks or oscillations. The volume fraction correctly localizes the water--gas interface, and neither scheme introduces anomalous density extrema near the interface. \Rfour{A zoomed view of the pressure profile in a narrow neighborhood of the material interface is used to verify that the pressure field remains smooth across the contact and that no interface-induced pressure defect is hidden by the full-domain scale.}

\begin{example}{Compressible Triple-Point Problem}\label
{ex:triplet}
\end{example}

The compressible triple-point problem is a two-dimensional Riemann problem involving three states and two distinct materials; it was used as a benchmark in~\cite{chamarthi2023gradient} for interface-capturing schemes precisely because it generates all three wave families simultaneously at the triple-point singularity. The primary objective here is to demonstrate that both the FC and SC characteristic reconstructions (i)~preserve mechanical equilibrium at the material interface, (ii)~correctly capture the vortical structures generated by Kelvin--Helmholtz instabilities along the contact discontinuity, and (iii)~confirm the advantage of computing vorticity waves with a central scheme. The computational domain is $(x,y) \in [0,7]\times[0,3]$, with reflective boundary conditions and a final simulation time of $t_{\mathrm{end}} = 5.0$. The initial conditions are:
\begin{equation}\label{eq:tc8}
(\alpha_1\rho_1,\;\alpha_2\rho_2,\;u,\;v,\;p,\;\alpha_1) = \begin{cases}
(1,\;0,\;0,\;0,\;1,\;1) & [0,1]\times[0,3],\\
(0,\;1,\;0,\;0,\;0.1,\;0) & [1,7]\times[0,1.5],\\
(0.125,\;0,\;0,\;0,\;0.1,\;1) & [1,7]\times[1.5,3],
\end{cases}
\end{equation}
with fluid properties $\gamma_1 = 1.5$, $\pi_1 = 0$ and $\gamma_2 = 1.4$, $\pi_2 = 0$. Simulations are carried out on \textcolor{black}{a grid of} $3584\times1536$. 

\begin{figure}[H]
    \centering
    \includegraphics[width=0.7\textwidth]{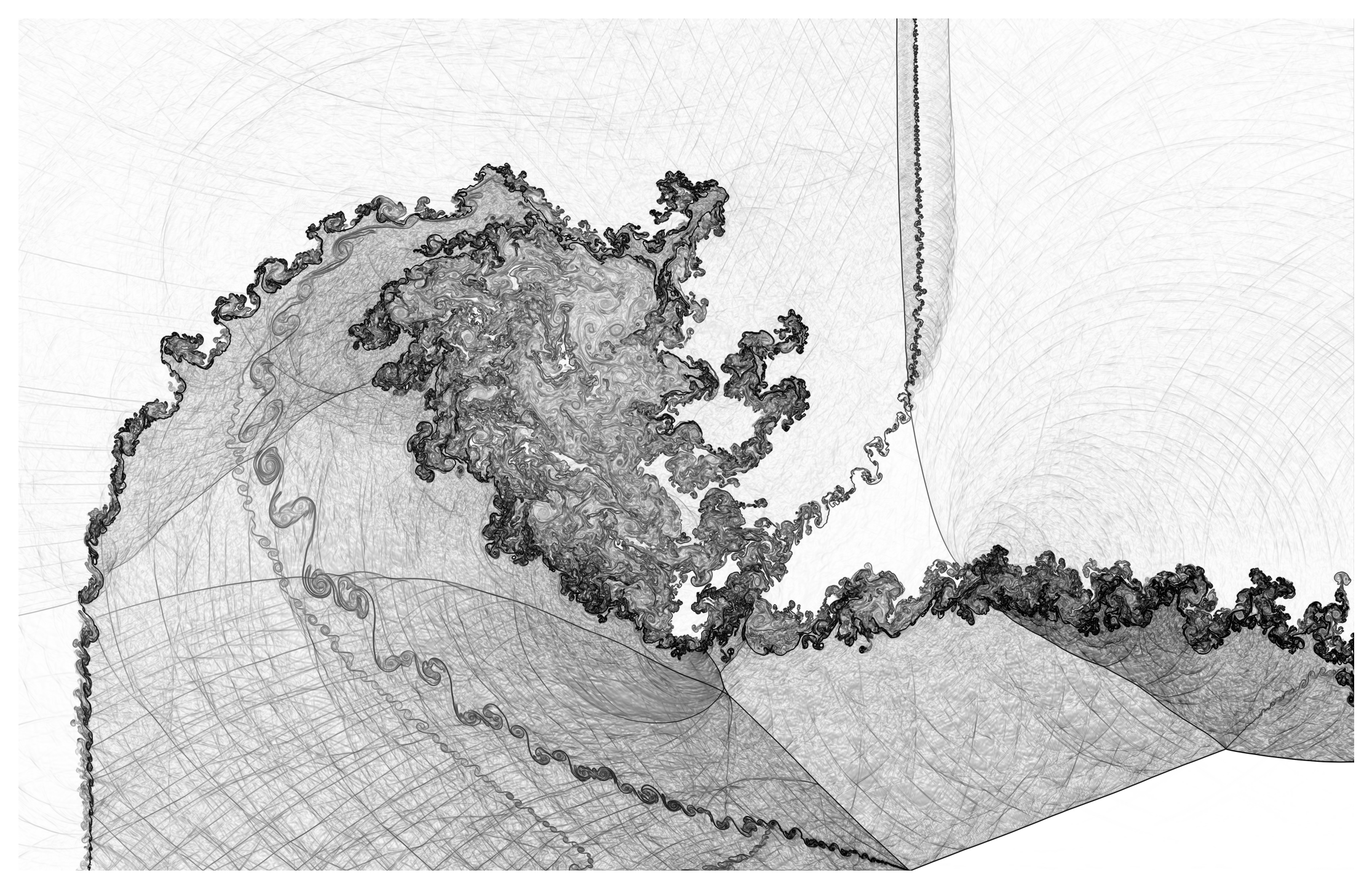}
    \caption{Density gradient contours at time $t=5$, high-resolution reference ($7168 \times 3072$), Example~\ref{ex:triplet}.}
    \label{fig:fhl}
\end{figure}

\begin{figure}[H]
    \centering
    \subfigure[Wave-MUSCL-FC.]{\includegraphics[width=0.48\textwidth]{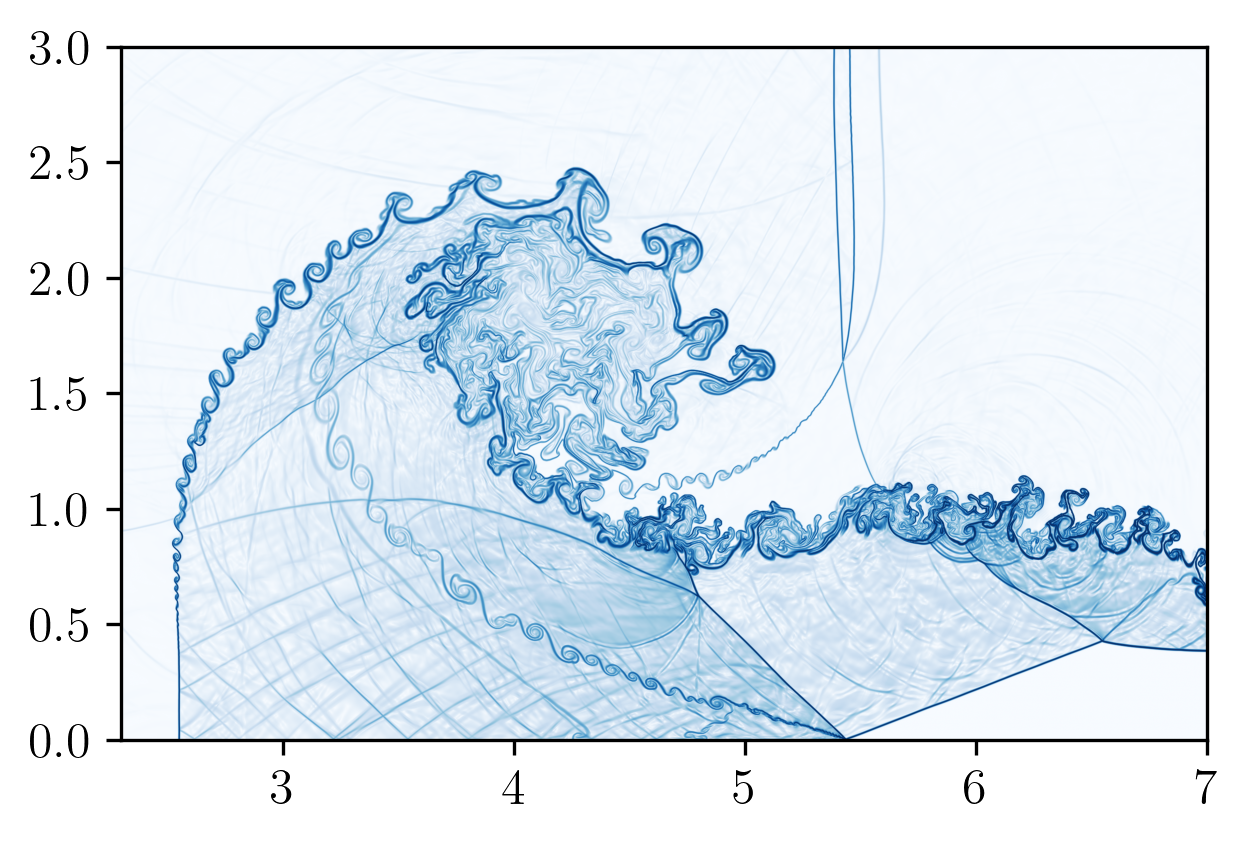}
    \label{fig:fmu}}
    \subfigure[Wave-MUSCL-SC.]{\includegraphics[width=0.48\textwidth]{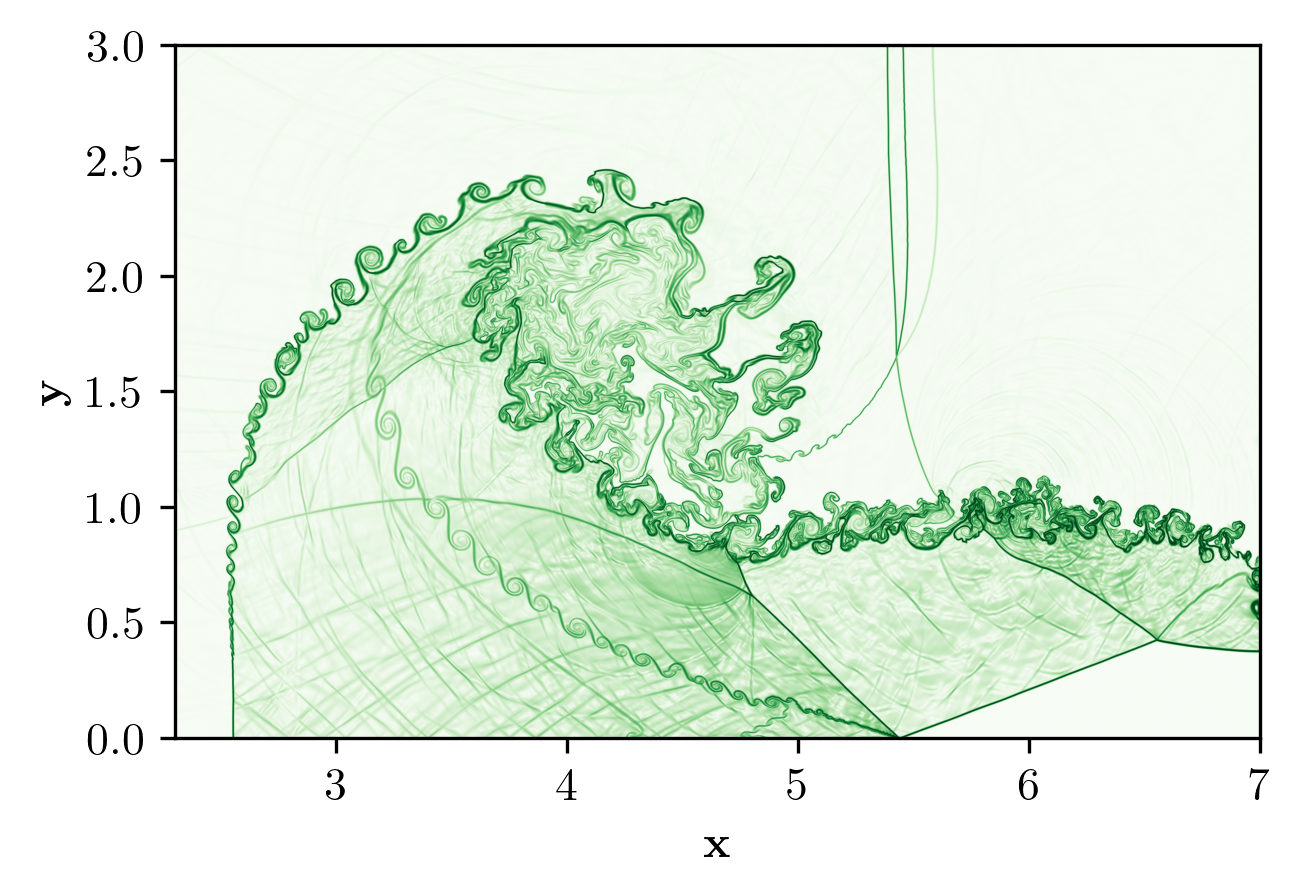}
    \label{fig:mp}}
    \subfigure[Wave-MUSCL-SC without central scheme for vorticity wave.]{\includegraphics[width=0.48\textwidth]{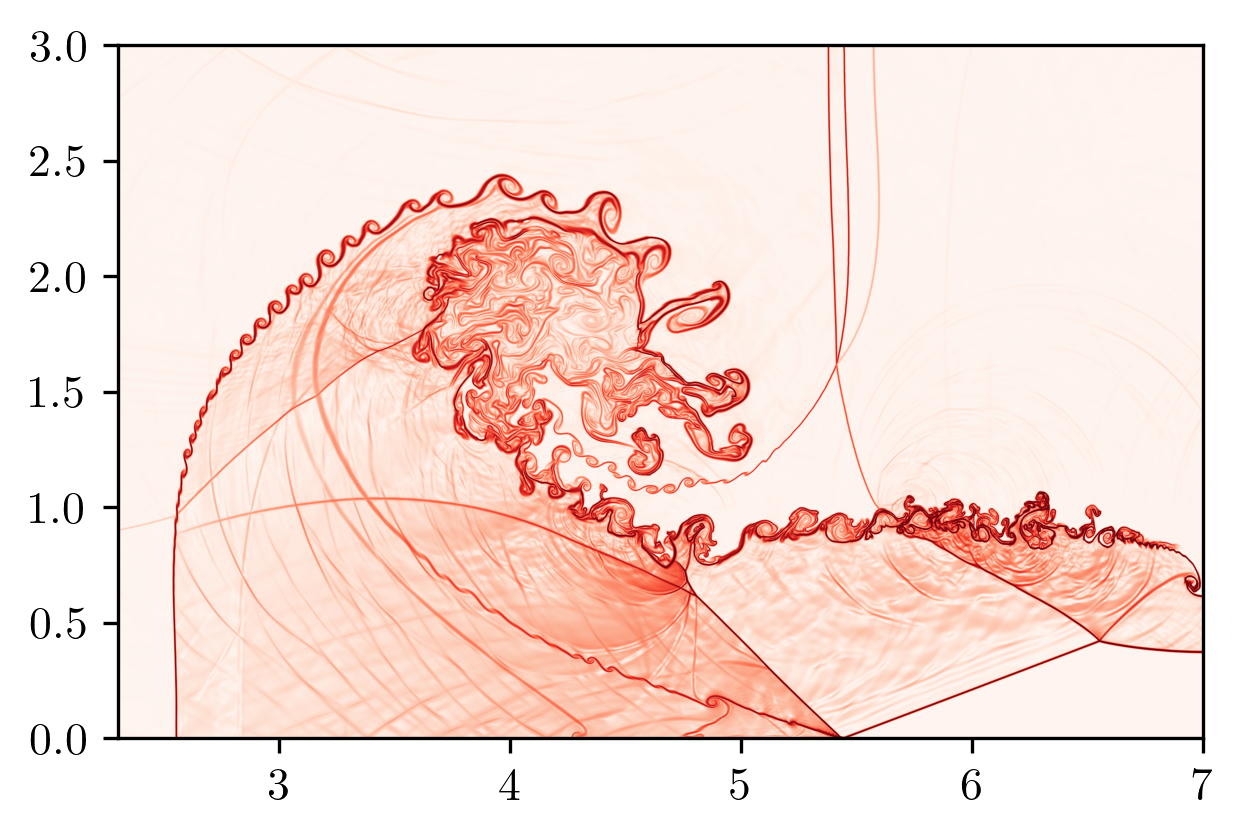}
    \label{fig:mv}}
    \subfigure[Right eigenvector, SC.]{\includegraphics[width=0.48\textwidth]{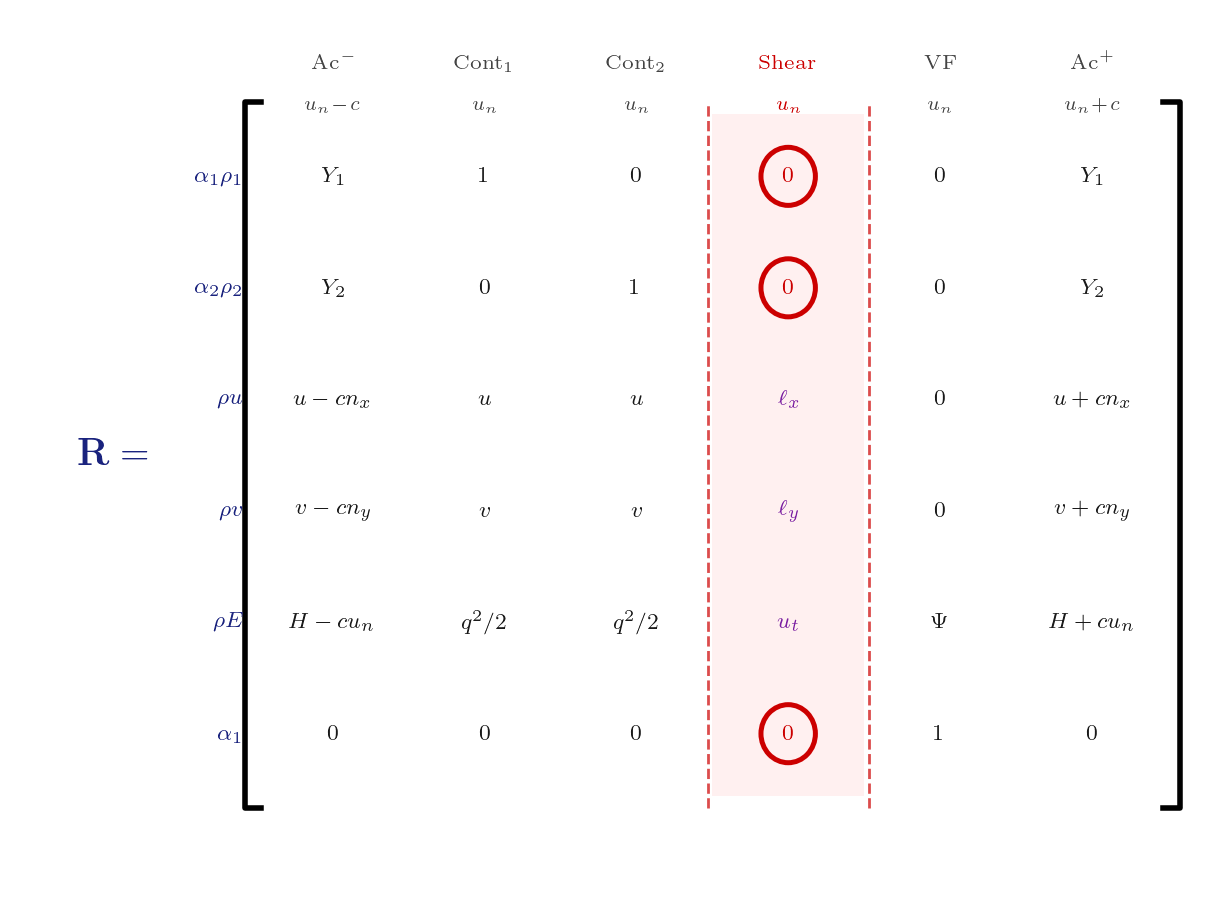}
    \label{fig:matrix}}
    \caption{Density gradient contours at time $t=5$ using various schemes, Example~\ref{ex:triplet} along with the right eigenvector. The central-scheme treatment of the vorticity wave (panels a and b) produces richer Kelvin--Helmholtz roll-up structures compared to the fully upwind variant (panel c).}
    \label{fig_fivetriple}
\end{figure}

The setup places a high-pressure driver ($p=1$, fluid~1) in the left region, which is separated from two lower-pressure regions ($p=0.1$) by a vertical shock/material interface at $x=1$ and a horizontal contact discontinuity at $y=1.5$. Regions~2 and~3 on the right differ only in their fluid identity ($\gamma_2 = 1.4$ below, $\gamma_1=1.5$ above), so the transmitted shock travels at different speeds in each region, generating a Kelvin--Helmholtz-unstable shear layer emanating from the triple point. Figure~\ref{fig:fhl} shows the results obtained by the high-resolution scheme on a reference grid size of $7168 \times 3072$ using the SC variable set. Figure~\ref{fig_fivetriple} presents density gradient contours at the final time $t=5$ for three scheme variants. Both Wave-MUSCL-FC and Wave-MUSCL-SC produce well-resolved vortical roll-up structures along the contact discontinuity, consistent with the fine grid reference. The Wave-MUSCL-SC scheme without the central-scheme treatment for the vorticity, Figure~\ref{fig:mv}, wave exhibits visibly reduced vortical content in the Kelvin--Helmholtz roll-up region, confirming that the central reconstruction of the shear characteristic field is responsible for capturing the instability-driven structures rather than suppressing them through numerical dissipation. This behavior is consistent with the discussion in~\cite{chamarthi2025wave}: an upwind-biased reconstruction of the vorticity wave introduces excessive diffusion of tangential velocity gradients, thereby damping the Kelvin--Helmholtz roll-up. Unlike that of \cite{chamarthi2025physics}, where primitive variables are used for reconstruction, here, even in the inviscid scenario, one can use a central scheme for shear wave reconstruction with both the FC and SC variables. These results also indicate that reconstructing conservative variables may, in fact, be superior to reconstructing primitive variables, as observed in \cite{hoffmann2024centralized,chamarthi2026wave} for the single-species case (Figure~\ref{fig:matrix} shows the right eigenvector matrix with zeroes in the corresponding slots, similar to the single species).

\begin{figure}[H]
    \centering
    \subfigure{\includegraphics[width=0.8\textwidth]{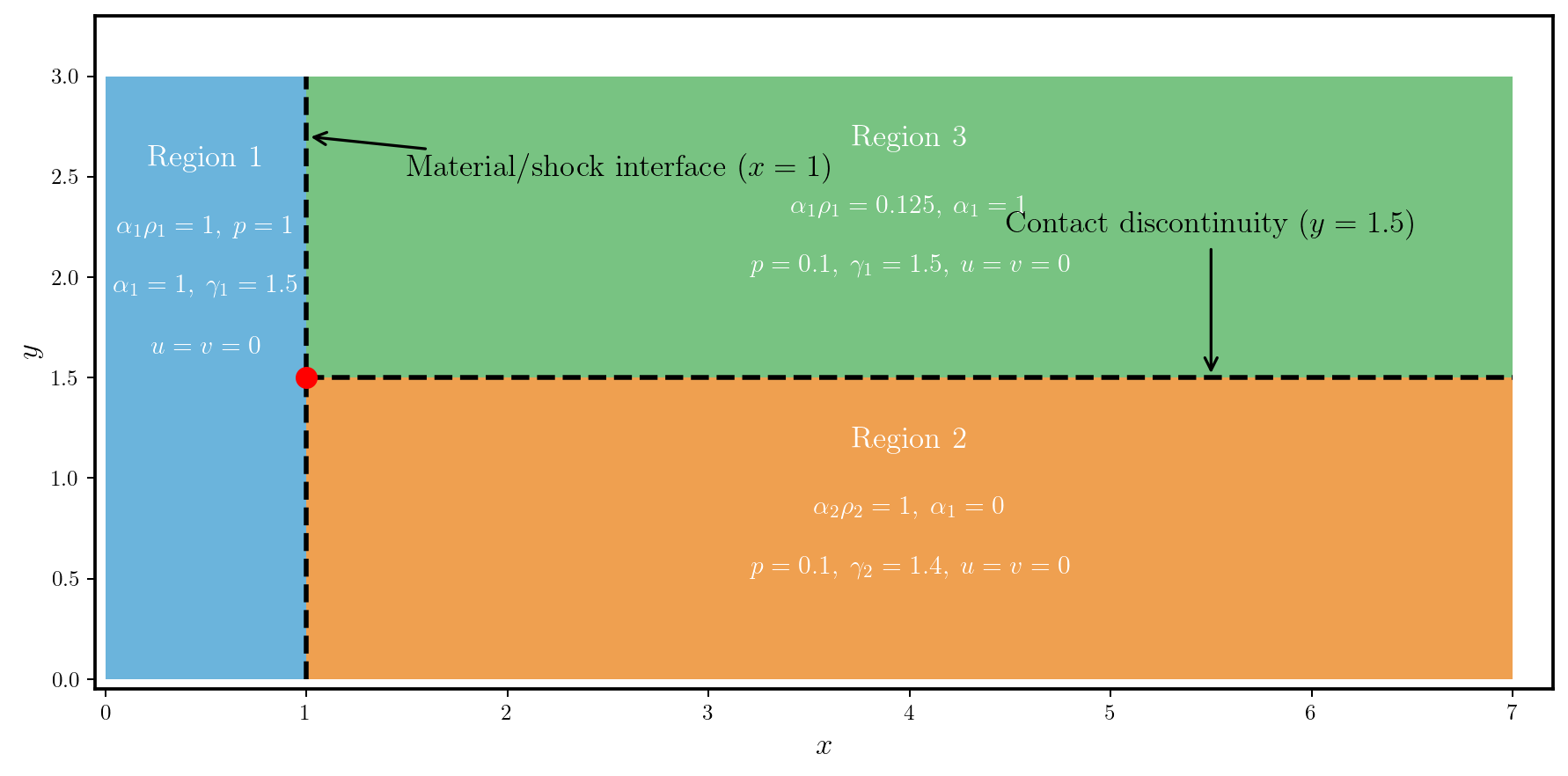}
    \label{fig:ic}}
    \subfigure[Density.]{\includegraphics[width=0.48\textwidth]{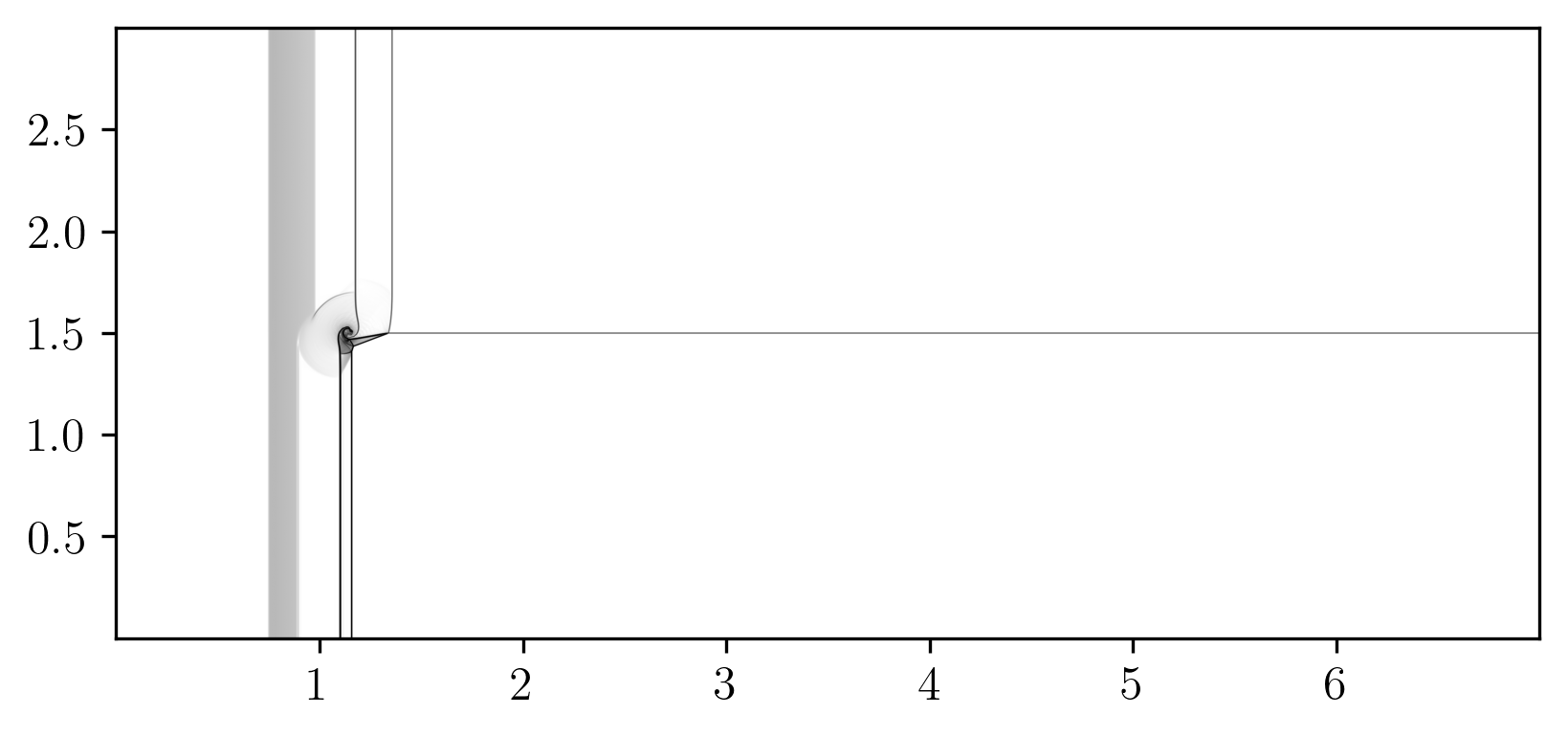}
    \label{fig:density}}
    \subfigure[Pressure.]{\includegraphics[width=0.48\textwidth]{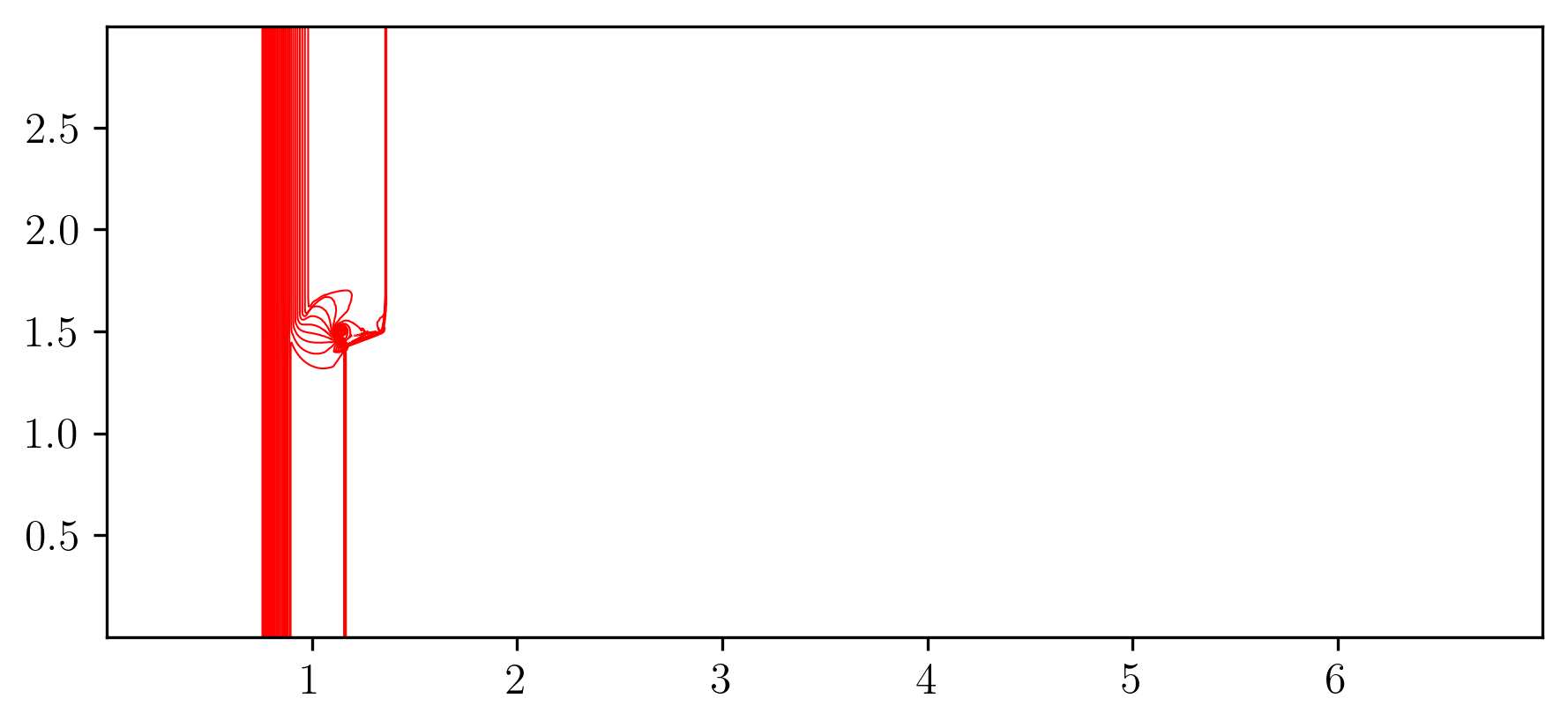}
    \label{fig:pressure}}
    \subfigure[Normal velocity $u$.]{\includegraphics[width=0.48\textwidth]{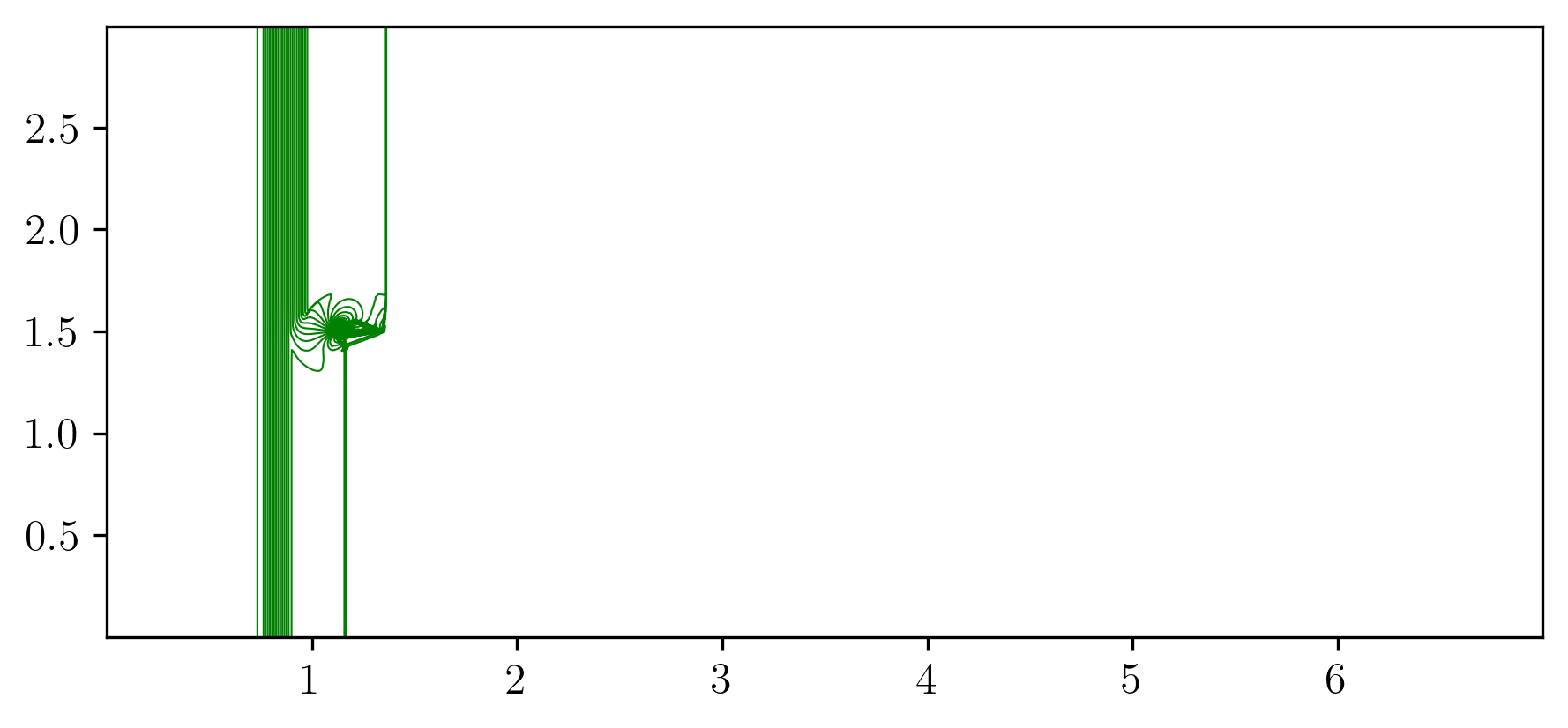}
    \label{fig:un}}
    \subfigure[Tangential velocity $v$.]{\includegraphics[width=0.48\textwidth]{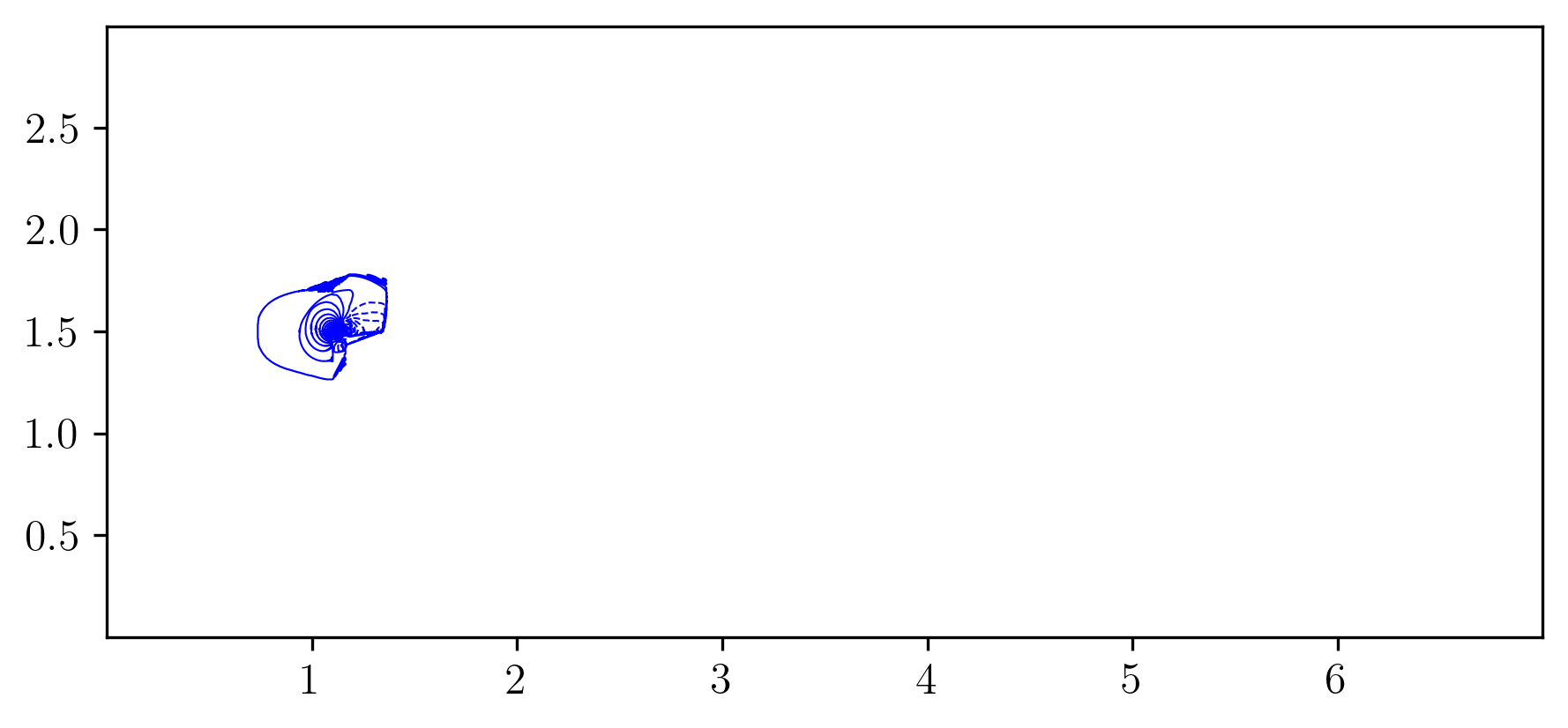}
    \label{fig:vn}}
    \caption{Triple-point problem at $t=0.2$. Contours of density (b), pressure (c), normal velocity $u$ (d), and transverse velocity $v$ (e). Pressure and normal velocity remain continuous across the material interface at $y=1.5$, confirming mechanical equilibrium preservation for both eigensystem formulations.}
    \label{fig:triple_point_fields}
\end{figure}

Figure~\ref{fig:triple_point_fields} shows density, pressure, and velocity fields at $t=0.2$. Figure~\ref{fig:ic} shows the initial condition for this case. Figure~\ref{fig:density} shows the density gradient contours. The pressure contours in the subsequent figure confirm mechanical equilibrium along the horizontal contact discontinuity at $y=1.5$: no spurious pressure jump is visible there, consistent with the proofs in Sections~\ref{sec:fc_proof} and~\ref{sec:sc_proof}. The normal velocity contours similarly show no unphysical jump at the contact, verifying velocity equilibrium. The tangential velocity $v$ contours reveal the physically correct slip layer at $y=1.5$ and the spiral vortical structure at the triple point; the vortical activity is confined to the physical triple-point location rather than spreading diffusely along the entire interface. The tangential velocity figure also indicates that it is continuous across the shock wave.

\begin{example}{Shock--Helium Cylinder Interaction}\label
{ex:SB}
\end{example}

\begin{figure}[H]
\centering\offinterlineskip
\subfigure[Wave-MUSCL-SC]{\includegraphics[width=0.42\textwidth]{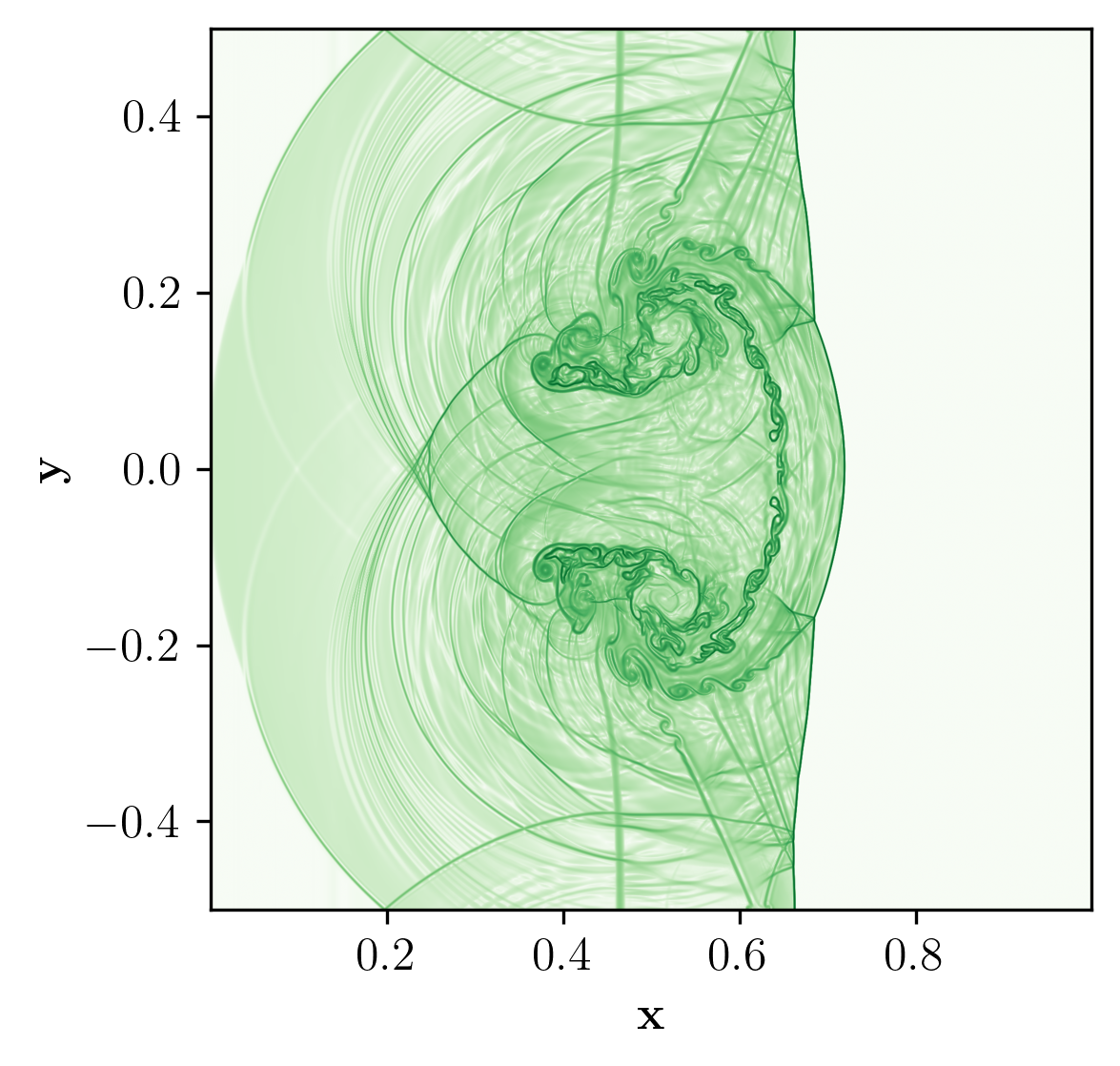}
\label{fig:mp6_SB-LLFM}}
\subfigure[Wave-MUSCL-FC]{\includegraphics[width=0.42\textwidth]{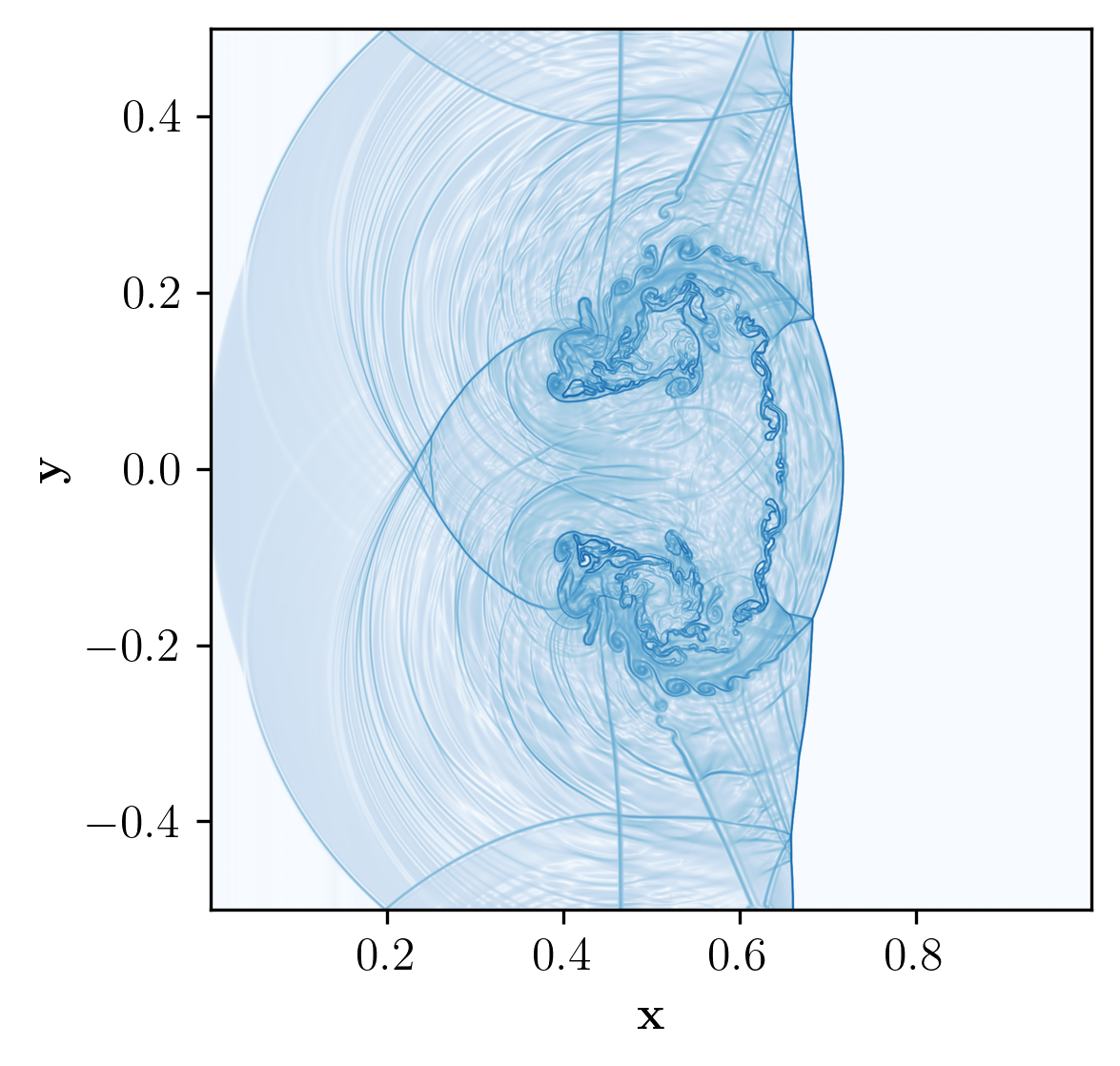}
\label{fig:Meg8c_base-GLF}}
\subfigure[High resolution scheme, $2400^2$.]{\includegraphics[width=0.42\textwidth]{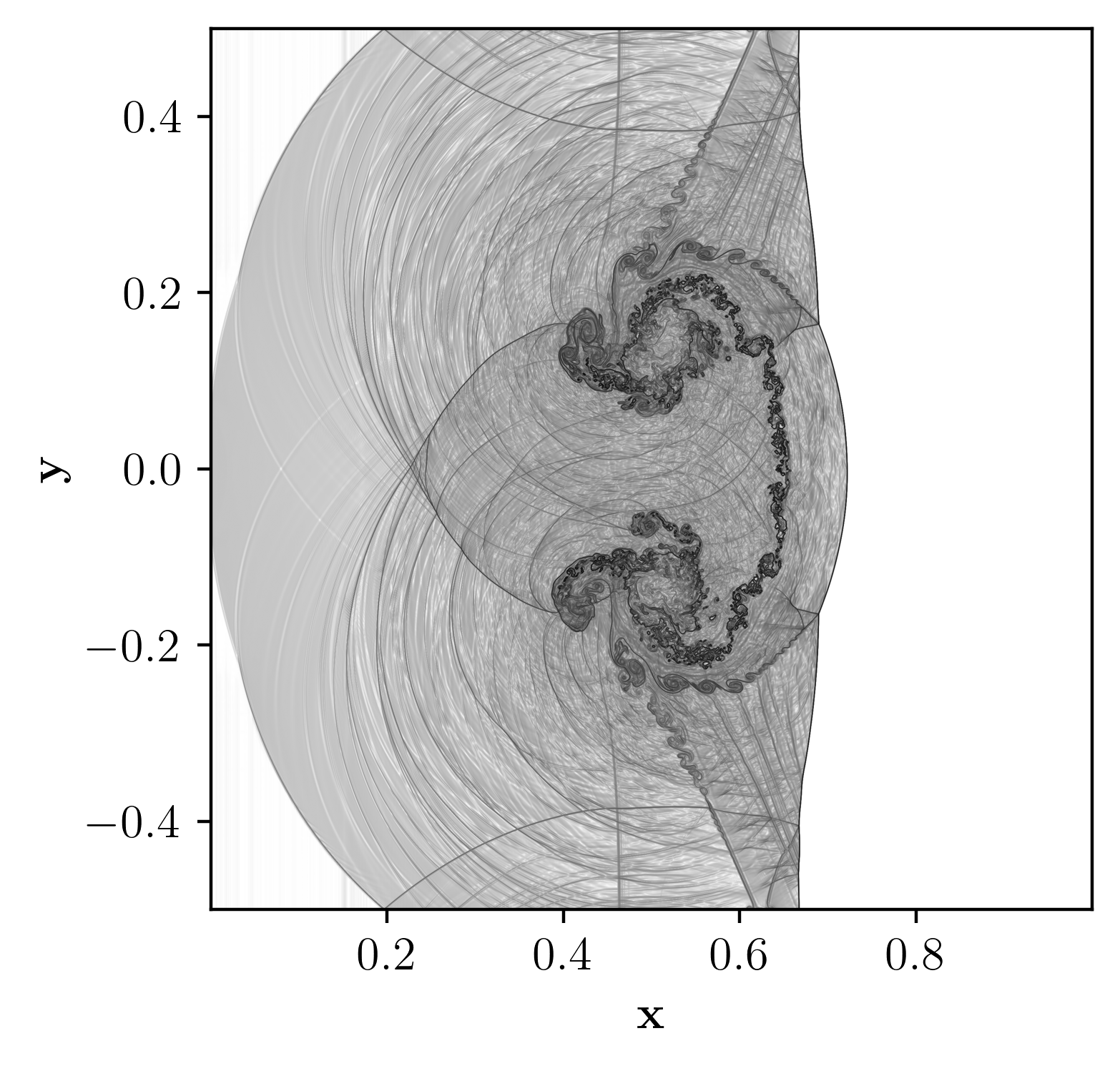}
\label{fig:teno_SB-LLFM}}
\caption{Shock-Bubble interaction for example \ref{ex:SB} using various schemes.}
\label{fig_allbubble}
\end{figure}

This test simulates the interaction of a planar shock with a cylindrical helium bubble embedded in air, a widely studied configuration in compressible multiphase flow. The problem is cast in a reference frame moving with the shock so that the shock is stationary, and a uniform free-stream velocity drives the flow. The helium bubble and the surrounding air initially have the same pressure and velocity, making this a pure interface-capturing test in the presence of shock-induced deformation, transmitted waves, and diffracted shocks at the bubble periphery. The final simulation time is $t_{\mathrm{end}} = 0.15$. The helium cylinder has radius $R = 0.15$ and is centered at $(x_c, y_c) = (0.25, 0)$. The planar shock is initially located at $x = 0.05$. The initial conditions are:
\begin{equation}\label{eq:tc9}
(\alpha_1\rho_1,\;\alpha_2\rho_2,\;u,\;v,\;p,\;\alpha_1) = \begin{cases}
\left(\frac{216}{41},\;0,\;\frac{1645}{286}-3,\;0,\;\frac{251}{6},\;1\right) & x < 0.05\;\text{(post-shock air)},\\
(0,\;0.138,\;-3,\;0,\;1,\;0) & (x-0.25)^2+y^2 \leq 0.15^2\;\text{(helium)},\\
(1,\;0,\;-3,\;0,\;1,\;1) & \text{otherwise (pre-shock air)},
\end{cases}
\end{equation}
with fluid properties $\gamma_1 = 1.4$, $\pi_1 = 0$ for air and $\gamma_2 = 1.648$, $\pi_2 = 0$ for helium. Figure~\ref{fig_allbubble} compares the density gradient (numerical Schlieren) fields computed by the Wave-MUSCL-FC and Wave-MUSCL-SC schemes on an $1600\times1600$ grid against a high-resolution reference, computed on a grid size of $2400^2$. Both formulations accurately capture the transmitted shock accelerating through the helium bubble, the regular and Mach reflection structures on the upstream side, and the diffracted shock wrapping around the bubble. The two formulations yield visually indistinguishable results at this resolution, consistent with the theoretical equivalence established in Section~\ref{sec:comparison}.

No spurious pressure or density oscillations are observed at the air–helium interface, confirming that the eigensystem accurately accounts for the $\gamma$ mismatch between air and helium throughout the shock-interaction process. Additionally, employing a central scheme for shear waves does not induce oscillations, even in inviscid scenarios, as demonstrated in single-species cases~\cite{hoffmann2024centralized,chamarthi2026wave}. The standard primitive shear characteristic is not the same density-weighted momentum-space quantity reconstructed by the FC and SC bases in variable-density flow; earlier primitive treatments therefore used viscous or sensor-controlled settings~\cite{batchelor1967introduction,meng2018numerical,chamarthi2025physics,chamarthi2025wave}. In contrast, the FC and SC shear characteristics provide the structural decoupling used here in inviscid conditions. Readers can refer to~\cite{chamarthi2026wave}. For this test case, the HLL Riemann solver is used \cite{harten1983upstream}.

\begin{example}{Underwater explosion}\label{underoos}
\end{example}

\begin{figure}[H]
\centering\offinterlineskip
\subfigure[Wave-MUSCL-FC, with THINC]
    {\includegraphics[width=0.33\textwidth]{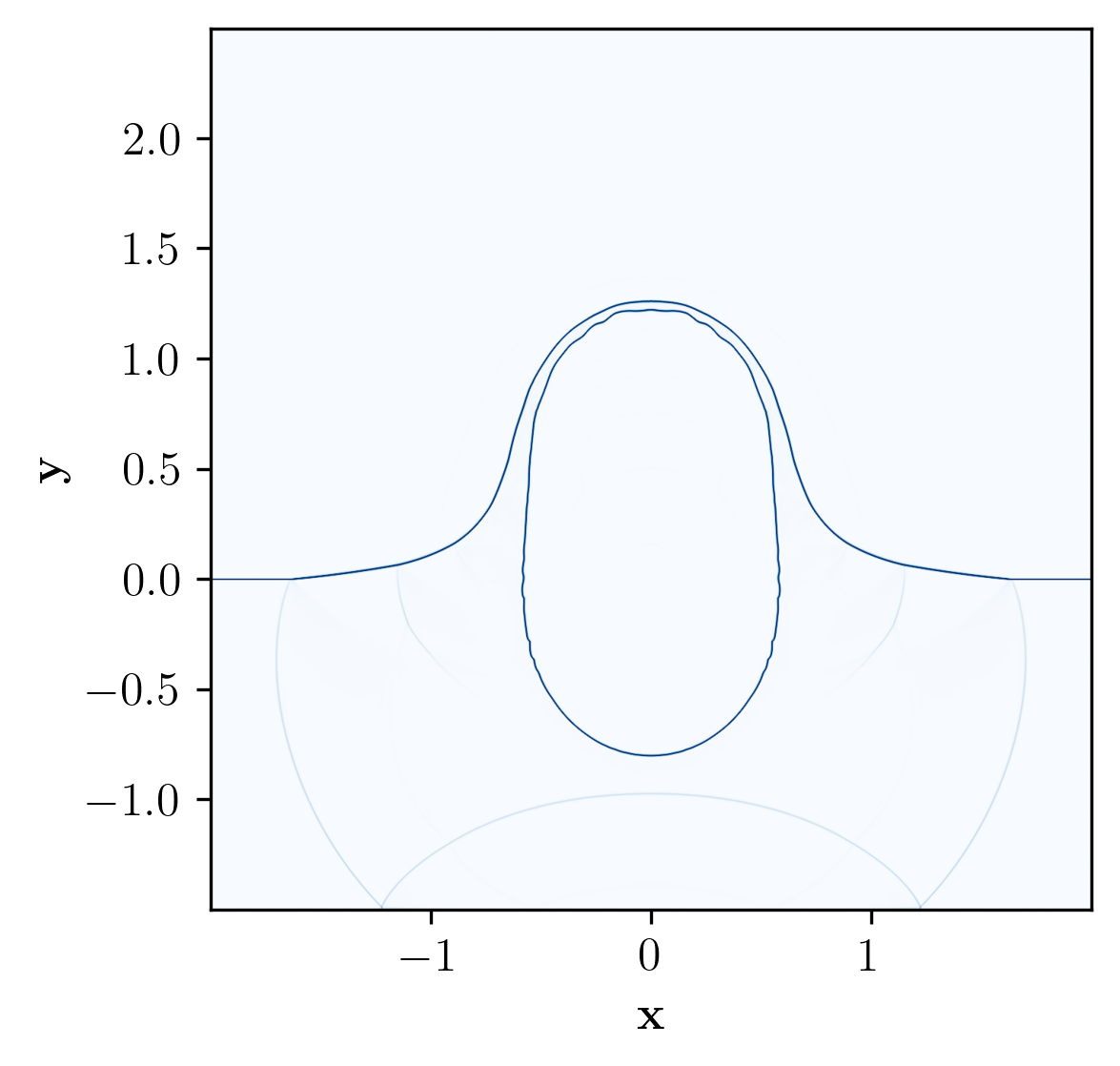}
    \label{fig:uw_muscl}}
\subfigure[Wave-MUSCL-SC, with THINC]
    {\includegraphics[width=0.33\textwidth]{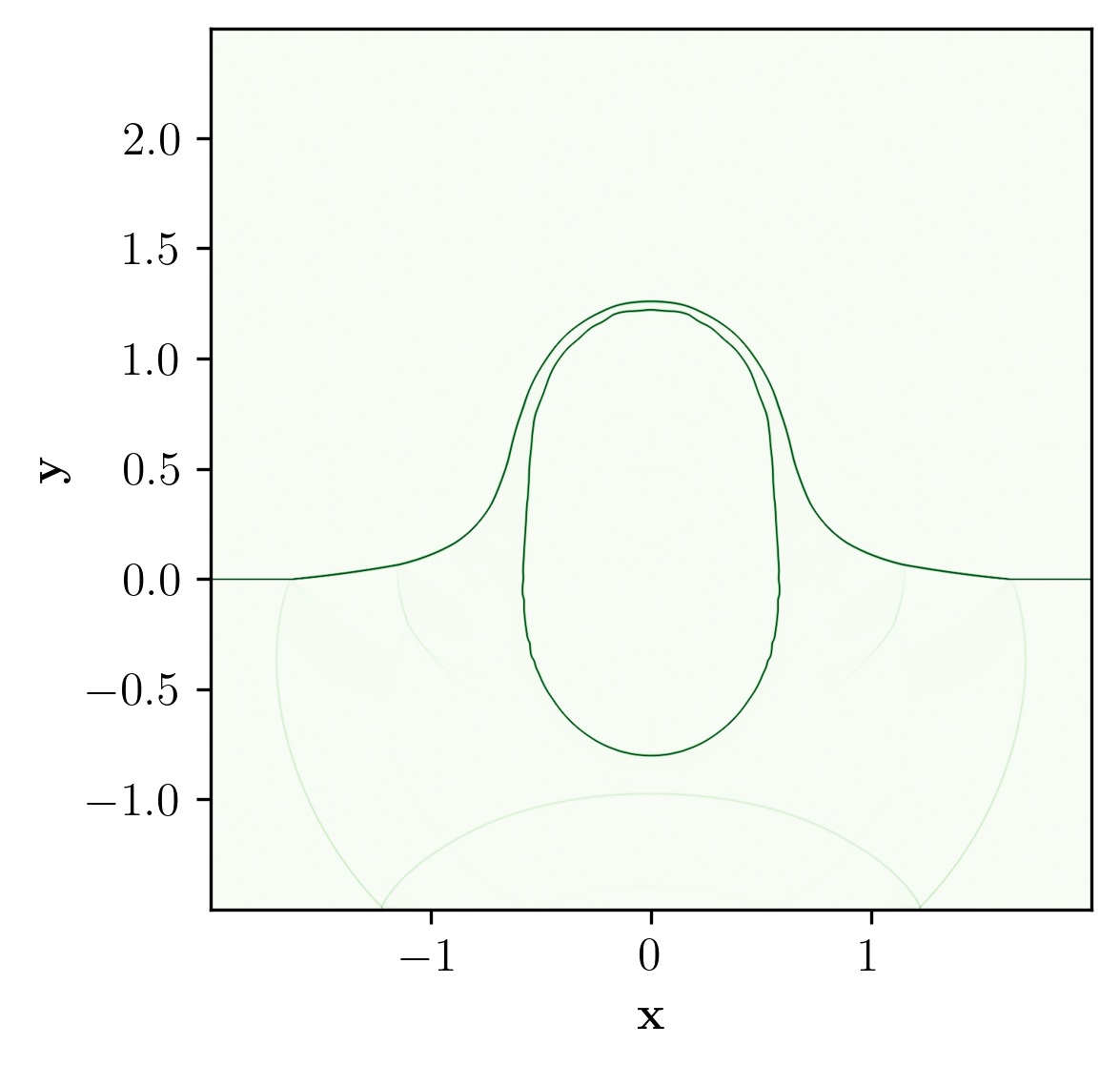}
    \label{fig:uw_weno}}
\subfigure[Wave-MUSCL-SC, with central scheme for shear wave]
    {\includegraphics[width=0.33\textwidth]{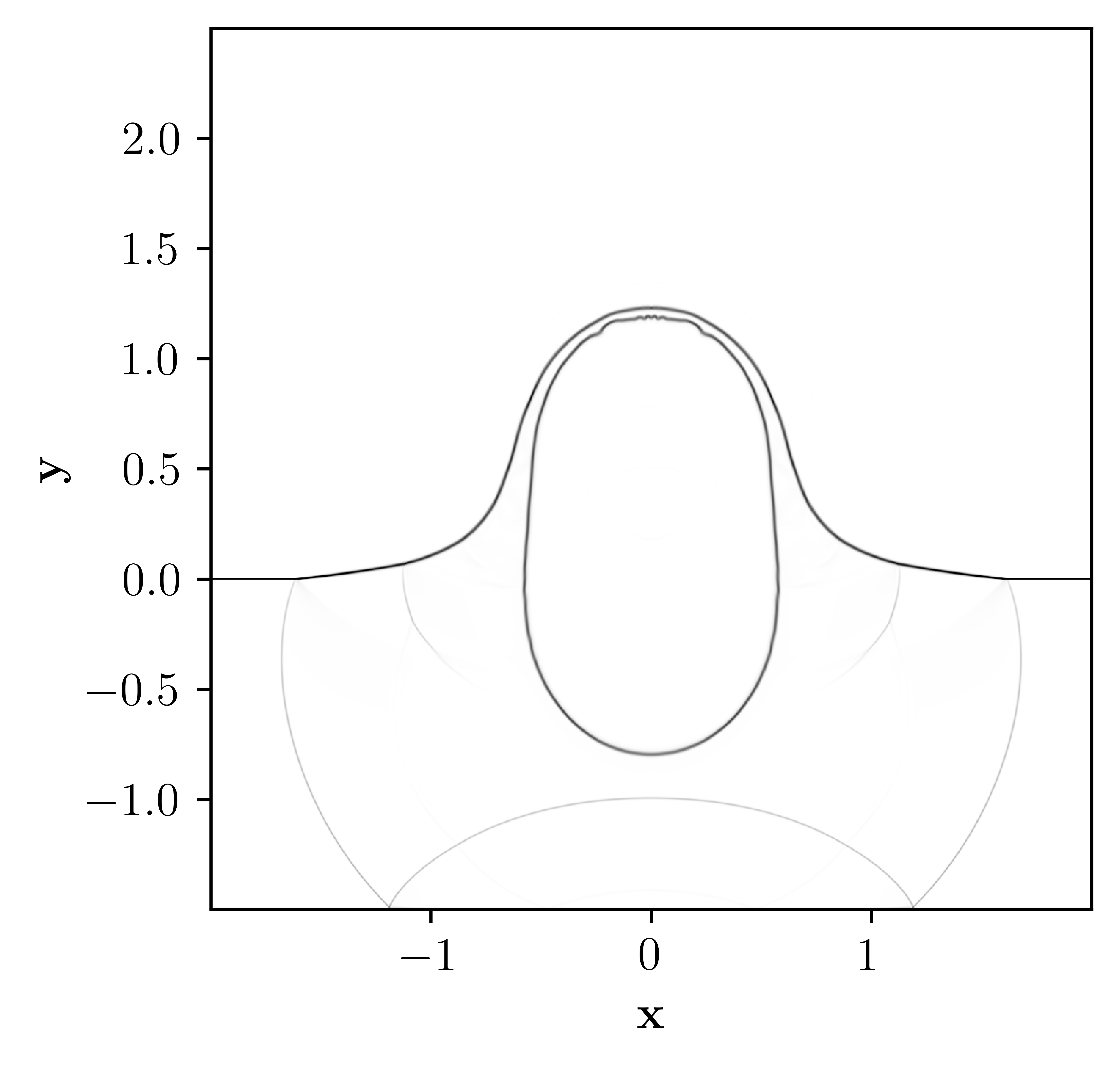}
    \label{fig:uw_cen_SC}}
\subfigure[Wave-MUSCL-FC, with central scheme for shear wave]
    {\includegraphics[width=0.33\textwidth]{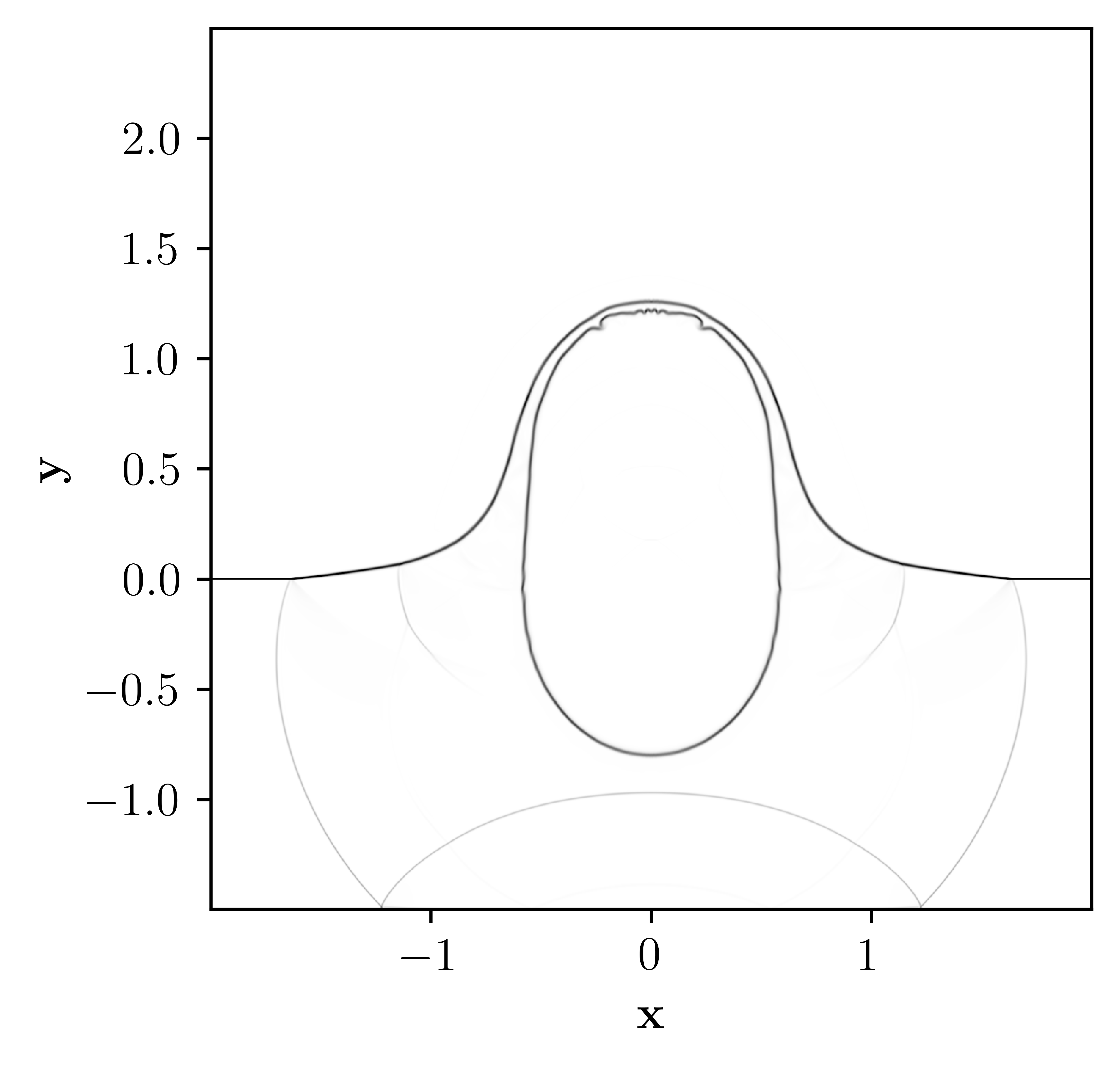}
    \label{fig:uw_cen_FC}}
\caption{Nonlinear function of normalised density gradient magnitude, 
$\phi = \exp\!\left(|\nabla\rho|/|\nabla\rho|_{\max}\right)$, computed by 
various schemes for the underwater explosion problem, Example~\ref{underoos}, 
on a $1200\times1200$ grid at $t = 0.19$. Contours are from 1.0 to 1.7.}
\label{fig:under00}
\end{figure}

This test case investigates the explosion of a highly compressed cylindrical air bubble in water beneath a free surface, following the configuration studied by Deng et al.~\cite{deng2018high} and Shukla, Pantano, and Freund~\cite{shukla2010interface}. It combines three distinct material regions (air above, water below, and a compressed bubble), large pressure ratios, a free surface subject to complex reflection and transmission, and fine-scale interfacial dynamics, including water-jet formation. The computational domain spans $(x,y) \in [-2,2]\times[-1.5,2.5]$, with reflective boundary conditions. The air--water free surface lies at $y = 0$, and the center of the compressed air bubble is at $(x_c, y_c) = (0, -0.3)$ with radius $r = 0.12$. The initial conditions are:
\begin{equation}\label{eq:tc10}
(\alpha_1\rho_1,\;\alpha_2\rho_2,\;u,\;v,\;p,\;\alpha_1) = \begin{cases}
(0,\; 1.225,\; 0,\; 0,\; 1.01325\times10^5,\; \epsilon) & y > 0\;\text{(air)},\\
(0,\; 1250,\; 0,\; 0,\; 10^5,\; \epsilon) & r < 0.12\;\text{(compressed gas)},\\
(1000,\; 0,\; 0,\; 0,\; 1.01325\times10^5,\; 1-\epsilon) & \text{otherwise (water)},
\end{cases}
\end{equation}
where $\epsilon = 10^{-8}$, with fluid properties $\gamma_1 = 1.4$, $\pi_1 = 0$ for air and $\gamma_2 = 4.4$, $\pi_2 = 6\times10^3$ for water. Simulations are performed on a $1200\times1200$ grid up to $t_{\mathrm{end}} = 0.19$. Figure~\ref{fig:under00} shows the normalised density gradient magnitude $\phi = \exp(|\nabla\rho|/|\nabla\rho|_{\max})$ computed by the Wave-MUSCL-FC and Wave-MUSCL-SC schemes with THINC, and by the corresponding central-scheme variants for the shear wave. Both formulations capture the transmitted shock propagating outward through the water, the reflected rarefaction wave retreating into the bubble interior, and the complex free-surface deformation. The thin water bridge separating the expanding bubble from the surrounding water remains intact in both simulations, a feature that is sensitive to the amount of numerical dissipation applied at the interface~\cite{chamarthi2025wave}. The results obtained using central reconstruction for the shear wave with both FC and SC formulations are also shown. These remain free of spurious oscillations and correctly capture the overall wave structure; however, the material interfaces appear more diffuse compared to the THINC results, as the central scheme does not provide interface sharpening. This is consistent with the expected behaviour: the central scheme reduces dissipation for the shear characteristic but does not replace the role of THINC in compressing the volume-fraction profile.

\begin{example}{Interaction between $Ma$=2.4 shockwave and a water cylinder}\label{water-24}
\end{example}

\begin{figure}[H]
%\begin{halfspacing}
\centering
\subfigure[Experimental result \cite{sembian2016plane}.]{\includegraphics[width=0.9\textwidth]{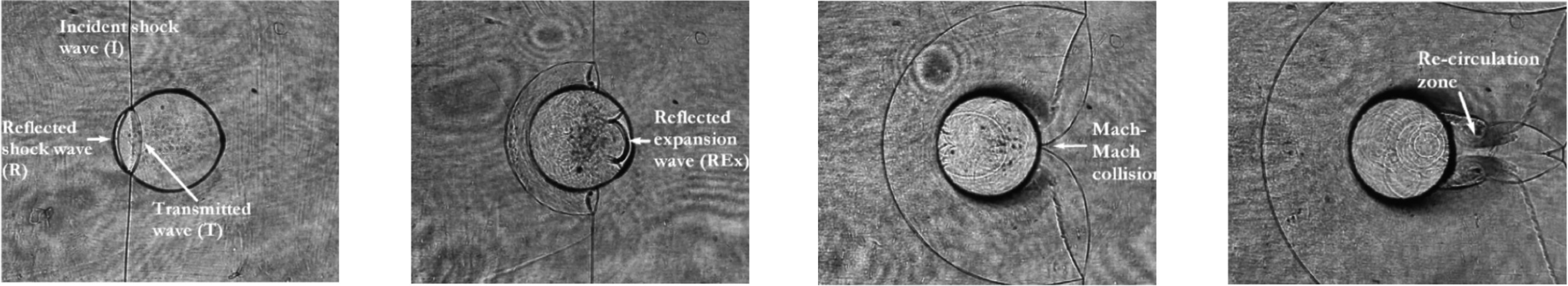}
\label{fig:water_exp}}
\subfigure[Numerical simulation using Wave-MUSCL-FC scheme with THINC.]{\includegraphics[width=0.9\textwidth]{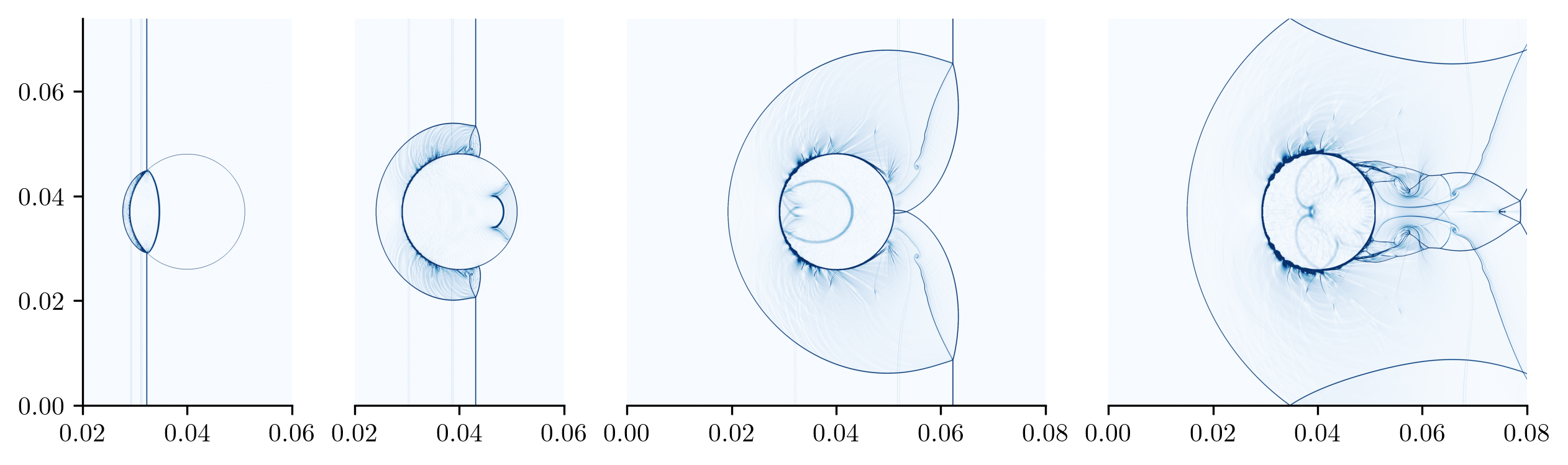}
\label{fig:water_num_fc}}
\subfigure[Numerical simulation using Wave-MUSCL-SC scheme with THINC.]{\includegraphics[width=0.9\textwidth]{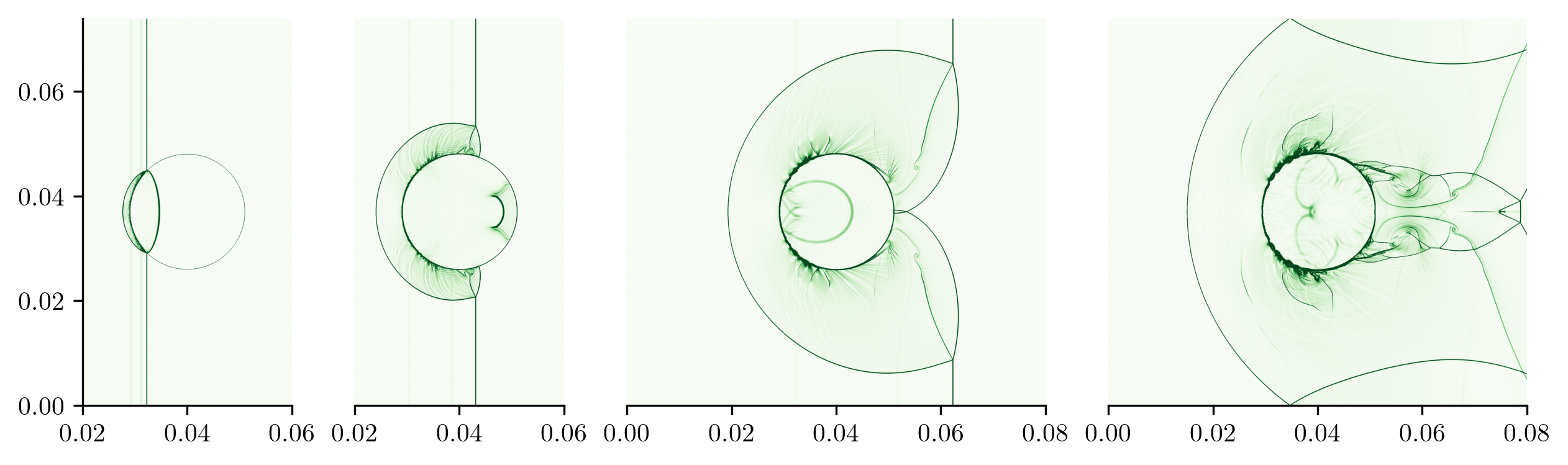}
\label{fig:water_num_sc}}
    \caption{\textcolor{black}{Shock--water cylinder interaction,} Example \ref{water-24}, comparison of numerical results with experiment.}
    \label{fig:w_exp2}
%\end{halfspacing}
\end{figure}

This test case examines the interaction of a Mach 2.4 planar shock with a water cylinder in air, replicating the experimental configuration of Sembian et al.~\cite{sembian2016plane}. It constitutes a challenging validation scenario for compressible multiphase solvers. The shock impinges on an air-water interface with an acoustic impedance ratio of approximately 3700, resulting in a highly asymmetric Mach reflection, a rapidly transmitted shock within the water cylinder, and a pair of vortices downstream of the cylinder generated by the converging Mach stems. Both the FC and SC formulations are evaluated against experimental Schlieren images. The computational domain is defined as $(x,y) \in [0, 0.111\,\mathrm{m}]\times[0, 0.074\,\mathrm{m}]$, with a water cylinder of diameter $D = 0.022$ m centered at $(0.04, 0.037)$ m and the incident shock initially positioned at $x = 0.029$ m. Reflective boundary conditions are imposed at the top and bottom boundaries. The simulation utilizes either $3072\times2048$ mesh or $4608 \times 3072$ mesh and is advanced to $t_{\mathrm{end}} = 67\,\mu$s. The initial conditions outside and inside the water cylinder are:
\begin{equation}\label{eq:tc11}
(\alpha_1\rho_1,\;\alpha_2\rho_2,\;u,\;v,\;p,\;\alpha_1) = \begin{cases}
\text{post-shock air (Rankine--Hugoniot, } M_s = 2.4\text{)} & \text{left of shock},\\
(\rho_{\mathrm{air}},\; 0,\; 0,\; 0,\; p_0,\; 1-\epsilon) & \text{pre-shock air},\\
(0,\; \rho_{\mathrm{water}},\; 0,\; 0,\; p_0,\; \epsilon) & \text{water cylinder},
\end{cases}
\end{equation}
with fluid properties $\gamma_{\mathrm{air}} = 1.4$, $\pi_{\mathrm{air}} = 0$ and $\gamma_{\mathrm{water}} = 6.12$, $\pi_{\mathrm{water}} = 3.43\times10^8$. Figure~\ref{fig:w_exp2} presents a comparison between experimental Schlieren images and density gradient contours computed using the Wave-MUSCL-FC and Wave-MUSCL-SC schemes.

Both formulations successfully reproduce the qualitative wave topology observed in experiments, including the reflected expansion wave in the upstream air, the rapidly propagating transmitted shock within the water cylinder, the Mach reflection developing at the cylinder surface as the angle of incidence increases, and the re-circulation zone downstream of the cylinder where the converging Mach stems generate a pair of contact waves. The sharp interface of the cylindrical water droplet is preserved throughout the simulation by the THINC reconstruction of the volume-fraction characteristic amplitude, thereby preventing the boundary smearing that would otherwise result from polynomial reconstruction alone. 

\begin{figure}[H]
%\begin{halfspacing}
\centering
\subfigure[Wave-MUSCL-FC scheme with central scheme for shear wave, higher resolution.]{\includegraphics[width=0.9\textwidth]{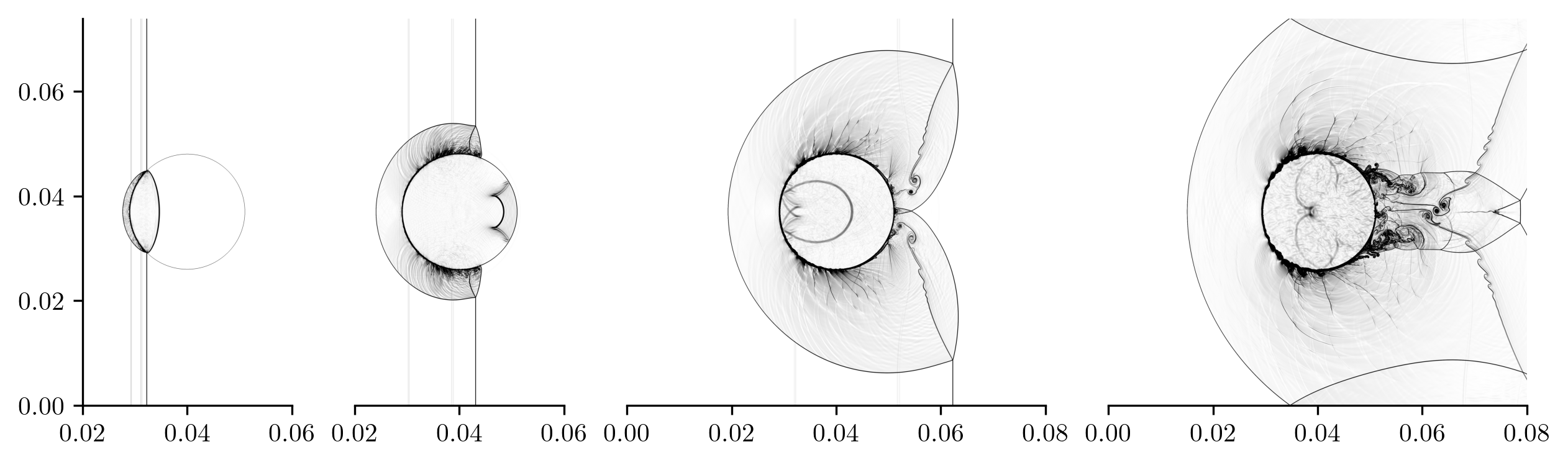}
\label{fig:water_num_fc_cen}}
\subfigure[Wave-MUSCL-SC scheme with central scheme for shear wave.]{\includegraphics[width=0.9\textwidth]{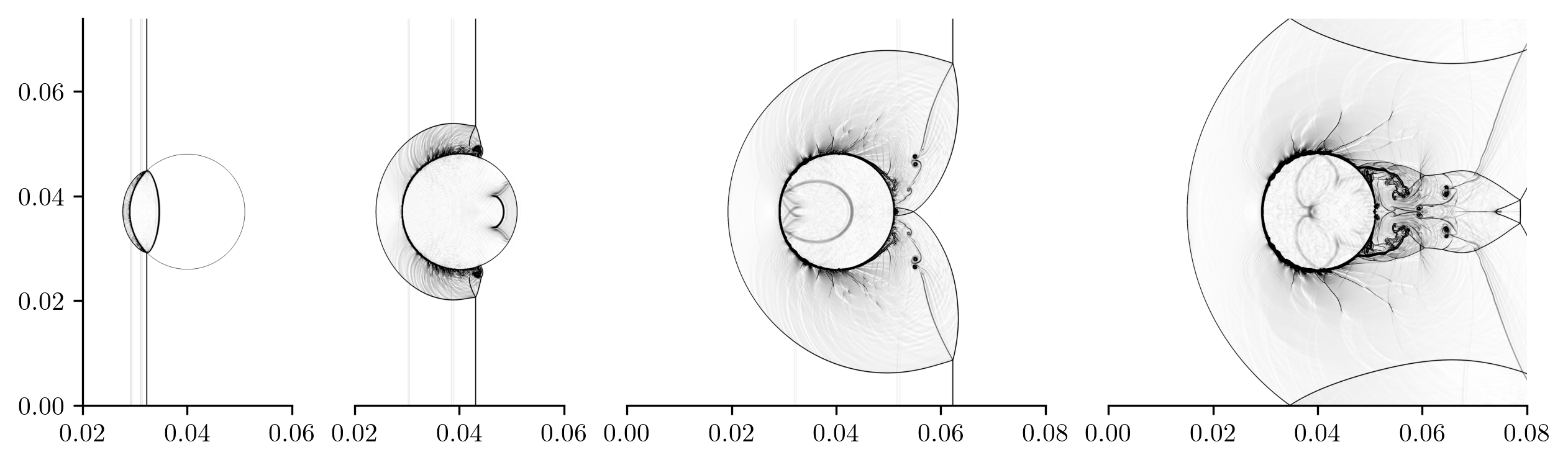}
\label{fig:water_num_sc_cen}}
    \caption{\textcolor{black}{Shock--water cylinder interaction,} Example \ref{water-24}.}
    \label{fig:w_cen}
%\end{halfspacing}
\end{figure}

Figure~\ref{fig:w_cen} shows the corresponding results obtained using central reconstruction for the shear wave with both FC and SC variable sets. Both formulations remain free of spurious oscillations and reproduce the experimental wave topology with comparable accuracy to the THINC results, confirming that central reconstruction of the shear characteristic is viable for gas-liquid configurations when FC or SC variables are employed. The FC result is computed on a finer grid of $4608\times3072$, as the simulation on the standard $3072\times2048$ grid failed to complete beyond $t = 4.8\,\mu$s due to a loss of positivity in the low-density wake region behind the cylinder. The SC formulation did not exhibit this issue on the standard grid.

\begin{example}{Shock interaction with multiple droplets with air pockets}\label{multiple-air}
\end{example}

We consider a Mach $2.4$ planar shock interacting with five water droplets arranged in a staggered two-column layout, each containing a concentric air pocket. This configuration extends the canonical single-droplet benchmark of Sembian et al.~\cite{sembian2016plane} to a multi-droplet setting and introduces an additional level of complexity through the presence of internal gas cavities. The arrangement generates inter-droplet shock reflections and transmitted wave interactions simultaneously at the outer liquid--gas interfaces and the inner pocket boundaries, providing a stringent test of the algorithm's ability to maintain interface sharpness and positivity across multiple co-existing material interfaces under strong compressive loading. The computational domain is $\Omega = [0, 0.111] \times [0, 0.074]$ m$^2$,  discretised on a uniform Cartesian grid of $N_x \times N_y = 3072 \times 2048$ cells. Five liquid droplets of radius $R_d = 6$ mm are arranged in a staggered  two-column layout, each containing a concentric air pocket of radius $R_p = 2$ mm. Column~1 ($x = 0.038$ m) holds three droplets centred at  $(x, y) = (0.038, 0.018)$, $(0.038, 0.037)$, and $(0.038, 0.056)$ m. Column~2 ($x = 0.058$ m) holds two droplets centred at $(0.058, 0.027)$ and $(0.058, 0.047)$ m. The initial conditions are:
\begin{equation}
(\rho_1\alpha_1,\, \rho_2\alpha_2,\, u,\, v,\, p,\, \alpha_1) =
\begin{cases}
(10^{-8},\ 3.758,\ 574.6,\ 0,\ 6.619\times10^{5},\ 10^{-8})
  & x < x_s, \\[4pt]
(10^{-8},\ 1.17,\ 0,\ 0,\ 1.013\times10^{5},\ 10^{-8})
  & (x,y) \in \mathcal{P}_k, \\[4pt]
(1000,\ 10^{-8},\ 0,\ 0,\ 1.013\times10^{5},\ 1-10^{-8})
  & (x,y) \in \mathcal{D}_k \setminus \mathcal{P}_k, \\[4pt]
(10^{-8},\ 1.17,\ 0,\ 0,\ 1.013\times10^{5},\ 10^{-8})
  & \text{otherwise},
\end{cases}
\end{equation}
where units are kg\,m$^{-3}$, m\,s$^{-1}$, and Pa respectively. The shock is initialised at $x_s = 0.032$ m, corresponding to a Mach $2.4$  shock propagating in air ($\gamma_2 = 1.4$, $\pi_2 = 0$) into a stationary ambient. The liquid droplet regions $\mathcal{D}_k$ and air pocket regions 
$\mathcal{P}_k$ are defined as
\begin{equation}
\mathcal{D}_k = \bigl\{(x,y) : (x - x_k)^2 + (y - y_k)^2 \leq R_d^2 \bigr\},
\quad
\mathcal{P}_k = \bigl\{(x,y) : (x - x_k)^2 + (y - y_k)^2 \leq R_p^2 \bigr\},
\end{equation}
for $k = 1,\ldots,5$, with centres $(x_k, y_k)$ as listed above. The liquid phase (water) is governed by the stiffened gas equation of state  with $\gamma_1 = 6.12$ and $\pi_1 = 3.43\times10^8$ Pa.

\begin{figure}[H]
    \centering
    \includegraphics[width=0.6\textwidth]{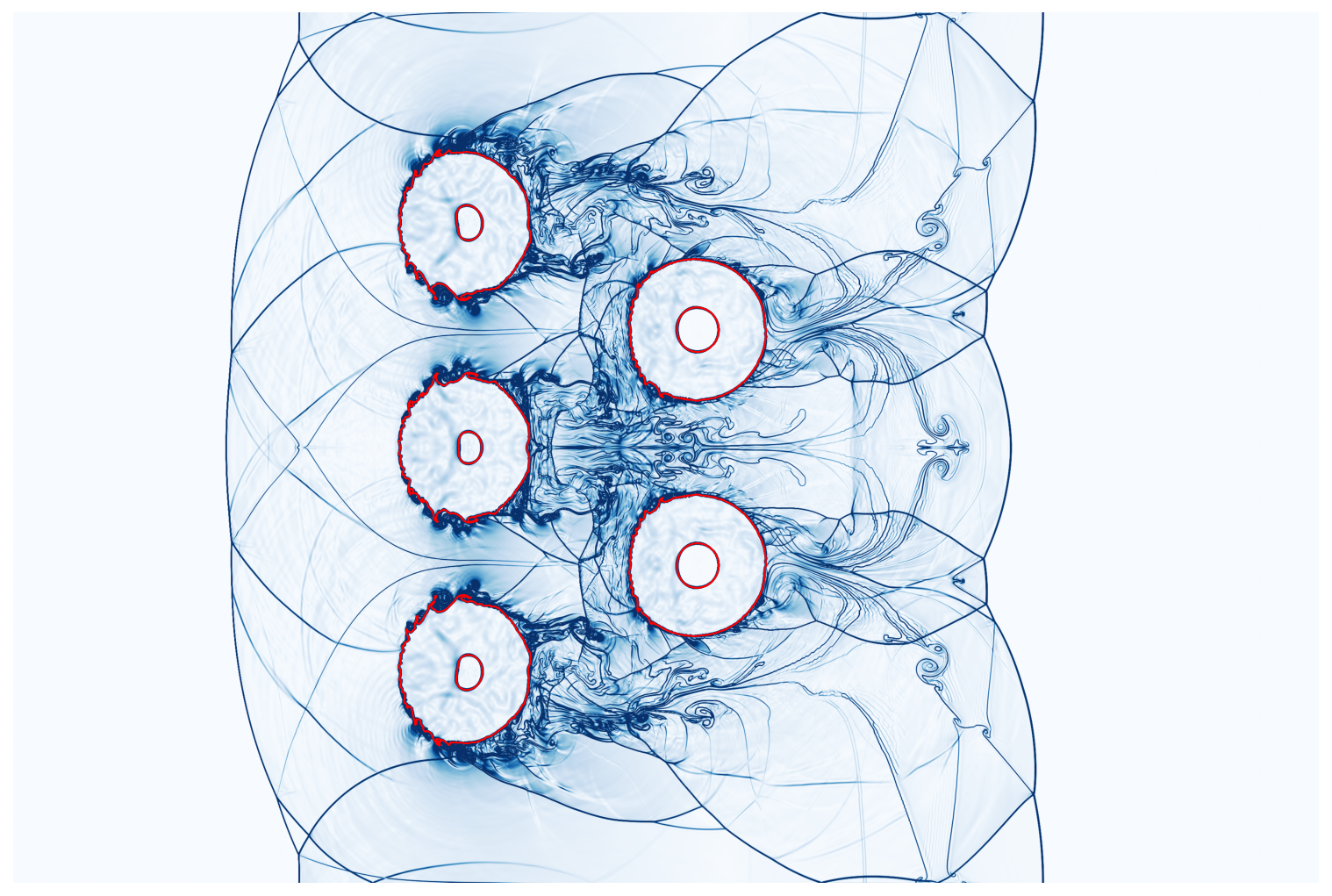}
    \caption{Density gradient contours for Example~\ref{multiple-air} using Wave-MUSCL-SC with THINC.}
    \label{fig:mul_thinc} 
\end{figure}

\begin{figure}[H]
    \centering
    \includegraphics[width=0.6\textwidth]{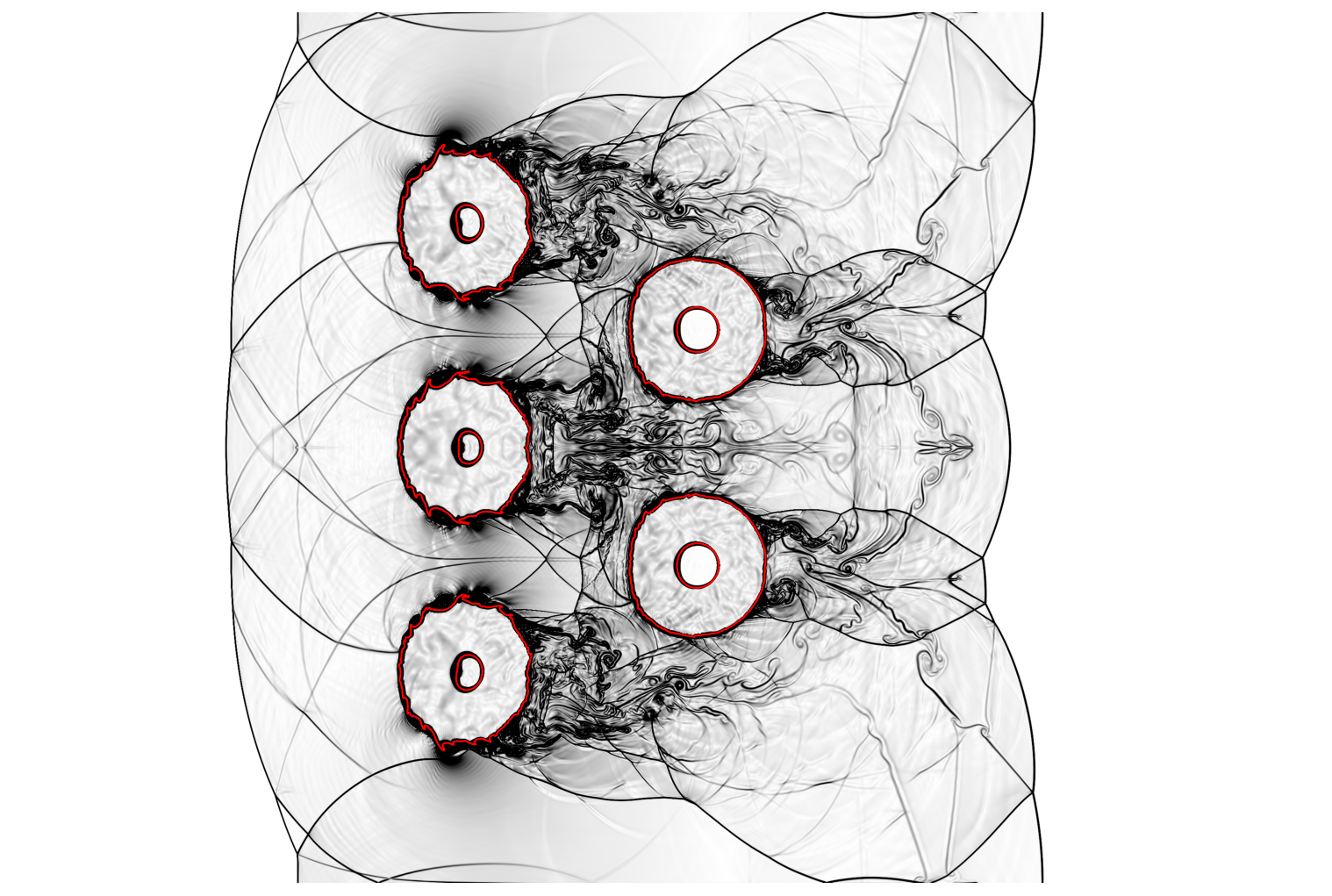}
    \caption{Density gradient contours for Example~\ref{multiple-air} using Wave-MUSCL-SC with central scheme for shear wave.}
    \label{fig:mul_cen}  
\end{figure}

Figures~\ref{fig:mul_thinc} and~\ref{fig:mul_cen} display numerical schlieren images of a Mach $2.4$ shock interacting with five water droplets at $t = 67\,\mu$s, generated using two distinct reconstruction strategies. The first strategy applies THINC interface sharpening to the volume fraction and phase densities, while the second utilizes a fourth-order central reconstruction for the shear characteristic wave in smooth-flow regions. Both methods effectively resolve inter-droplet shock reflections, transmitted waves within the liquid shells, and internal air-pocket boundaries across all five droplets. The Kelvin--Helmholtz instability developing on the downstream faces of the three front-column droplets is captured in both cases, with the central scheme yielding slightly richer vortical structures in the wake region, consistent with its lower numerical dissipation for smooth shear waves. When THINC sharpening is employed, the interfaces are more sharply defined; in particular, the internal air pockets appear significantly less diffuse than in the unsharpened case. Ideally, combining both strategies, as is done for pure gas--gas configurations, would provide optimal results. However, the current findings indicate that the central scheme for the shear characteristic wave remains effective in this complex multi-material context. The primary numerical challenge in this configuration is to maintain positivity of density and pressure in the low-pressure regions that develop in the wake behind each droplet. The FC formulation did not complete this test case beyond  $t = 4.7\,\mu$s for either reconstruction strategy, failing in the same  low-density wake regions; all results shown are therefore obtained with the SC formulation.

\begin{example}\label{ex:m6multibubble}{Mach-6 Shock Interaction with a Three-Species Bubble}
\end{example}

\Rthree{This inviscid test follows the Mach-6 air-shock interaction with a concentric R22--helium bubble of Pan et al.~\cite{pan2018conservative}. The domain is $(x,y)\in[0,0.356]\times[0,0.089]$, and the bubble is centred at $(x_c,y_c)=(0.150,0.0445)$ with helium-core and R22-shell radii $r_i=0.015$ and $r_o=0.030$, respectively, where $r=[(x-x_c)^2+(y-y_c)^2]^{1/2}$. The initial states are}
\begin{equation}\label{eq:m6multibubble}
\Rthree{
(\rho,u,v,p,\gamma)=
\begin{cases}
(5.268,\;5.752,\;0,\;41.83,\;1.400), & x<0.100 \quad \text{(post-shock air)},\\
(0.138,\;0,\;0,\;1.00,\;1.667), & r\leq r_i \quad \text{(helium)},\\
(3.154,\;0,\;0,\;1.00,\;1.249), & r_i<r\leq r_o \quad \text{(R22)},\\
(1.000,\;0,\;0,\;1.00,\;1.400), & \text{otherwise} \quad \text{(pre-shock air)}.
\end{cases}}
\end{equation}
\Rthree{Symmetry conditions are imposed at the upper and lower boundaries, with inflow and outflow conditions at the left and right boundaries. The calculation uses Wave-MP5-SC on an $8192\times2048$ grid to $t=0.03$. Acoustic and material characteristics are reconstructed with the MP5 upwind-biased and THINC procedures, respectively, while the SC shear characteristic is centrally reconstructed wherever the shock sensor is inactive. A positivity-preserving high-/low-order flux-blending approach of Wong et al.~\cite{wong2021positivity} is applied at every Runge--Kutta stage when required to retain positive partial densities and pressure and admissible volume fractions. Earlier inviscid three-species calculations with primitive/primitive-characteristic reconstruction used the MP admissibility criterion to restrict central treatment~\cite{chamarthi2025wave,chamarthi2025physics}. For a primitive tangential velocity $q$, the central candidate $q^C_{i+1/2}$ was accepted only when
\begin{equation}\label{eq:prior_mp_admissibility}
\left(q^C_{i+1/2}-q_i\right)
\left(q^C_{i+1/2}-q^{\rm MP}_{i+1/2}\right)
\leq 10^{-40},\qquad
q^{\rm MP}_{i+1/2}=q_i+\operatorname{minmod}\!\left(q_{i+1}-q_i,\,4(q_i-q_{i-1})\right);
\end{equation}
otherwise, the nonlinear MP reconstruction was used, so the earlier treatment was only conditionally central. In the present SC calculation, no MP admissibility test is applied to the shear state: sixth-order central reconstruction is used outside shock-marked regions and MP5 within them. This difference follows from the SC shear amplitude derived above rather than direct central reconstruction of a primitive tangential velocity.}

\begin{figure}[H]
    \centering
    \subfigure[$t=5.0\times10^{-3}$]{\includegraphics[width=0.29\textwidth]{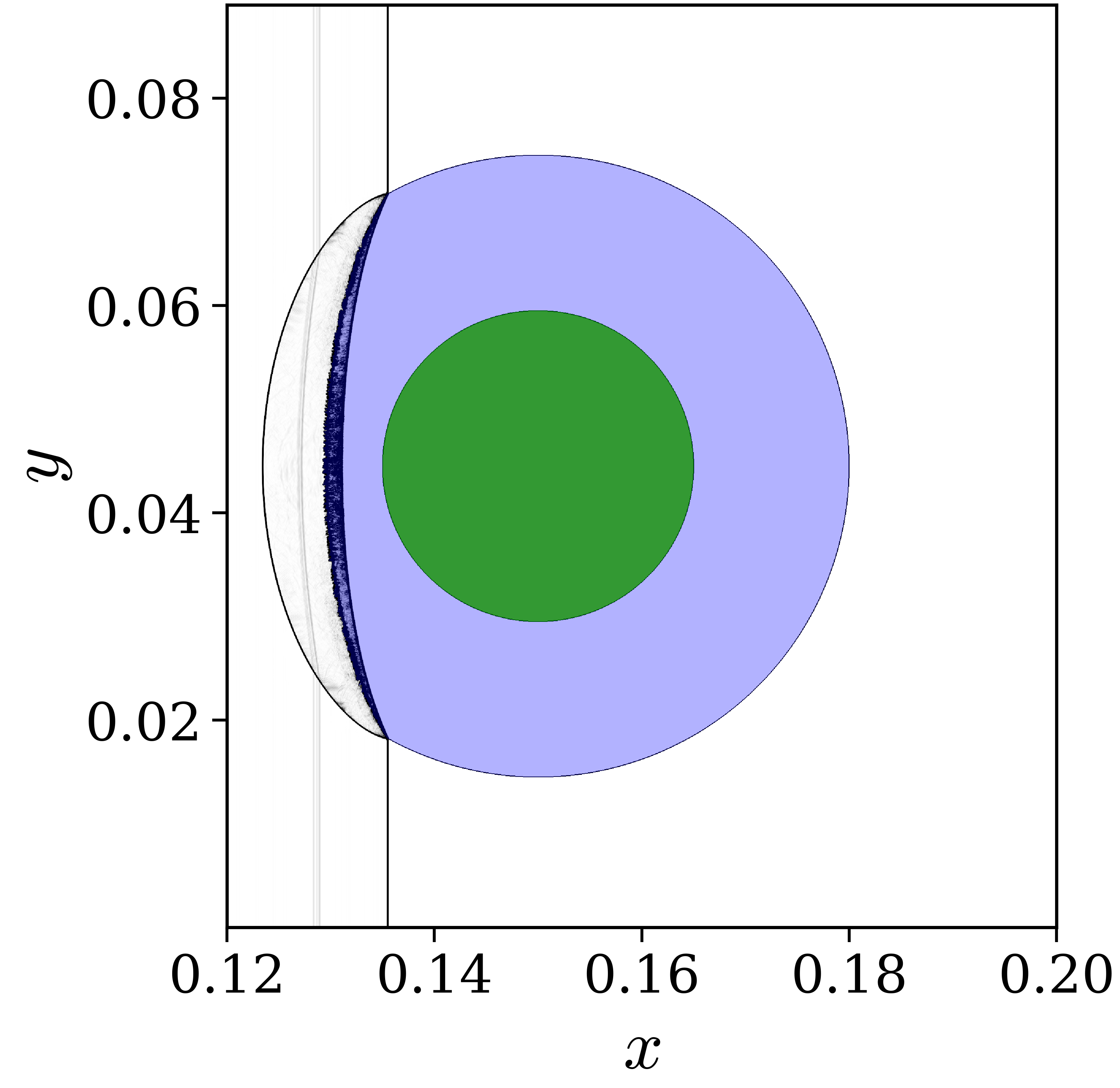}
    \label{fig:m6multibubble-t005}}
    \subfigure[$t=1.0\times10^{-2}$]{\includegraphics[width=0.29\textwidth]{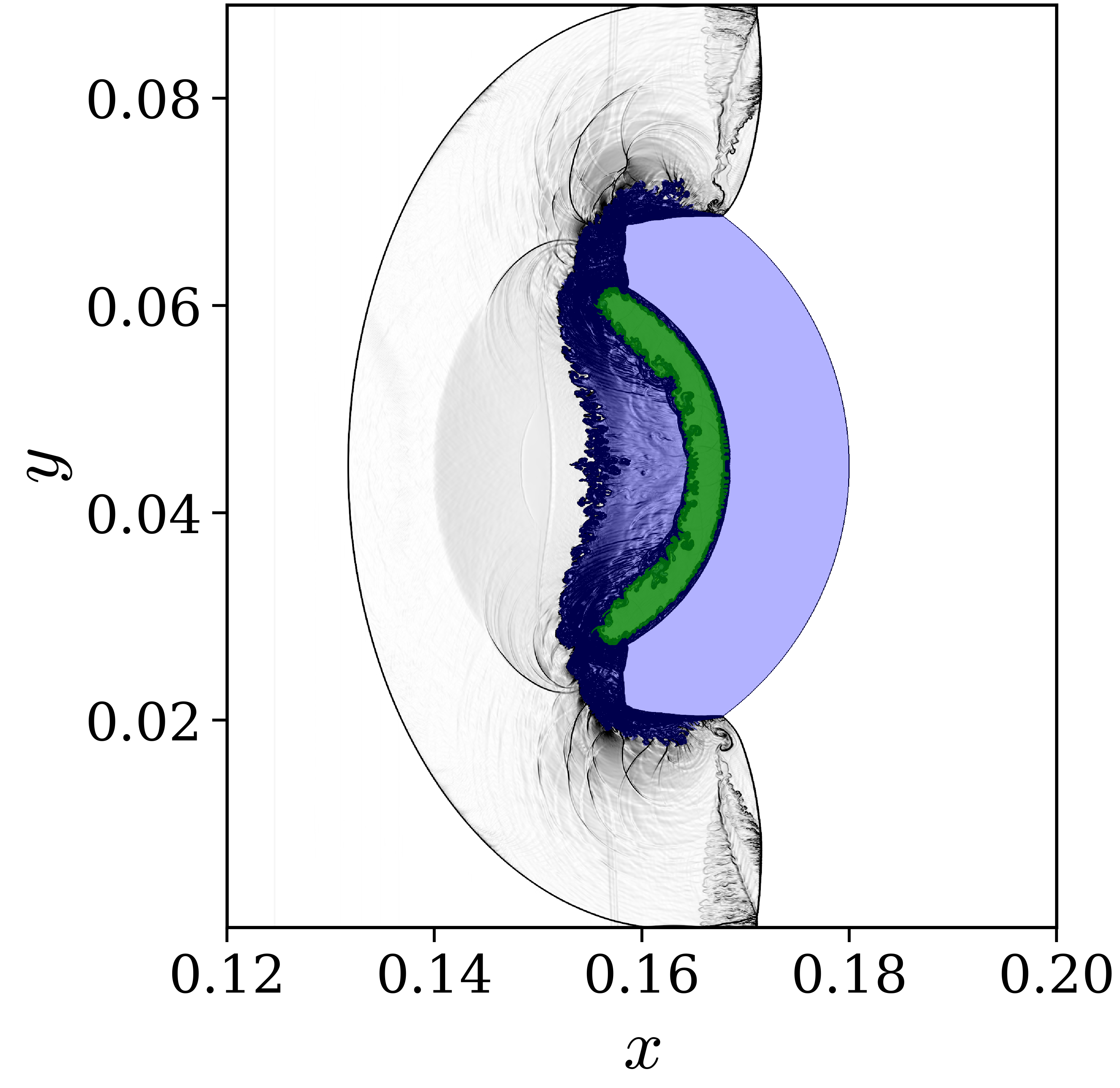}
    \label{fig:m6multibubble-t010}}
    \subfigure[$t=1.5\times10^{-2}$]{\includegraphics[width=0.29\textwidth]{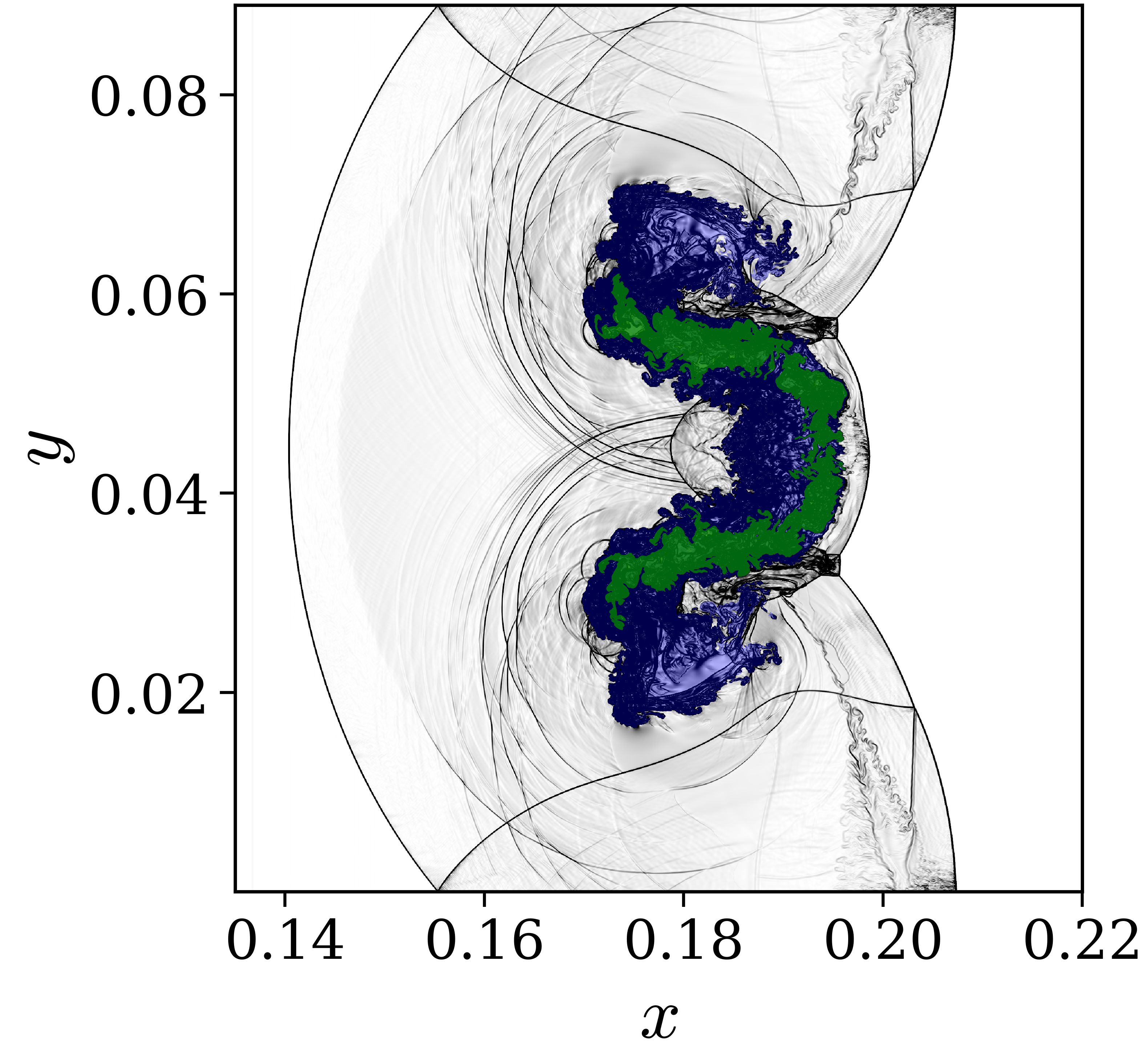}
    \label{fig:m6multibubble-t015}}
    \caption{\Rthree{Numerical Schlieren for the Mach-6 three-species bubble interaction, Example~\ref{ex:m6multibubble}, computed with Wave-MP5-SC on an $8192\times2048$ grid. Green denotes helium and blue denotes R22.}}
    \label{fig:m6multibubble-early}
\end{figure}

\begin{figure}[H]
    \centering
    \subfigure[$t=2.0\times10^{-2}$]{\includegraphics[width=0.60\textwidth]{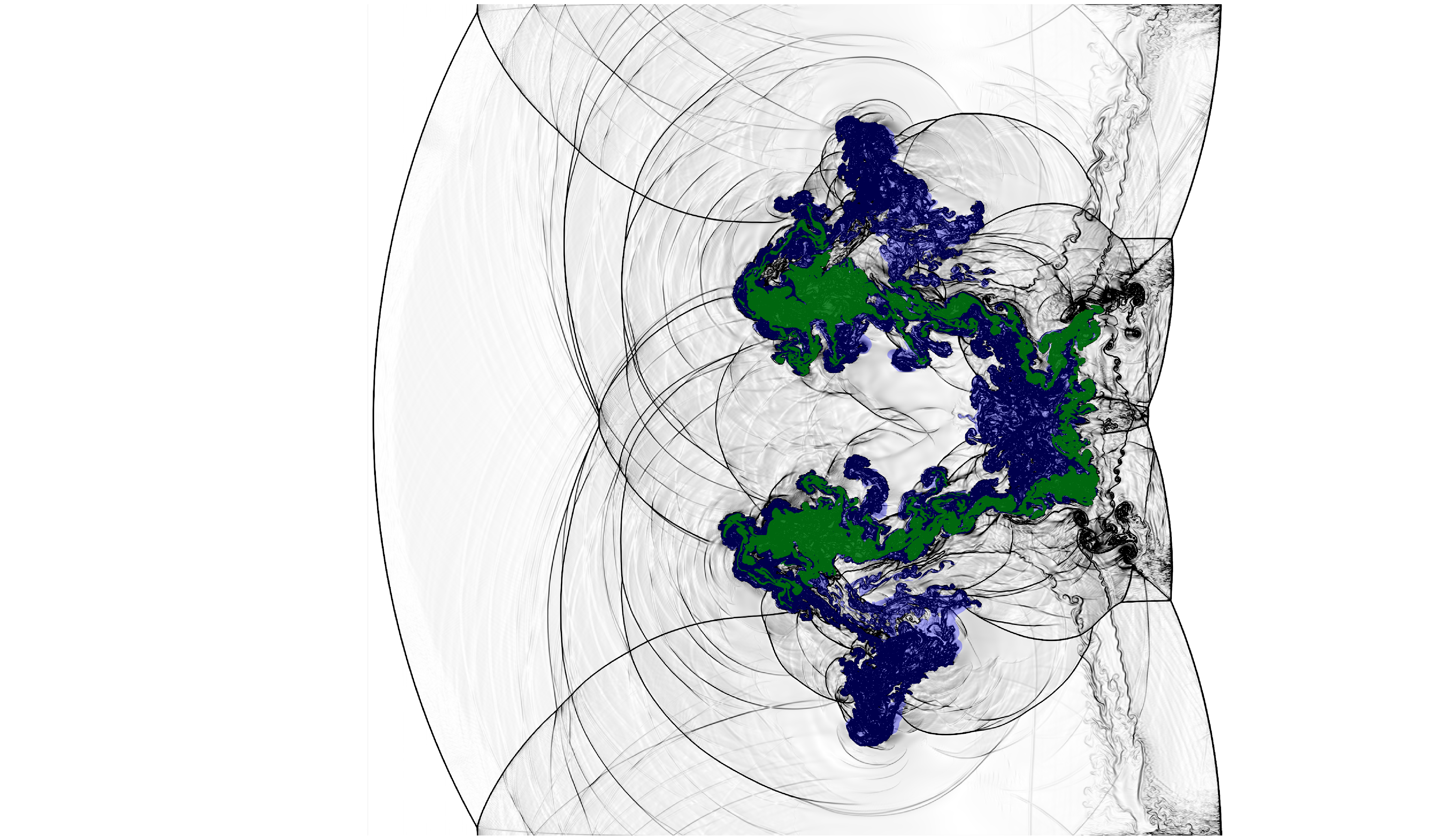}
    \label{fig:m6multibubble-t020}}\\
    \subfigure[$t=2.5\times10^{-2}$]{\includegraphics[width=0.60\textwidth]{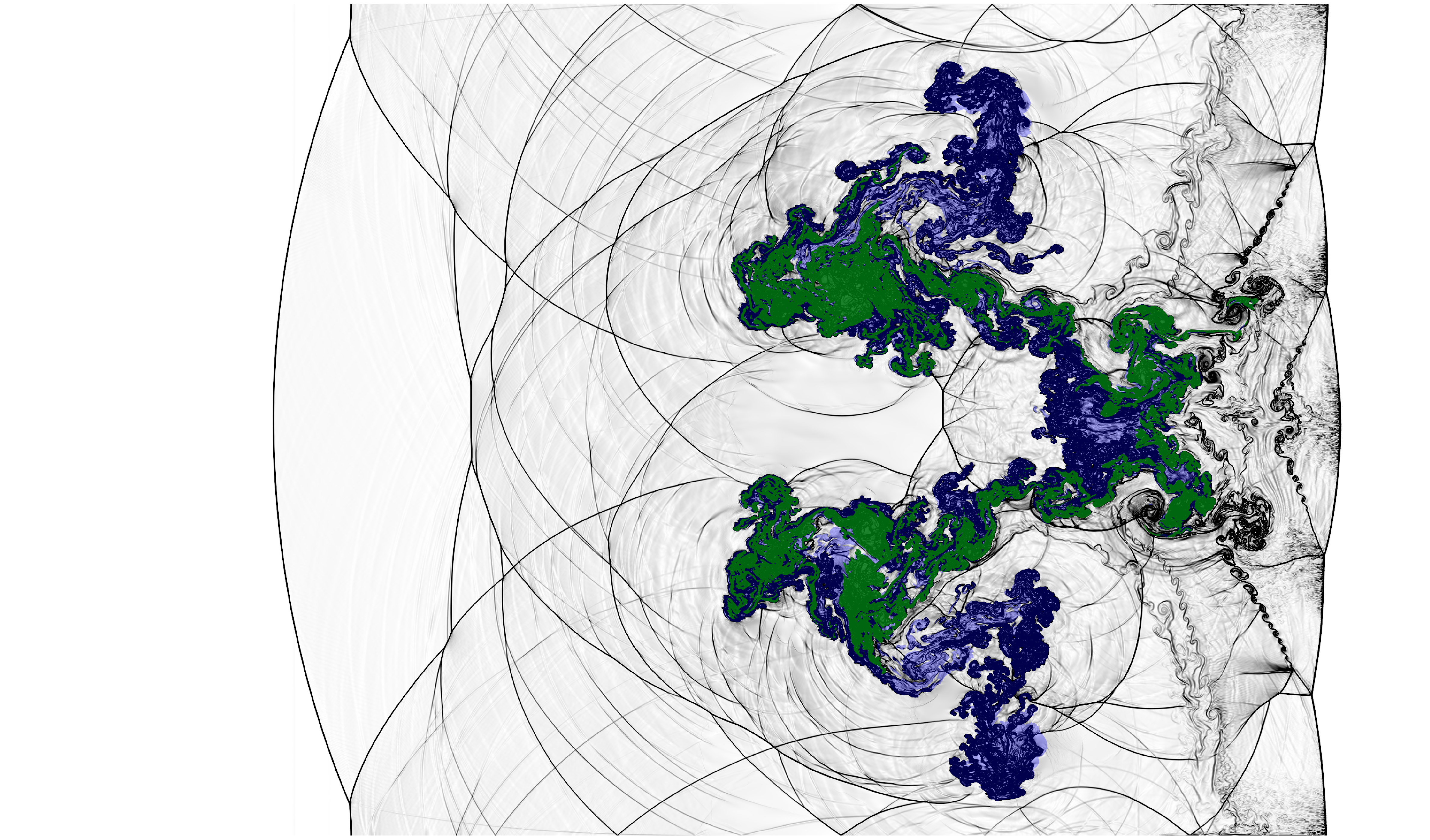}
    \label{fig:m6multibubble-t025}}\\
    \subfigure[$t=3.0\times10^{-2}$]{\includegraphics[width=0.60\textwidth]{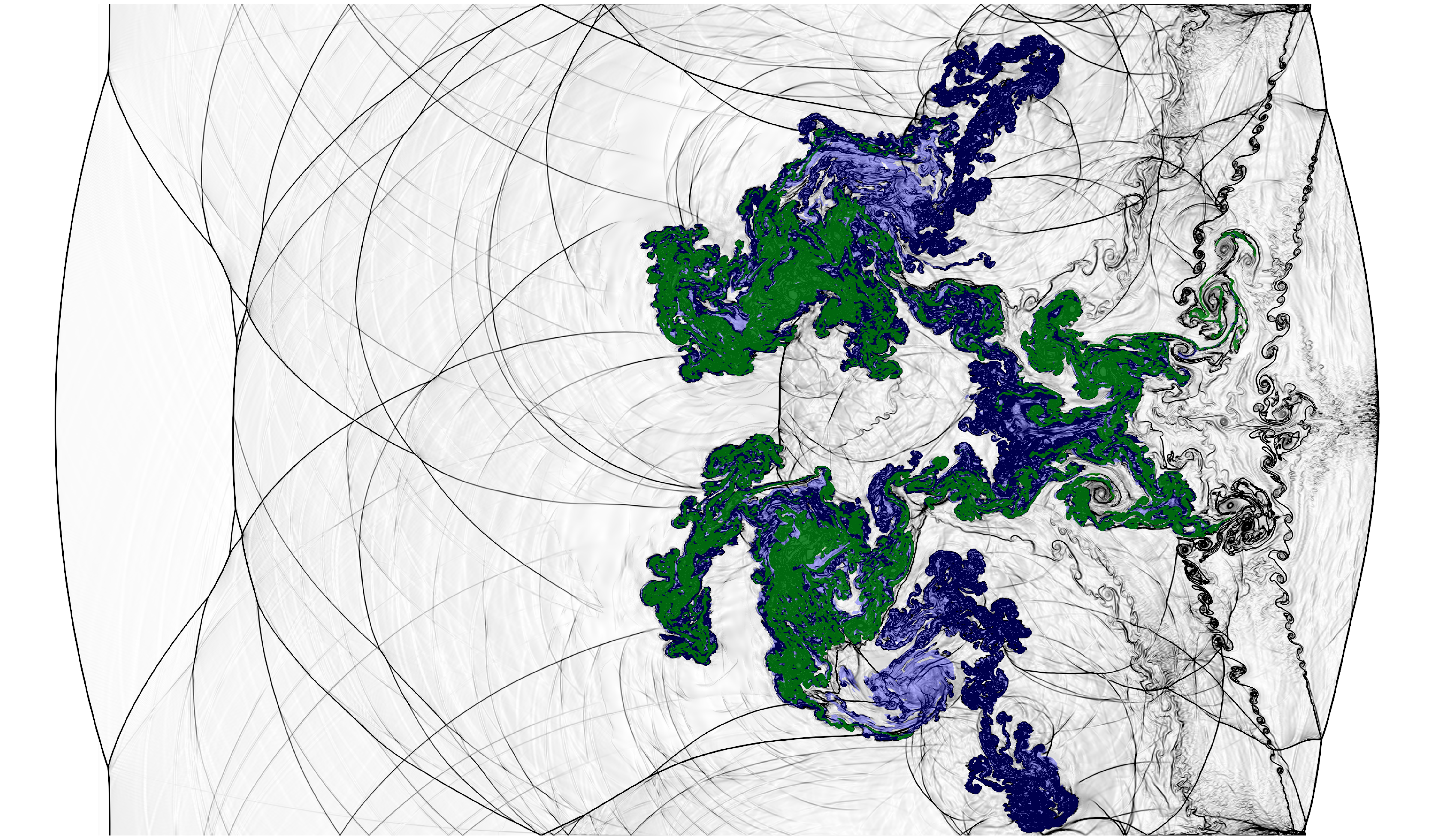}
    \label{fig:m6multibubble-t030}}
    \caption{\Rthree{Late-time numerical Schlieren for Example~\ref{ex:m6multibubble}. Green denotes helium and blue denotes R22.}}
    \label{fig:m6multibubble-late}
\end{figure}

\Rthree{Figures~\ref{fig:m6multibubble-early} and~\ref{fig:m6multibubble-late} show the early and later stages of the interaction. The incident shock refracts through the R22 shell, producing a concave transmitted wave; after reaching the helium core, the transmitted wave becomes convex because of the acoustic-impedance change. The later fields show the resulting deformation and roll-up of both material interfaces.}

\section{Conclusions}
\label{sec:conclusions}

The eigenstructure of the Allaire five-equation model has been derived for both fully conservative (FC) and semi-conservative (SC) variable sets, with explicit eigenvectors provided for one- and two-dimensional stiffened-gas flows. Both formulations are constructed to satisfy Abgrall's equilibrium condition: the FC formulation achieves this through the thermodynamic jump term $\Psi$ and its algebraically matched counterpart in the acoustic left eigenvectors, while the SC formulation accomplishes this via the structural zero in the pressure slot of the volume-fraction eigenvector. The principal finding is that characteristic-space reconstruction is a necessary condition for either eigensystem to preserve equilibrium at the discrete level. Reconstruction of any non-primitive variable in physical space introduces $\mathcal{O}(1)$ pressure and velocity errors, regardless of the variable set used. This result supports and generalizes the arguments of Johnsen and Colonius~\cite{johnsen2006implementation} and Coralic and Colonius~\cite{coralic2014finite}, demonstrating that characteristic-space reconstruction is essential for oscillation-free results and for equilibrium preservation at material interfaces, irrespective of the variable set employed.
 
A direct consequence of the eigenvector structure is that the shear wave is decoupled from all thermodynamic and interface fields in both FC and SC formulations. This decoupling enables the use of a central reconstruction scheme for the shear characteristic. The standard primitive shear characteristic reconstructs the tangential velocity itself, whereas the FC and SC shear characteristics reconstruct the corresponding momentum-space perturbation; these are not equivalent when density varies and must be assessed separately. This finding extends and validates the central shear-wave argument of~\cite{hoffmann2024centralized,chamarthi2026wave} from single-species to multiphase flows. The Kelvin--Helmholtz roll-up observed in the triple-point and gas-liquid benchmarks confirms that central reconstruction recovers vortical structures that are suppressed by fully upwind schemes. One- and two-dimensional benchmarks further demonstrate that FC and SC characteristic reconstruction are both viable options for the simulations. The choice between THINC and central reconstruction for the shear wave in gas-liquid configurations depends on the physical quantity of interest. THINC provides sharper material interfaces and is preferable when interface resolution is the primary concern. Central reconstruction reduces dissipation of the shear characteristic and is better suited for flows where mixing and vortical structures are of interest. Both options are viable within the FC and SC frameworks, and the selection is left to the practitioner based on the application. However, SC provides greater simplicity, robustness, and equation-of-state generality. 

These results indicate that the wave-appropriate reconstruction framework for compressible flows is valid across a wide range of flow physics~\cite{chamarthi2023wave,hoffmann2024centralized,chamarthi2025wave,chamarthi2025physics,chamarthi2026wave}.

\appendix
\section{Supplementary Code}

\Rone{Two-dimensional solver implementations for selected calculations in this work are provided as supplementary material. The codes are written using NVIDIA Warp~\cite{macklin2022warp} and run on CUDA-capable GPU hardware. After installing the required dependency using \texttt{pip install warp-lang}, the supplied scripts reproduce the corresponding triple-point and three-species bubble calculations reported in this paper. For the $8192\times2048$ three-species calculation on two GPUs, the command is \texttt{python three\_species.py --nx 8192 --ny 2048 --gpus 2}. Run instructions and case-specific GPU requirements are included with the code.}

\bibliographystyle{elsarticle-num}
\bibliography{lasty_fixed-2}

\end{document}